\documentclass[11pt,preprint]{aastex}

\usepackage{epstopdf}

\shorttitle{{\it Sptizer} IRS Survey of NGC 1333}
\shortauthors{Arnold et al.}

\begin{document}

\title{ A {\it Spitzer} IRS Survey of NGC 1333: Insights into disk evolution from a very young cluster}
\author{L. A. Arnold\altaffilmark{1}, Dan M. Watson\altaffilmark{1}, K. H. Kim\altaffilmark{1}, 
P. Manoj\altaffilmark{1}, I. Remming\altaffilmark{1}, P. Sheehan\altaffilmark{1}, 
L. Adame\altaffilmark{2}, W. J. Forrest\altaffilmark{1}, E. Furlan\altaffilmark{3,6}, 
E. Mamajek\altaffilmark{1}, M. McClure\altaffilmark{2}, C. Espaillat\altaffilmark{4,7}, 
K. Ausfeld\altaffilmark{5}, V. A. Rapson\altaffilmark{5} }

\altaffiltext{1}{Dept. of Physics and Astronomy, University of Rochester, Rochester, NY 14627; laa@pas.rochester.edu, dmw@pas.rochester.edu} 
\altaffiltext{2}{Department of Astronomy, The University of Michigan, 500 Church St., Dennison Bldg., Ann Arbor, MI 48109}
\altaffiltext{3}{Jet Propulsion Laboratory, California Institute of Technology, Mail Stop 264-723, 4800 Oak Grove Drive, Pasadena, CA 91109}
\altaffiltext{4}{Harvard-Smithsonian Center for Astrophysics, 60 Garden Street, MS-78, Cambridge, MA 02138, USA}
\altaffiltext{5}{Center for Imaging Science, Rochester Institute of Technology, 54 Lomb Memorial Drive, Rochester, NY 14623, USA}
\altaffiltext{6}{{\it Spitzer} fellow}
\altaffiltext{7}{NSF Astronomy and Astrophysics Postdoctoral Fellow}

\begin{abstract}
We report on the $\lambda = 5-36 \mu$m {\it Spitzer} Infrared Spectrograph spectra of 79 young 
stellar objects in the very young nearby cluster NGC 1333. NGC 1333's youth enables the study of 
early protoplanetary disk properties, such as the degree of settling as well as the formation of 
gaps and clearings. We construct spectral energy distributions (SEDs) using our IRS data as well 
as published photometry and classify our sample into SED classes. Using ``extinction-free" spectral
 indices, we determine whether the disk, envelope, or photosphere dominates the spectrum. We analyze
 the dereddened spectra of objects which show disk dominated emission using spectral indices and 
properties of silicate features in order to study the vertical and radial structure of 
protoplanetary disks in NGC 1333. At least nine objects in our sample of NGC 1333 show signs of 
large (several AU) radial gaps or clearings in their inner disk. Disks with radial gaps in NGC 1333
 show more-nearly pristine silicate dust than their radially continuous counterparts. We compare 
properties of disks in NGC 1333 to those in three other well studied regions, Taurus-Auriga, 
Ophiuchus and Chamaeleon I, and find no difference in their degree of sedimentation and dust 
processing. 
\end{abstract}


\section{Introduction} \label{intro}
NGC 1333 was first identified as an optical reflection nebula at the western end of the Perseus 
molecular cloud. The moniker NGC 1333 is also used to identify the molecular cloud and young 
stellar cluster associated with the optical reflection nebula along with the dark cloud L1450 
\citep{herbig72, herbig74, strom76, walawender08}.
NGC 1333 is one of the best studied very young clusters of active low- to intermediate-mass 
star formation \citep{aspin94, lada96, wilking04}. Imaging surveys in the near- and mid-infrared 
have revealed NGC 1333 contains both a northern and southern young stellar cluster 
\citep{aspin94, lada96, gut08}. Previous studies have also shown NGC 1333 is heavily 
extinguished and contains a substantial fraction of deeply embedded young stellar objects (YSOs) as
 well as many prominent outflows \citep{lada96,knee00,winston09}. The {\it Spitzer Space Telescope}
 \citep{werner04} infrared view of NGC 1333 in \citet{gut08} is streaked with emission at 4.5$\mu$m,
 which is dominated by H$_2$ lines  arising in shocks from YSO outflows colliding with ambient cloud 
material. These energetic outflows originate from actively accreting protostars with infalling 
envelopes (Class 0, I). As they are surrounded by dusty envelopes, these protostars as well as more 
evolved pre-main-sequence stars with circumstellar disks (Class II) emit in the mid-infrared and are 
well detected by {\it Spitzer}. Pre-main-sequence stars with little to no dusty disk or envelope 
(Class III) are not as easily detected in the mid-infrared and have been found in NGC 1333 through 
X-ray surveys \citep{getman02,winston09,winston10}. 

Many previous studies have addressed the age of NGC 1333.
In a near-infrared (NIR) spectrographic and X-ray study of NGC 1333, \citet{winston09} estimated 
most objects in NGC 1333 to be $<$3 Myr, and found an age spread from 1 to 10 Myr based on 
isochrones by \citet{baraffe98}. An optical and NIR study of NGC 1333 by \citet{aspin03} found 
similar results using isochrones by \citet{dantona97} namely an age spread from 7$\times$10$^4$ yr 
to 10 Myr with many sources around 3 Myr. They interpreted this spread in ages to be evidence of 
multiple epochs of star formation potentially spurred by turbulence generated by the many prominent
 outflows in NGC 1333. Others have also suggested turbulence driven star formation could have taken
 place in NGC 1333 \citep[e.g.][]{quillen05}. Studies of the low mass (0.25 - 0.02 M$_{\odot}$) 
population of NGC 1333 estimate a median age of $\sim$0.3 Myr for that population using the low 
mass stellar evolution tracks by \citet{dantona97} \citep{greissl07, wilking04}. However, for 
objects less than 1 Myr old age estimates based on isochrones are more uncertain than for older 
objects as the location of the birth line (t = 0) is ambiguous \citep{dantona97, siess01}. For this
 reason, class fractions are a common tool used to get an idea of the relative youth of a cluster.
\citet{gut08} show NGC 1333 is very young due to its large disk fraction (83\%$\pm$11\%) which
 they define as the fraction of total objects which show a large infrared excess at certain 
wavelengths indicative of emission from a circumstellar disk. \citet{lada96} estimate the age of 
NGC 1333 to be very young based on the similarity of its $K$-band luminosity function to that of 
the Trapezium ($\la$ 1 Myr), as well as its large fraction of infrared sources (61\%) which suggest
 an age of $\le$ 1-2 Myr. The protostar and Class II populations also have the same peak nearest 
neighbor distance (0.045 pc), meaning stars do not appear to have moved far from where they formed, 
another indication of youth \citep{gut08}. By these measures, NGC 1333 is one of the youngest 
clusters observed comprehensively by {\it Spitzer}. Based on disk fraction, NGC 1333 is younger than
 three well-studied star forming regions also observed by {\it Spitzer}, namely Taurus ($\sim$ 1-2 
Myr; \citet{furlan06}), Ophiuchus ($\sim$ 0.8 Myr; \citet{mcclure10}), and Chamaeleon I ($\sim$ 2 Myr;
 \citet{manoj11}). We adopt an age of $\la$ 1 Myr for this study.

The age of a cluster, while difficult to determine, is necessary to understand the timescales of 
star formation and disk evolution processes. Observations show disk dissipation is well 
underway in 3-5 Myr, as on this timescale infrared excess around young stellar objects drops by 
half \citep{haisch01,her...dez08} and accretion rate falls below 10$^{-10}$ M$_{\odot}$ yr$^{-1}$ 
for low mass stars \citep[e.g.][]{muzerolle03}. Therefore the study of regions significantly 
younger than this allow for investigation of early disk evolution and dissipation processes and can
 further constrain the timescales of these phenomena. Circumstellar dust and gas predominately emit
 in the infrared making it the ideal wavelength range to investigate processes in disk evolution.
Models predict \citep{weidenschilling97,harker02, gail04,dullemond05,dalessio06} and observations 
confirm \citep{furlan05,kessler06,sicilia07, sicilia08, sicilia09,bouwman08, sargent09, olofsson09, olofsson10} 
that grain growth, dust processing and dust settling in disks proceed on timescales $<$1 Myr. 

Giant planets are predicted to form within a few million years through collisional growth of grains
 into multi-M$_\oplus$ cores which may then accrete gas from the disk \citep{pollack96,weidenschilling08, dodson08}.
 They may also form through gravitational instability of gas and dust in the disk in $\la$10$^5$yr
 \citep{gold73,boss01}. The gravitational interaction between a disk and a giant planet would clear
 out material in its orbit thereby opening a gap or forming an inner clearing \citep[e.g.][]{borderies89}.
 Disks with radial gaps and holes have been detected mostly via deficits in their spectral energy
 distributions (SEDs) compared to the median for their population 
\citep{strom89,rice03b, dalessio05, brown07, calvet05,espaillat07,espaillat08, kim09, merin10}, but
 in several cases have been confirmed by $mm$-wave interferometric imaging \citep[e.g.][]{pietu06, hughes09}.
 The leading explanation for such gaps and clearings in disks is the gravitational and tidal effect
 of companions with the central star. In the paradigm ``central clearing" disk around CoKu Tau/4 
the companion is a star \citep{ireland08}, but in most others it is likely to be a gas-giant planet
 \citep{artymowicz94,calvet05,kim09,pott10}. However, several disk-dissipation mechanisms which
 clear disks from the inside out have been invoked to explain central clearings, among them 
photoevaporation \citep{clarke01, alexander06} and MRI draining \citep{BH91, chiang07}. These 
methods generally fail to account for gaps separating inner and outer disks, which outnumber the 
central clearings \citep{furlan09,kim09,mcclure10,manoj11}. Disks with holes or gaps are 
historically called transitional disks (TDs) as they appear to be in transition between a star with
 a full disk of dust and gas around it (Class II object) and a star where little to no emission is 
seen from a disk(Class III object) \citep{strom89, skrutskie90,gauvin92}. Disks with radial gaps 
are a type of transitional disk known as `pre-transitional disks' \citep[PTDs;][]{espaillat07, espaillat08,espaillat09, espaillat10}.
 The interaction of one or more planets with the disk is key to understanding the timescales and 
processes of planet formation as well as disk dissipation.

Here we present the spectra of 81 objects in NGC 1333. In \S~\ref{obs} we discuss the observations, selection, 
completeness, and data reduction for our sample. We correct our YSO sample for interstellar extinction in 
\S~\ref{EC} and using their observed spectral indices we classify our sample into SED classes and 
evolutionary states in \S~\ref{sam}. In \S~\ref{Anal} we characterize the settling in different 
regions of our disks as well as identify objects with interrupted radial disk structure. We also 
compare the distribution of various spectral properties and the median spectrum of disks in NGC 
1333 to those in Taurus, Ophiuchus and Chamaeleon I. Further discussion of the classification of 
TDs and PTDs in NGC 1333 is found in \S~\ref{disc}. Finally, in \S~\ref{conclusion} we summarize 
our results and state our conclusions. 

\section{Observations and data reduction} \label{obs}

\subsection{IRS observations}

As part of a {\it Spitzer} GTO program (PID 40525) we observed 78 objects in NGC 1333 on 2007 
October 5-6 \& 9-10, 2008 February 21 to March 2 and 2008 March 30, in IRS campaigns 44, 48 and 49 
respectively. Our sample of YSOs was chosen based on their excess in photometric observations from 
both the {\it Spitzer} Infrared Array Camera \citep[IRAC;][]{fazio04} 8.0$\mu$m band and Multi-band
 Imaging Photometer \citep[MIPS;][]{reike04} 24.0$\mu$m band, published in 
\citet{gut08}. We observed all objects with detections ($\sigma$ $<$ 0.2 mag) in the IRAC 8.0$\mu$m
 band and MIPS 24$\mu$m band which are brighter than 11.54 mag at 8.0$\mu$m and 8.5 mag at 24$\mu$m.
 For this reason we adopt their identifiers for our sample \citep[these identifiers were also used in][]{winston09, winston10}.
 The basic data for this sample, namely, [GMM08] SED index, RA, Dec, AOR ID and alternate names, 
are listed in Table \ref{tbl1}.

This selection excluded a total of 55 objects from the \citet{gut08} YSO sample: 39 had no MIPS 
24$\mu$m photometry, 2 had no IRAC 8.0$\mu$m photometry and 14 were too faint at either 8.0$\mu$m 
or 24$\mu$m. Of the \citet{gut08} sample, 4 objects were already scheduled to be observed with IRS 
and therefore these objects were not observed in program 40525. One of these conflicts was the 
class 0 protostar, IRAS 2A, which was observed in {\it Spitzer} program 2. The 
other three, ASR 118, SSTc2d J032924.1+311958, and SSTc2d J032929.3+311835, were observed in 
program 30843 on 2006 September 18, 2007 March 9 and 21. These objects were 
selected from their photometry as possible `cold disks' which have inner holes and are presented in
 \citet{merin10} indexed there by \#2, \#4 and \#5 respectively. \citet{gut08} identifies these 
objects as \#67, \#122 and \#137 and we include these three observations with IRS in our sample 
using the \citet{gut08} identifiers. This addition brings the total number of observations to 81.

Selecting for objects with excess in the 8.0$\mu$m IRAC and 24$\mu$m MIPS bands, biases our sample 
toward earlier star formation stages (Class 0/I and II) where massive dusty envelopes and disks are
 present. \citet{gut08} find the 137 cluster members identified with IRAC and MIPS to represent 
83\%$\pm$11\% of the total cluster membership. Our sample consists of over half of these sources (N = 81) which is $\sim$49\% of the extrapolated total cluster membership.

We observed each object from $5-38\mu$m using the Staring Mode of the \textit{Spitzer} IRS in both 
orders of the Short-Low (SL; $5.2-14.5\mu$m) and Long-Low (LL; $14-38\mu$m) resolution modules 
($\lambda/{\Delta\lambda} \sim 60-120$). From the Spitzer Science Center (SSC)'s basic calibrated 
data (BCD) products using pipeline S18.7, we extracted a spectrum using the IRS instrument team's 
Spectral Modeling, Analysis and Reduction Tool \citep[SMART;][]{higdon04}. We corrected for bad or 
`rogue' pixels by interpolating the flux from two adjacent good pixels in the dispersion direction.

For each observation, we took two images with the object at different positions, or nods, in the 
slit, each one third of the width of the slit away from the edge. For most images in our sample, we
 followed the standard procedure for extracting spectra as is done in \citet{furlan06,furlan08, kim09, mcclure10}.
 We removed background emission by subtracting the image with the source at one nod from the image 
with the source at the opposite nod. We then extracted the spectrum from the sky-subtracted 2D 
image using a column which tapers as the width of the point-spread function does. Spectral extraction
 and removal of background emission for 60 of these objects proved troublesome because of a 
combination of bright cloud emission, multiple sources in the slit and/or the extinguished, faint 
nature of the source. For these observations with particularly difficult background emission, we 
individually inspected and carefully extracted a spectrum using the SMART optimal extraction tool, 
AdOpt \citep{lebout10}. AdOpt allows the user to narrow the extraction window as well as fit 
background emission in that window with an n-th order polynomial. Using AdOpt, we fit the background
 emission in most cases with a polynomial of 0th or 1st order, though a few cases required as high 
as 3rd order. 

With AdOpt, tapered column extraction or a combination of the two methods, we extracted a spectrum 
for each object from each module, order and nod. To calibrate the flux of the spectra, we created 
relative spectral response functions (RSRFs) by dividing a template spectrum, $\alpha$ Lac 
\citep[A1 V;][]{cohen03} for SL (5.2-14.5$\mu$m) and $\xi$Dra (K2 III) for LL (14.0-36$\mu$m) by 
its tapered column or AdOpt extracted IRS spectrum. We multiplied the target spectra by the 
similarly prepared RSRF for each module, order and nod to calibrate the flux from the IRS. For each
 nod, we combined the calibrated spectra from each module and order. To produce a final spectrum, 
we averaged these spectra from the two nods and calculated uncertainty as half of the nod 
difference. Occasionally, slight mispointing resulted in a decrease in flux in the modules which 
were mispointed, causing a mismatch between the SL and LL spectra. We corrected for this by scaling
 the fainter module, which presumably lacks flux due to mispointing, up by a scalar factor. A 
typical scaling factor was 1.1, with a maximum value of 3.5 (\#34). As an additional check on our 
reduction, we extracted a spectrum for each object with the optimal extraction software, UR Optimal
 (W. Forrest and C. Tayrien, personal communication) which extracts a spectrum using data from 5, 7
 or 9 pixels around the nod position. The spectrum extracted with tapered column extraction and/or 
AdOpt had smaller nod-difference uncertainties than that of UR Optimal, therefore we use the 
tapered column and/or AdOpt extracted spectrum for our analysis. We estimate the 
absolute photometric uncertainty in flux
 in the final spectra to be $\sim$5\%.

\subsection{Ancillary Data}
We supplement the IRS spectra with broadband photometry from the 
literature in the $J$, $H$ and $K_s$ bands of the Two Micron All Sky Survey \citep[2MASS;][]{2MASS}.
NGC 1333 was also observed as part of the 2MASS extended mission, 2MASS 6X. Data from this mission 
were taken from observations with six times the integration time of standard 2MASS observations. 
This resulted in $\sim$ 1 mag improvement in sensitivity limit and reduced the uncertainty for 
faint objects near the detection limit of regular 2MASS data \citep{carpenter01}. Where available,
 we use the 2MASS 6X data over the standard 2MASS from the literature. We also supplement our data,
 where available with $R$ and $I$ broadband photometry from \citet{aspin94}. We used 2MASS $K_s$ 
broadband photometry in our SEDs and where 2MASS $K_s$-band photometry was not available we used 
$K$-band photometry from \citet{aspin94} or \citet{lada96}. Every object was observed 
photometrically with {\it Spitzer} and we supplement our data with broadband photometry from 
\citet{gut08} in the IRAC 3.6, 4.5, 5.8 and 8.0$\mu$m bands as well as the MIPS 24$\mu$m band.

\subsubsection{Notes on Individual Objects} \label{notes}
Spatially variable, bright sky emission resulted in careful reduction of 60 of the 81 images 
using off nod and off order sky subtraction as well as polynomial fit sky subtraction with AdOpt. 
Below are the details of particularly troublesome objects. 
We omit objects from our sample which 
have a spectrum that is not representative of a YSO, namely \#8 and \#38.

{\it \#5}:~ Bright emission lines in the SL spectrum could not be removed through various sky 
subtraction methods; this results in a seemingly discontinuous spectrum.

\begin{figure}
\figurenum{1}
\epsscale{1}
\plotone{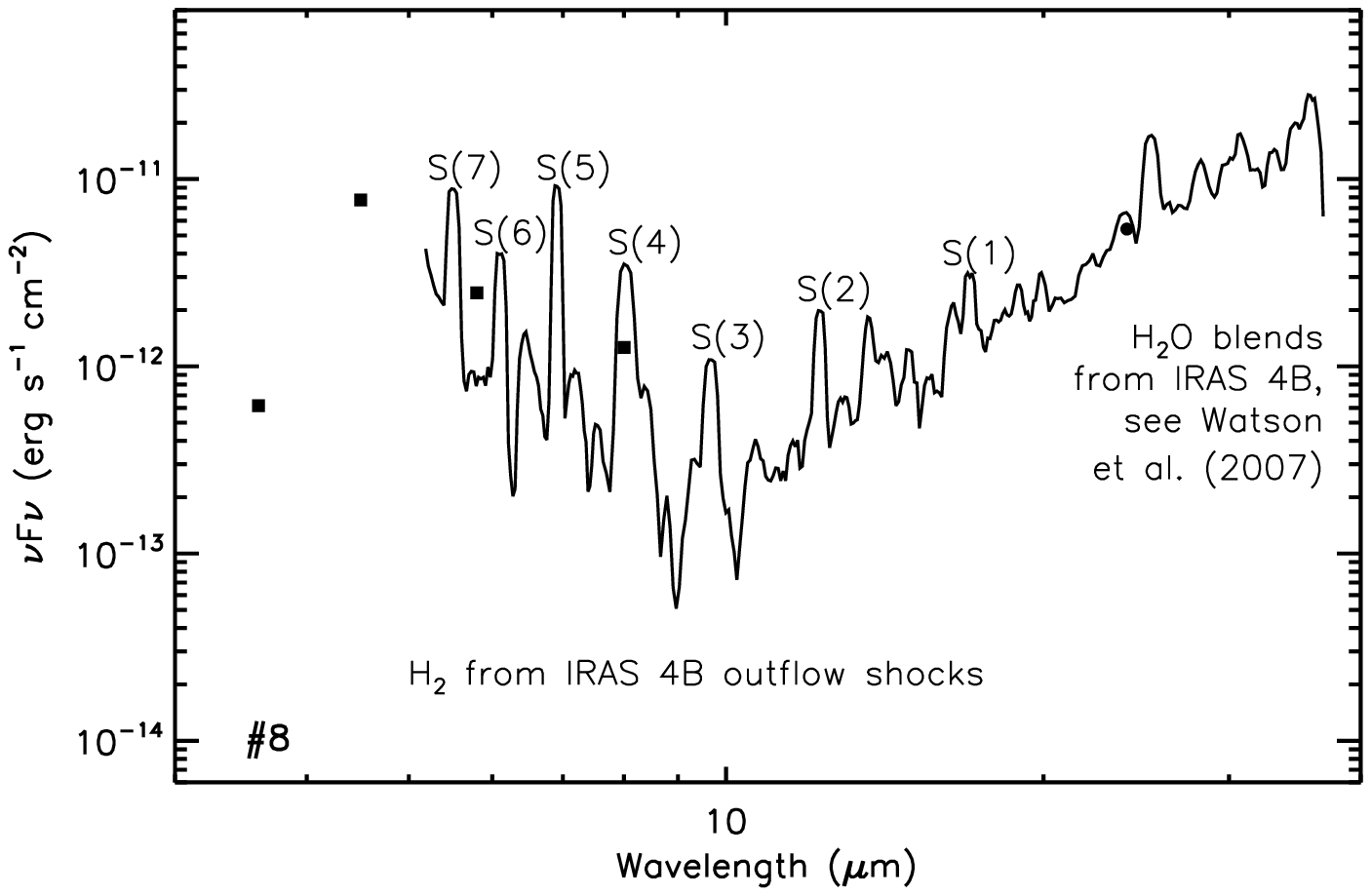}
\caption{{\it Spitzer} IRAC, MIPS and IRS data on emission from the southern outflow lobe of IRAS 4B. The S(7) to S(1) H$_2$ lines are identified. The LL spectrum contains many attenuated blends of H$_2$O emission lines from IRAS 4B, which lies 5" north of this position. The IRS spectrum of IRAS 4B was presented and studied in detail in \citet{watson07} and therefore it is left out of our analysis. \label{IRAS4b}}
\end{figure}

{\it \#8}:~ Identified as IRAS 4B in \citet{gut08}, \#8 is not identical with the Class 0 object but corresponds instead to one of IRAS 4B's outflow lobes. The 
observation for \#8 was pointed at the bright emission originating from one of IRAS 4B's 
outflow cavities to the south of the embedded object's sub-mm coordinates. The SED is shown in 
Figure~\ref{IRAS4b} and it is left out of our sample. The SL spectrum is dominated by the emission 
from the outflow as evidenced by the collisionally excited molecular hydrogen emission lines. The LL
 spectrum contains a forest of attenuated, blended water emission lines from IRAS 4B itself 
\citep{watson07} though it lies about 5" away, at the edge of the slit.

{\it \#23}:~ Careful extraction was unable to remove H$_2$ and polycyclic aromatic hydrocarbon 
(PAH) emission features which do not originate from \#23.

{\it \#34}:~ The SL spectrum was very difficult to disentangle from the sky emission and had to be 
scaled up by 3.5 to match the LL spectrum. Also, the SL spectrum possibly suffers from an offset 
in the dispersion direction.

{\it \#37}:~ The LL spectrum is dominated by a bright source near the target coordinates. The flux 
contribution of this nearby bright source dominated the spectrum and could not be removed. For this
 reason we only present the SL spectrum of \#37 and only use the SL spectrum in our analysis.

\begin{figure}
\figurenum{2}
\epsscale{1}
\plotone{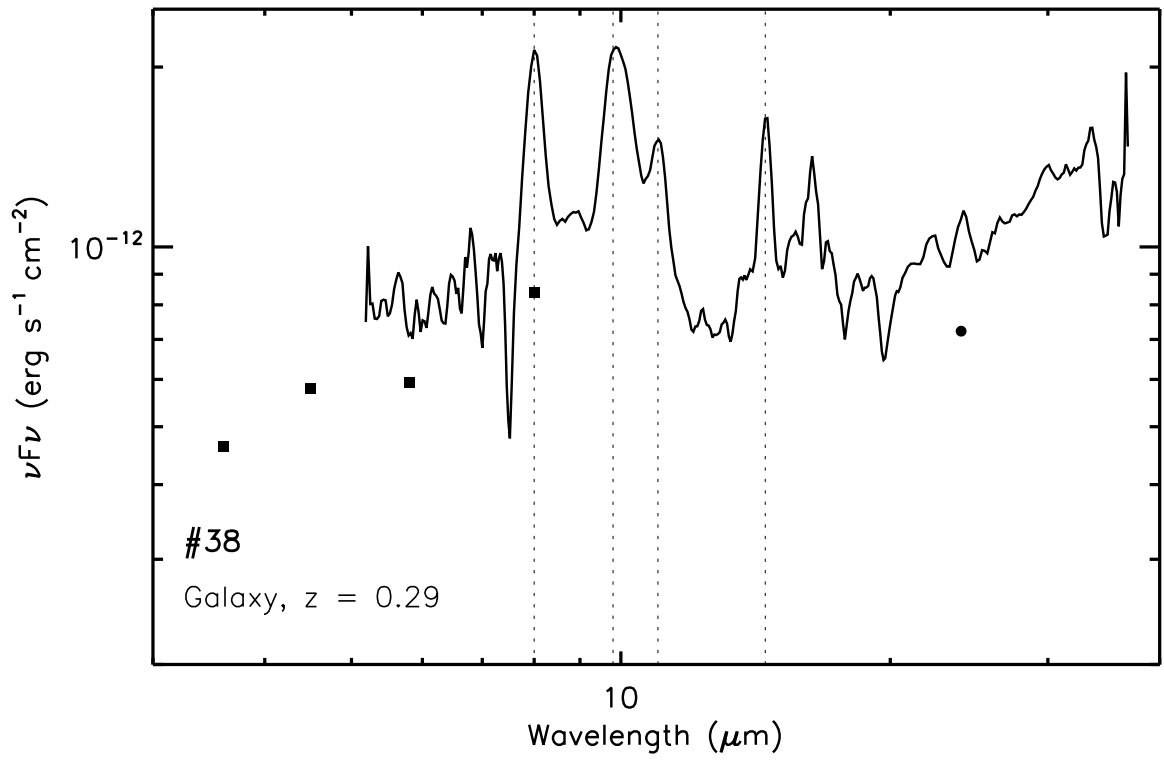}
\caption{SED of a background galaxy redshift of 0.29. The dotted lines indicate the 6.2, 7.7, 8.6 and 11.2$\mu$m PAH features at their redshifted wavelengths. \label{galaxy}}
\end{figure}

{\it \#38}:~ Based on the presence of 
redshifted PAH emission \#38 is a galaxy. Using the redshifted 6.2, 7.7, 8.6 and 11.2$\mu$m PAH 
emission features we calculate a redshift from each feature and obtain an average redshift of 0.29. 
The SED of the galaxy labeled here as \#38 is shown in Figure~\ref{galaxy} and it is left out of 
our sample.

{\it \#52}:~ The 2nd module of SL nod 1 had data missing shortward of 5.43$\mu$m as a result of the
 SSC pipeline trying to correct for rows in the detector which were altered by peak-up camera 
saturation. This resulted in an unreliable nod 1 spectrum and for this reason we use only the nod 2 
spectrum for SL2.

{\it \#58}:~ Two close sources in the LL slit provided some difficultly. 
Narrow window extraction at nod position with AdOpt yielded the best result.
LL was scaled by 0.7 to match SL. 
We include \#58 in our analysis with the caveat that the LL spectrum is likely not solely representative 
of \#58. 

{\it \#63}:~ Both nods of the 1st module of SL displayed a striping effect in the dispersion 
direction, deemed 'jail-bars,' likely due to correction for a bright source in the peak-up array. 
Careful extraction using AdOpt was able to recover the source emission and we use this spectrum in our analysis. This effect as well as the 
faint nature of the object is the likely cause of the noise in the spectrum of \#63.

{\it \#67}:~ Both nods of the 1st module of LL were severely affected by stray emission in the 
detector from a very bright source just outside the LL slit; for this reason we exclude the LL1 
data from the SED. This effect is exemplified in its 
spectrum in \citet{merin10} as a wide false 30$\mu$m feature, also seen in the IRS spectrum of Hn 
10 E in Chamaeleon I \citep{manoj11}. \citet{merin10} classify this source (called therein ASR118) as a 'cold disk,' a term equivalent to our TD. However, due to lack of reliable data in LL1, we exclude \#67 from our analysis and cannot definitively address its radial structure.

{\it \#68}:~ The emission lines in the SL module at 5.5, 6.1, 6.9, 8.0 and 9.6$\mu$m are highly 
spatially-variable molecular hydrogen lines from the sky, which could not be removed in the 
reduction process.

{\it \#75, \#94, \#110 \& \#133}:~ The spectrum of each of these objects contains one or more 
spikes due to difficult to remove background emission and/or possible remnant bad pixels.

{\it \#84}:~ Sky subtraction in LL proved difficult, the strong dip around 30$\mu$m is not a real 
feature but could not be removed.

{\it \#92 \& \#102}:~ Variable, bright sky emission and the faint nature of the source cause these 
carefully extracted spectra to be noisy and uncertain.

\section{Extinction Correction} \label{EC}
\begin{figure}
\figurenum{3}
\epsscale{1}
\plotone{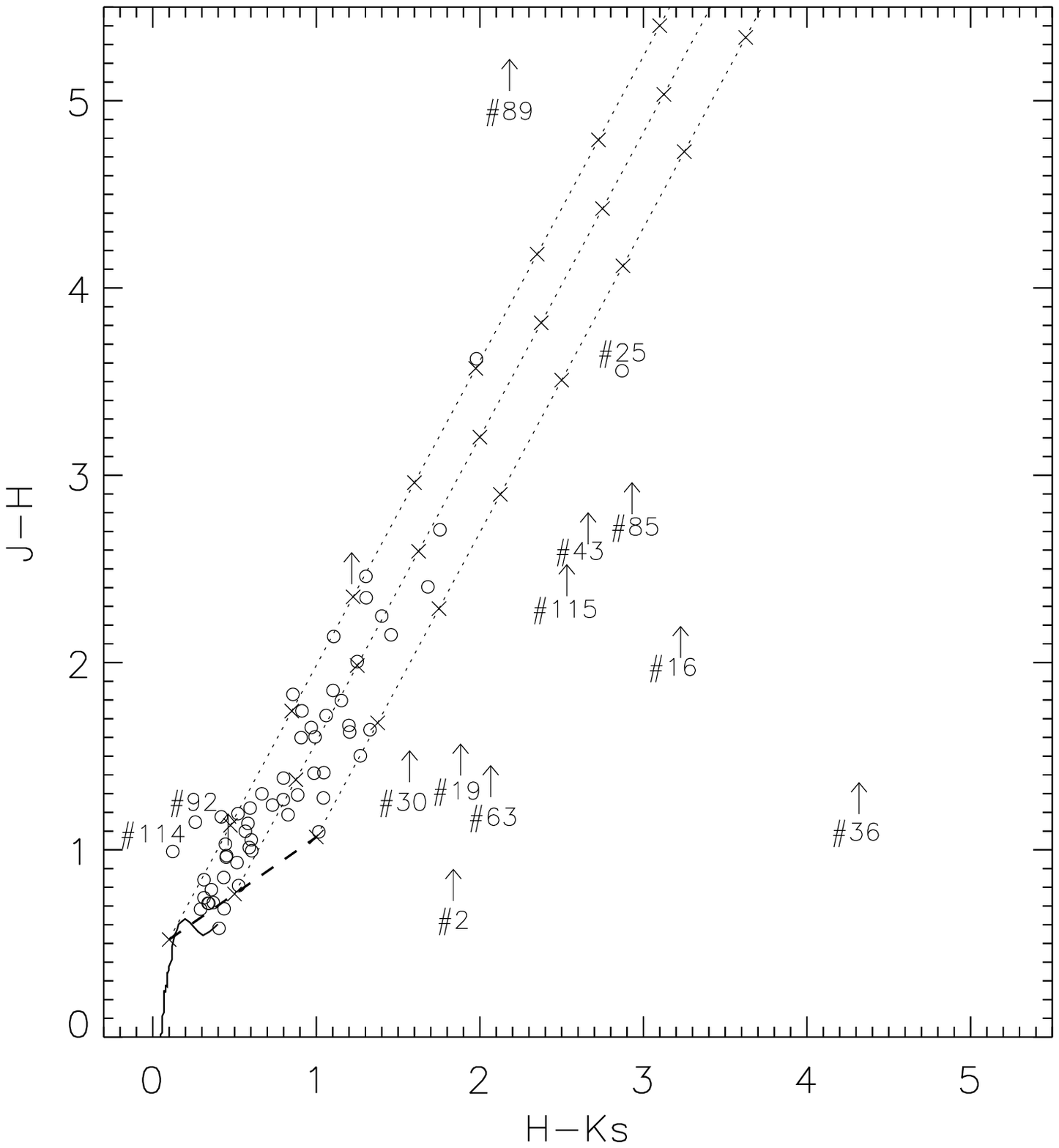}
\caption{$J-H$ vs. $H-K_s$ color-color diagram for our sample with known 2MASS measurements, arrows indicate objects with upper limit at $J$, $H$ and/or $K_s$. Long dashed line indicates the classical T Tauri locus of \citet{meyer97}, dotted lines with crosses at intervals of $A_V$=10 show how reddening would affect the colors of the classical T Tauri locus. The solid line indicates the colors of stars from \citet{kenyon95}. \label{CTTLocus}}
\end{figure}

As discussed in \S~\ref{intro}, NGC 1333 is heavily reddened; 
therefore in order to understand the intrinsic properties of this 
sample we determined the amount of reddening toward each object 
in a uniform way. To determine reddening, or extinction, in terms 
of $A_V$, we utilized 2MASS photometry because it 
is the most complete set of NIR photometry available for this sample. 
All but 12 objects have good detections (quality flags A, B or C) in 
$J, H$ and $K_s$. We adopt a method to determine the 
extinction toward these objects similar to that by \citet{mcclure09}. First, we 
plotted (Figure~\ref{CTTLocus}) the observed $J-H$ and $H-K_s$ colors of our sample against 
each other and compared their distribution to that of the classical 
T Tauri locus \citep{meyer97} and main sequence stars and dwarfs 
\citep{kenyon95}. 
Most of 
our sample's colors are consistent with a reddened classical T Tauri star. However, there are a few outliers: 10 objects are on the right and 3 objects fall to the left of the reddened T Tauri locus.
The colors of the objects to the right of the classical T Tauri locus could also be more heavily reddened due to local extinction as would be the case for Class 0/I objects which show envelope dominated emission.

\begin{figure}
\figurenum{4}
\epsscale{1}
\plotone{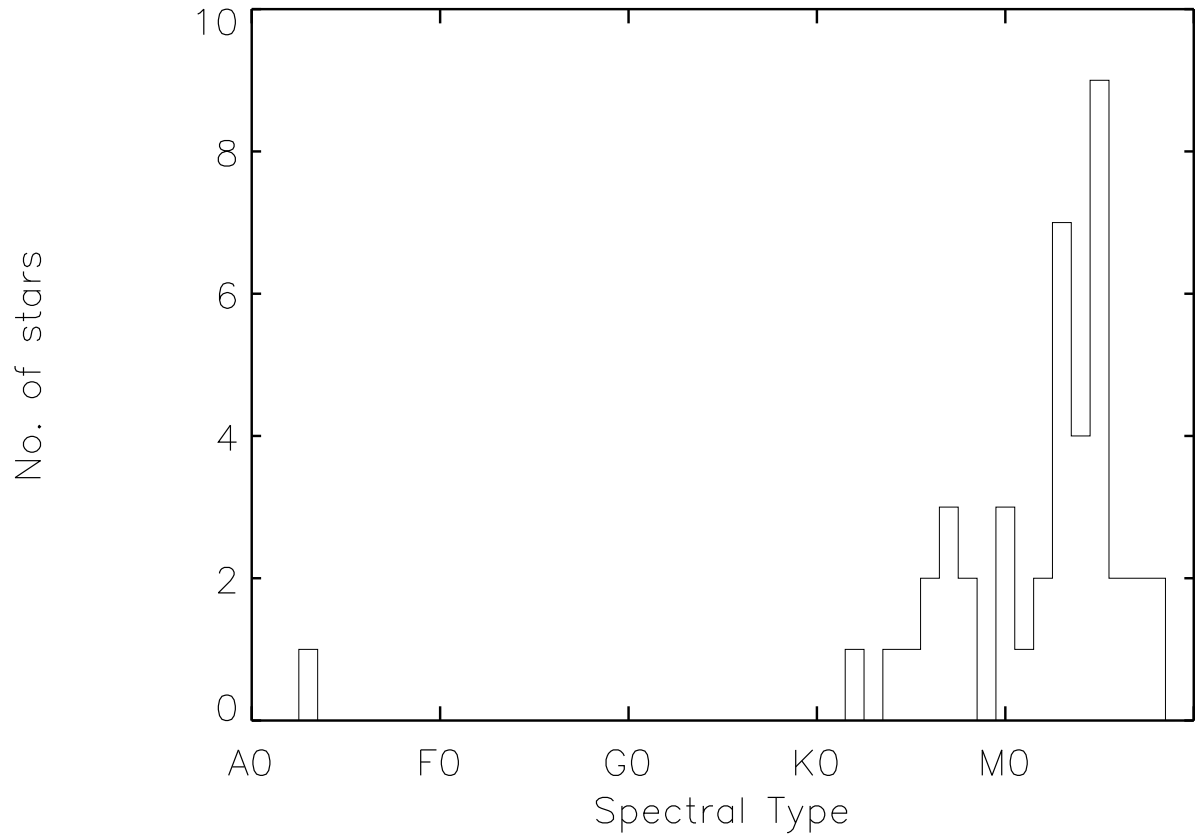}
\caption{Histogram of known spectral types from literature for our sample in NGC 1333. \label{SpT}}
\end{figure}

To determine the extinction towards these objects we first compiled a
 list of spectral types from the literature. The large extinction towards NGC 
1333 ($A_V >$ 10) makes optical and near infrared spectra and subsequent spectral 
type difficult to obtain. From a literature search, 43 objects have known spectral types which are listed in Table \ref{tbl2}. Figure~\ref{SpT} 
shows the distribution of known spectral types. For objects 
with known spectral types we adopted the intrinsic $J-H$ and $H-K$ 
photospheric colors for a given spectral type from \citet*{kenyon95} transformed to the 2MASS 
system to determine the color excess. As we assume photospheric colors for all objects with known spectral type, extinctions calculated will be overestimates for objects with $H$ or $K_s$ band excess. For objects with no spectral type information, we made the assumption 
that all the extinction is foreground 
extinction and determined the color excesses, $E(H-K_s)$ and $E(J-H)$, by extrapolating the 
observed color of each object along the reddening vector $\frac{E(H-K_s)}{E(J-H)}$ = 1.63, until 
their intrinsic $J-H$ color and $H-K_s$ color fell along the classical T Tauri locus 
\citep{meyer97}. Using the $J-H$ and $H-K_s$ color excess, generally expressed as $E(\lambda_1-\lambda_2)$, we calculated two values of $A_V$ for each object using 
\begin{equation}
A_V = 
\frac{\frac{A_V}{A_{\lambda_2}}}{\frac{A_{\lambda_1}}{A_{\lambda_2}}-1}\times [E(\lambda_1-\lambda_2)] 
\end{equation}
with a value of 
$\frac{A_{\lambda_1}}{A_{\lambda_2}}$ from the $R_V$ = 5 version of the Mathis
 \citeyearpar{mathis90} extinction law with $\frac{A_V}{A_{K_S}}$ = 
7.75. The $R_V$ = 5 extinction law is appropriate for dense molecular 
clouds \citep{mathis90}. This value of $R_V$ is in agreement with \citet{cernis90}'s findings for the
 L1450 dark cloud in NGC 1333. Those authors estimate to have 
$R_V$$\sim$5 toward the center, around SVS 3, and $R_V$$\sim$4 elsewhere.

\begin{figure}
\figurenum{5}
\epsscale{1}
\plotone{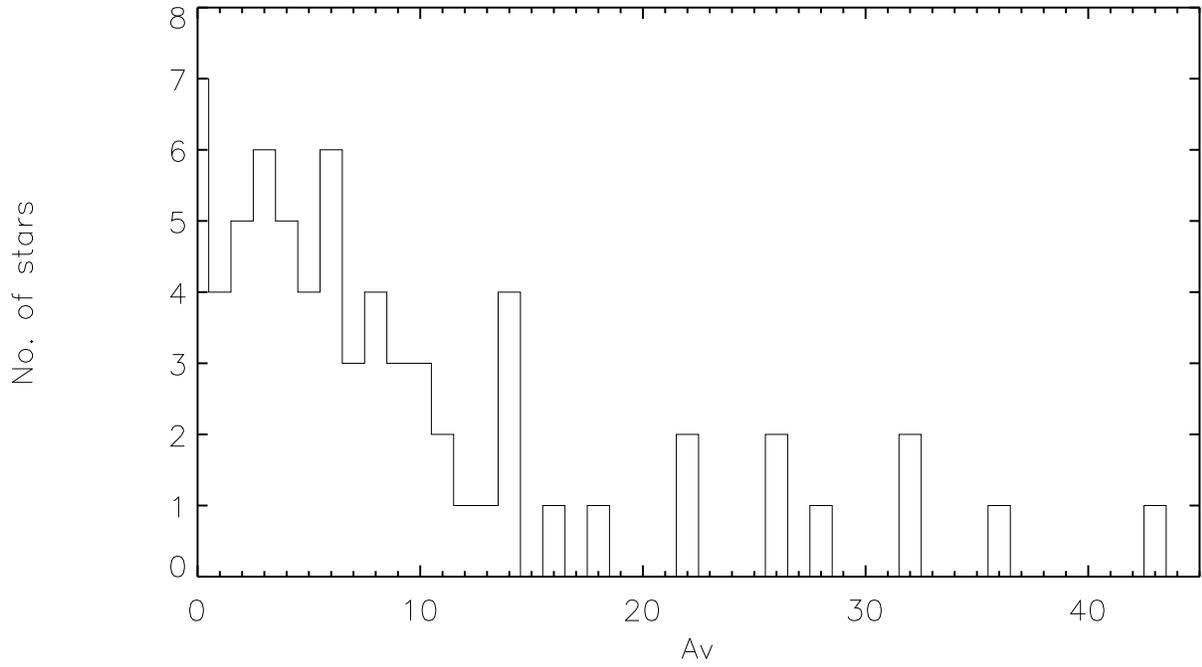}
\caption{Histogram of $A_V$ values derived through the method described in \S~\ref{EC} for our sample in NGC 1333. \label{Av}}
\end{figure}

We adopt either the $J-H$ or $H-K_s$ derived $A_V$ for our analysis.
 For objects with known spectral type, the $J-H$ and $H-K_s$ derived $A_V$
were generally within 2 mag with a maximum difference in one case 
of 7 mag. For objects with no spectral type, typical differences 
between the two vales of $A_V$ were around 1 mag, with a maximum 
difference of 9 mag. When the $A_V$ values were within 1 mag, we 
chose the smallest of the two for our analysis. For differences in derived $A_V$ 
greater than 2 mag, we compared the dereddened photometry to the 
appropriate photosphere scaled to $J$ band, for a given spectral 
type where available, and chose the $A_V$ which gave the most reasonable 
physical excess. In most cases the $A_V$ derived from the $J-H$ 
band excess was used. 
For the 8 objects without spectral type information and difference in derived $A_V $ greater than 2 mag, we compared the spectrum and photometry dereddened with the each $A_V $ value and chose the value which minimized the presence of silicate and ice absorption features.
Two objects had a negative $A_V$ based on $J-H$ or $H-K_s$ and two objects had a negative $A_V$ based on $J-H$ and $H-K_s$; for these objects we adopted an $A_V$ of 0.  
For the objects to the left and 
right of the classical T Tauri locus, the $J-H$ and $H-K_s$ derived 
$A_V$ tended to differ by 15 mag and in one case by 40 mag; it 
should be noted there is greater 
uncertainty in their $A_V$. The $A_V$ value for each object, 
derived from the $R_V$ = 5 version of \citet{mathis90} extinction 
curve, as well as the method with which they were derived is 
noted in columns 2 and 3 of Table \ref{tbl2} and a histogram showing the distribution of $A_V$ for our sample is plotted in Figure~\ref{Av}. The objects with 
no 2MASS $J$, $H$ or $K_s$ band measurements are noted with a null 
value. The median $A_V$ for NGC 1333 is 5.8 mag.

Overall, these values of $A_V$ differ on average about 2 mag from the equivalent values of 
$A_K$ in \citet{gut08}. In only one case do the two values
differ by as much as 12 mag. This object, \#25, falls to the right of the classical T Tauri locus and has large extinction. 
In order to characterize low level extinction, \citet{gut08} derived an $A_K$ from an extinction map generated for the 2MASS $H-K_s$ colors of field stars in the region using the method described in \citet{gut05}.
Because of the high galactic latitude of NGC 1333, extinction maps based on field stars are low resolution, in this case resolution ranges from 70" to 200". We attribute the discrepancies in our extinction values to this difference in methodology.

Of the 66 objects with reliable $J-H$ and $H-K_s$ broadband colors, 49 have an $A_V$ $>$ 3, where the 
molecular cloud extinction law is known to diverge from the Mathis \citeyearpar{mathis90} $R_V$=5 law for the 
diffuse interstellar medium \citep{mcclure09, chapman09}. For this reason, we deredden the 
broadband photometry and the IRS spectra with \citet{mcclure09}'s composite extinction law, which 
follows the Mathis \citeyearpar{mathis90} $R_V$=5 law for $A_V$ $<$ 3 and corrects objects with $A_V$ $>$ 3 with two different curves, one for 3 $<$ $A_V$ $<$ 8 and one for $A_V$ $>$ 8. This composite extinction law has also been used 
to deredden the IRS spectra of young stars in the nearby clusters of
 Ophiuchus \citep{mcclure10} and Chamaeleon I \citep{manoj11}.

\section{Characterization of the Sample} \label{sam}
\subsection{SED Classification\label{SEDclass}} 
Classification of a YSO is typically accomplished using the slope of its SED spanning 
from the near- to mid-infrared \citep{lada87}. 
We find the logarithmic slope, $n$, or spectral index, in a given wavelength range, henceforth expressed as $n_{\lambda_{1}-\lambda_{2}}$, which is given as
\begin{equation} \label{eq1}
n_{\lambda_1 - \lambda_2} =
\frac{\log(\lambda_2 F_{\lambda_2})- \log(\lambda_1 F_{\lambda_1})}{\log(\lambda_2) - \log(\lambda_1)}.
\end{equation}
An IR slope, typically from 2 to 25$\mu$m, greater than 0.3 corresponds to that of a {\it Class I} object, 
$n_{2-25}$ between -0.3 and 0.3 is that of {\it Flat Spectrum} object, $n_{2-25}$ between 
-1.6 and -0.3 defines a {\it Class II} object and $n_{2-25}$ less than -1.6 classifies a 
{\it Class III} object.
The empirical 
classification of the SED of YSOs has become 
inextricably tied with an evolutionary sequence \citep{adams87} as discussed in \S~\ref{intro}, 
therefore proper classification 
is important in order to understand the evolutionary state 
of these systems.

We empirically classify the observed SEDs of the objects in our sample with 2MASS $K_s$ band photometry using $n_{2-25}$. The flux at 2$\mu$m is the 2MASS $K_s$ photometric flux ($\lambda$=2.17$\mu$m) and the flux at 25$\mu$m is the integrated flux of the observed IRS spectrum from 23.5-26.5$\mu$m.
SED classification from 2 to 25$\mu$m could not be done for 8
objects because they lack detection in the $K_s$ band of 2MASS and for 1 object (\#67) because of unreliable data near 25$\mu$m.
We find 16 \emph{Class 
I} objects, 14 \emph{Flat Spectrum} (henceforth, FS) objects and 40 \emph{Class II} objects based on $n_{2-25}$ SED classification.
No objects were classified as \emph{Class III} based on their $n_{2-25}$. This is not surprising as Class III objects would not be bright in the mid-infrared and we selected our sample based their flux at 8 and 24$\mu$m.

When interpreted as an indication of evolutionary state the $n_{2-25}$ index is at times known to be misleading \citep{robitaille06, crapsi08, mcclure10, manoj11}. 
 Edge on disks are known to have very red spectral indices 
\citep{pont05} and are frequently identified as Class I objects based on $n_{2-25}$.  The {\it Spitzer} 
IRS spectra reveal silicate emission and ice absorption 
features, which are not accounted for in an empirical $n_{2-25}$ classification and allow us to gain a better 
understanding of the evolutionary state of these objects.
 Also important to note is $K_s$ band measurements from the extended 2MASS mission were taken in 2000 and the IRS spectra in 2007 and 2008. As these T Tauri stars are known to be highly variable this difference in epochs of observation can affect the 2 to 25$\mu$m slope and subsequent SED classification. The empirical SED classification from 2 to 25$\mu$m gives us one look into the evolutionary state of these objects.

\subsection{Evolutionary State \label{extfree}} 
To better understand the evolutionary state of our sample, including 
those objects without known 2MASS $K_s$ band measurements, we examine 
the ``extinction-free" spectral indices of the observed spectra for the 
entire sample \citep{mcclure10}. These indices, $n_{5-12}$ and $n_{12-20}$, characterize the 
SED slope of the spectra before extinction correction between wavelengths with roughly the same value of 
$\frac{A_\lambda}{A_V}$ according to \citet{mcclure09}, namely 5.3, 
12.9 and 19.8$\mu$m. The flux at each of these wavelengths is 
calculated by averaging the flux over 3 channels, i.e. 0.3$\mu$m, 
centered at the anchor point. This index is truly extinction-free for objects with $A_V$$>$8 and varies slightly with wavelength for objects with $A_V$$<$8. 
Based on the $n_{5-12}$ spectral index, we classify the evolutionary 
state of the sample (N = 79)
 according to the following criteria: $n_{5-12}$ $\ge$ $-0.2$ - the spectrum is dominated 
by \emph{envelope} emission, $-0.2$ $>$ $n_{5-12}$ $\ge$ $-2.25$ - 
dominated by \emph{disk} 
emission , and $n_{5-12}$ $<$ $-2.25$ - dominated by 
\emph{photosphere} emission \citet{mcclure10}. Using $n_{5-12}$, 23 objects are 
classified as envelope dominated, 55 as disk dominated, and 1 as 
a photosphere dominated. Large values of $n_{12-20}$, greater than 
2.0, indicate typical values for TDs in Taurus and Ophiuchus \citep{mcclure10}.

\begin{figure}
\figurenum{6}
\epsscale{1}
\plotone{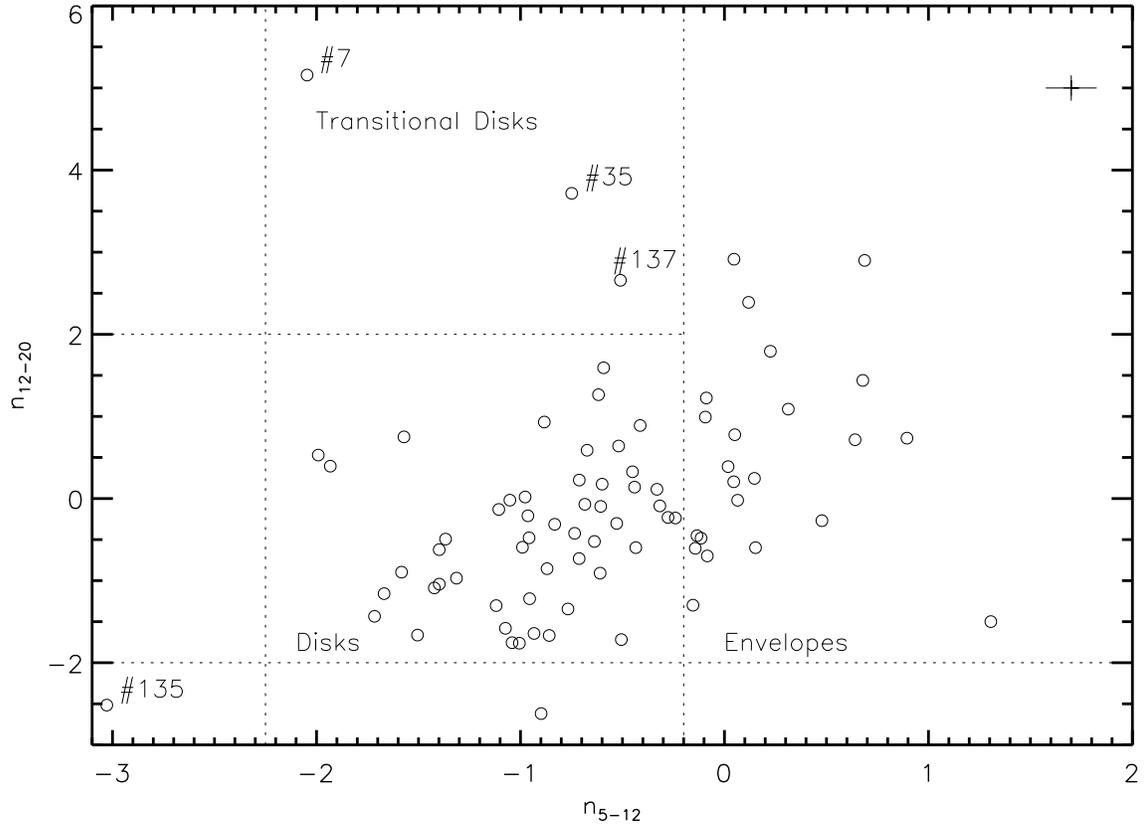}
\caption{Extinction-free indices $n_{5-12}$ and $n_{12-20}$ plotted against each other for all objects with a 2MASS $K_s$ band detection in our sample. Dashed lines indicate the boundaries in $n_{5-12}$ of envelope, disk and photosphere dominated objects, as well as a boundary in $n_{12-20}$ above which transitional disks are found. \label{n512_n1220}}
\end{figure}

Figure~\ref{n512_n1220} plots these two extinction-free indices 
against each other, allowing us to gain another perspective of
the evolutionary state of the objects in our sample. Three objects [\#7, 35 and 137] fall within the area of this plot
where transitional disks are found as they are characterized by a steeply rising SED beyond 12$\mu$m \citep{mcclure10, furlan11}. 
Upon examination of the SEDs of \#7 and
 \#35, the presence of 
deep silicate and CO$_2$ ice absorption features in their spectrum 
an absence of detection in the three bands of 2MASS indicates 
these objects are deeply embedded in an envelope.
Their decreasing SED 
from 5 to 12$\mu$m and steeply rising SED longer than $\sim$10$\mu$m 
suggests these deeply embedded protostars \citep{furlan08} and are not TDs. 
The extinction-free indices are misleading here as the steep excess at longer wavelengths is a result of an envelope rather than, as would be the case for TDs, the excess originating at the outer disk wall \citep{espaillat09}. 
\citet{furlan11} found embedded protostars seen edge-on are also found in this area of the plot.
 However, examination of the SED of \#137 shows no evidence of a significant envelope affecting the spectrum, therefore the steep rise picked out in a high $n_{12-20}$ is likely from emission originating in the outer layers of a heated outer disk wall. We will return to \#137 in our discussion of radial structure in \S~\ref{radial}.
 
\begin{figure}
\figurenum{7}
\epsscale{1}
\plotone{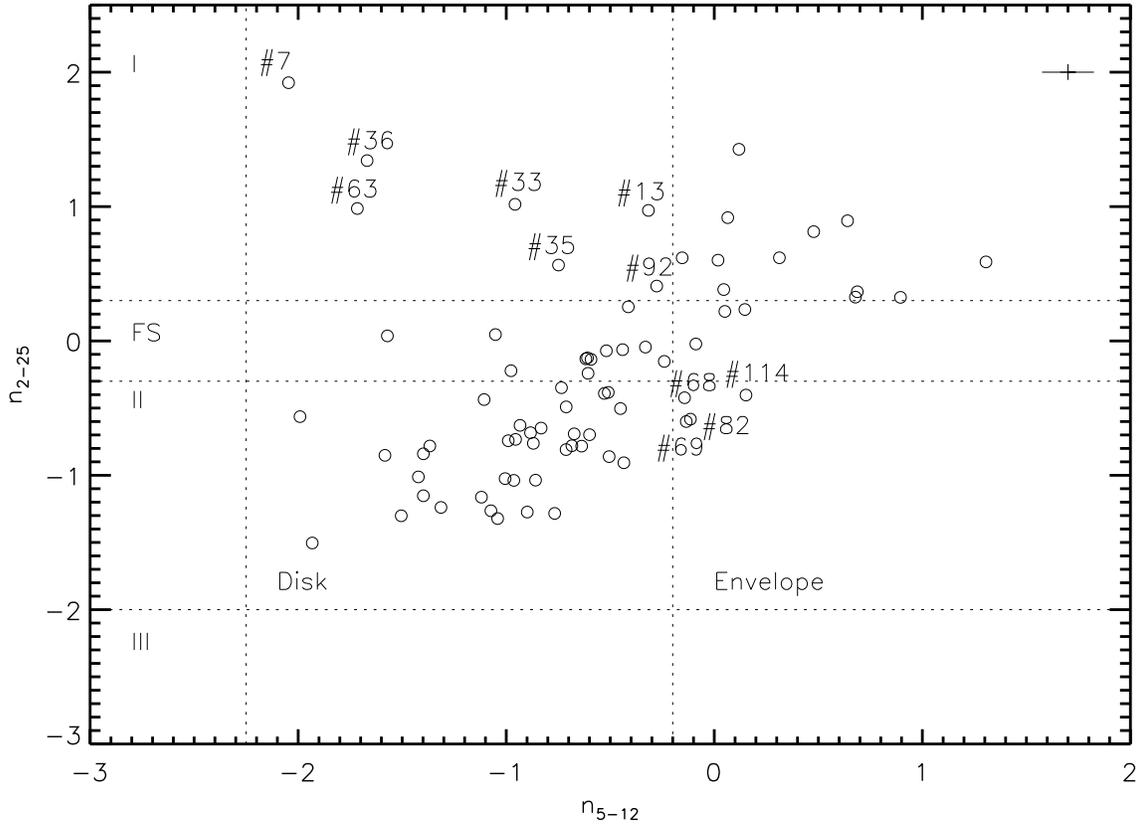}
\caption{Extinction-free index $n_{5-12}$ and SED index $n_{2-25}$ plotted against each other for all objects with a 2MASS $K_s$ band detection in our sample. Dashed lines indicate the boundaries of the different SED classifications namely Class I, FS, II and III and evolutionary stages which categorize the emission as envelope, disk and photosphere dominated.\label{n512_n225}}
\end{figure}

We examine the SEDs of our sample with respect to their $n_{5-12}$ 
classification and identify objects which require more careful 
consideration and possible reclassification. Figure~\ref{n512_n225} 
displays the SED classes (based on $n_{2-25}$) with respect to the extinction-free index 
$n_{5-12}$. Of the 23 envelope dominated 
objects, 12 are Class I, 3 are FS and 4 are Class II.  The other 4 lack data either at 25$\mu$m, as in the case of object \#67, or in the 
$K_s$ band and therefore could not be classified. Ice and silicate 
absorption features, as a result of local extinction, are evidence for a significant envelope 
affecting the SED \citep{boogert08,furlan08,zazowski09}. We use the presence or absence of these features 
to get an idea of how strongly an envelope dominates the emission of
 these objects as well as whether their SED might truly be considered
 as dominated by disk emission. The four Class II objects which are 
envelope dominated according to $n_{5-12}$ [\#68, 69, 82 and 
114] plus object \#67 which has no $n_{2-25}$ but is envelope dominated based on $n_{5-12}$, 
do not show ice or silicate absorption features in their 
observed or dereddened spectra; for these reasons we reclassify these
 objects as disk dominated though we do not include them in our disk analysis in \S~\ref{Anal}.
 If an envelope were present we would derive a large $A_V$ for these objects in \S~\ref{EC}. The relatively low extinction values calculated for \#68, 69, 82 and 
114 ($A_V$ = 5.4, 3.7, 5.7, and 3.0 respectively) signal an envelope does not strongly affect these systems. These objects are also close to the disk/envelope boundary in $n_{5-12}$, especially \#67, 68, 69, and 82 which have values between -0.20 and -0.09. This `boundary' is tentative and an approximation based on the properties and locations of objects in Taurus and Ophiuchus in this plot. 
The uncertainties of \#68, 69, and 82 span the disk/envelope boundary.
Five Class I envelope dominated objects [\#16, 19, 20, 113 and 115] display significant 10 $\mu$m silicate emission features in their dereddened spectrum. Silicate emission features can be seen from the dusty disk in a Class II object, from the warm envelope in a face-on Class I object \citep{watson04} and from both a disk and a warm residual envelope in a FS object. These objects appear more evolved than the canonical Class I stage and, though they were not classified observationally by $n_{2-25}$ as FS, these objects resemble
 the historical definition of a FS object as one which is in a transition phase 
between Class I and Class II, whose envelope is present yet tenuous and disk is
beginning to be seen in the SED \citep{casali92, greene96}. This explanation is consistent with the envelope dominated classification using $n_{5-12}$.

Worthy of note in the envelope dominated sample is object \#122, or SSTc2d J032924.1+311958, classified in \citet{merin10} as a likely edge-on disk due to ice and slicate absorption features in the IRS spectrum. Our analysis classifies \#122 as a Class I envelope. As stated in \S~\ref{SEDclass}, the $n_{2-25}$ classification of edge-on disks is known to be misleading. However the extinction free spectral index $n_{5-12}$ classifies 6 of 7 edge-on disks in the Taurus sample of \citet{furlan11} as disks, with the one outlier being HARO 6-5B with an $n_{5-12}$ = 0.26. Object \#122 has an $n_{5-12}$ of 0.69 as well as ice and slicate absorption features which suggest a significant envelope is still present in this system. Further, the spectrum also resembles that of the famous edge-on disk, HH 30 in Taurus \citep{furlan11}.

Of the 55 disks, 5 are Class I, 11 are FS and 36 are 
Class II according to $n_{2-25}$. Three disk dominated 
objects [\#3, 
7, and 35] have no SED class as they were not detected in the $K_s$ band of 2MASS ; as mentioned before, this is a strong sign these objects 
are deeply embedded protostars. The SEDs of \#3, 7, and 35 display strong 
10$\mu$m silicate and 15$\mu$m CO$_2$ ice absorption 
features as well as a steeply rising SED beyond $\sim$ 10$\mu$m. 
For these reasons we reclassify \#3, 7, and 35 as envelope dominated.
Three more objects classified as dominated by disk 
emission require closer examination [\#13, 33, 36, and 63].
Class I based on $n_{2-25}$, \#63 has strong ice absorption features, a steep drop in flux from 8-9$\mu$m and a flat SED $>$12$\mu$m which is reminiscent of the most deeply embedded ($A_V$ $>$ 30) Class II objects in Ophiuchus \citep{mcclure10}.
 The other objects, \#13, 33 and 36, are Class I 
and show strong silicate absorption 
and CO$_2$ ice absorption features. These objects also seem to suffer from high extinction and demonstrate how the extinction-free indices can give you a more accurate idea of evolutionary state, especially for objects with large $A_V$.
Based on a closer examination of the $n_{5-12}$ classification of objects in our sample with respect to their SEDs, our 
YSO sample in NGC 1333 consists of 21 objects dominated by 
envelope emission, 57 by disk emission, and 1 by photospheric emission; the SEDs of these objects are shown in Figure~\ref{envelope}, Figure~\ref{disk}, and Figure~\ref{phot}, respectively.

According to \citet{aspin94}, the photosphere dominated object \#135 (ASR 54) has $K$ = 15.74 for which we would assign it to Class I. The SED of ASR 54 (Figure~\ref{phot})
 shows 15$\mu$m CO$_2$ ice absorption as well as ice absorption around 
10$\mu$m, likely a result of foreground absorption and not the presence of an envelope.
ASR 54 was noted by \citet{gut08} as a candidate for a star which appears to be 
exciting PAH emission near the IRAS 4 filament in the southern NGC 1333. 
From IRAC photometry and the K band detection by \citet{aspin94}, \citet{gut08} 
suggest this is a heavily extinguished ($A_K$ $\sim$ 10) late B type star with 
some excess at 24$\mu$m, implying it might be a transitional 
or debris disk. The extinction-free spectral index $n_{5-12}$ agrees with \citet{gut08}
categorizing ASR 54 as a photosphere. There is certainly an excess at longer wavelengths, suggestive of a circumstellar accretion disk though we lack reliable extinction information which would allow us to determine whether this object is a transitional disk or debris disk.
We also cannot conclusively say whether this is the object
exciting the PAH emission seen in the southern end of NGC 1333.

\subsection{Class Fractions}
From our sample of full IRS spectra 
and 2MASS $K_s$ band photometry, the empirical $n_{2-25}$ Class I 
fraction is 23\%$\pm$12\% (16/70) and the fraction of sources dominated 
by 
envelope emission, as defined in the $n_{5-12}$ analysis in \S~{\ref{extfree}}, 
is 27\%$\pm$11\% (21/79). 
\citet{gut08} found 39 of the 137 members of NGC 1333 were protostars 
resulting in a protostar class fraction of 28\%.
Our findings are consistent with 
\citet{gut08} and reaffirms the protostar fraction is quite high 
 in NGC 1333 which implies the presence of a substantial young population.  

In comparison to a cluster of similar age analyzed in the same way, such as Ophiuchus 
\citep[$\sim$0.8 Myr, envelope fraction 9\% (10/106), 
Class I fraction 25\%;][]{mcclure10}, NGC 1333 has a significantly 
larger envelope fraction and a similar Class I fraction. 
Using Mathis's extinction curve with $R_V$ = 5, i.e. 
$\frac{A_V}{A_{J}}$ = 3.06, the median $A_V$ for Ophiuchus is 10.7 mag \citep{mcclure10}
and the median $A_V$ for NGC 1333 is 5.8 mag (see \S~\ref{EC}). 
While 
Ophiuchus is heavily extinguished, which skews age estimates 
from class fractions to the younger side, as extinction can make a disk have the SED class of an envelope \citep{pont05}, NGC 1333 is not as affected 
by extinction and therefore its large envelope fraction demonstrates NGC 1333 is genuinely young. 
Given this, NGC 1333's larger envelope fraction would imply it is even younger than Ophiuchus whose age is estimated to be $\sim$0.8 Myr\citep{mcclure10}. 

The fraction of our sample with disk dominated emission, 
as defined in the $n_{5-12}$ analysis in \S~{\ref{extfree}}, 
is 72\%$\pm$11\% (57/79). Our sample of 70 objects with an SED class yields a FS fraction of 20\%$\pm$12\% (14/70) and a Class II 
fraction of 57\%$\pm$12\% (40/70). \citet{gut08} identify 98 Class II objects in their 137 YSO sample, yielding a  Class II fraction of 72\%. Our results are consistent with \citet{gut08}'s findings as their Class II sample is comparable to our disk dominated sample rather than our Class II sample as they do not classify objects as FS. Comparison with Ophiuchus, which has a larger fraction of objects with disk dominated emission (85\%; 90/106), further demonstrates NGC 1333 is likely to be younger than Ophiuchus.

Our 
selection for objects with excess in the 8$\mu$m IRAC and 24$\mu$m MIPS bands does little to bias these class fractions as disk and envelope dominated objects typically have substantial IR excesses at 8 and 24$\mu$m. 
In the study of NGC 1333 X-ray sources by \citet{winston10},
 they found 41 Class III objects in their 95 object sample (43\%). \citet{winston10} selected for likelihood of X-ray activity and therefore SED class fractions would be biased toward later stages of star formation which have higher X-ray activity. \citet{gut08} calculated a disk fraction of NGC 1333 using sources within a circle of radius of 5.5' or 0.4 pc centered at the median right ascension and declination of the YSOs they identified. Those authors calculate a disk fraction of 83\%$\pm$11\% as the fraction of the sample which have $K_s$ $<$ 14 to all objects detected at $K_s$ after accounting for field star contamination. This method helps to mitigate biases against diskless members, which would make up the remainder of the cluster specifically 17\%$\pm$11\%. In a study of low-mass stars and substellar objects, \citet{wilking04} estimate a disk fraction for the northern cluster of NGC 1333 to be 75\%$\pm$20\%, implying a diskless population of 25\%$\pm$20\%. They also use $K$-band excess for this calculation and correct for their incomplete census of members with disks.

\section{Disk Analysis} \label{Anal}
In our analysis, we focus on the radial and vertical structure of the disk as well as the degree of processing of dust in the disks. We then statistically compare the structure of disks in NGC 1333 to Taurus, Chamaeleon and Ophiuchus. The sample for our disk analysis consists of all disk dominated objects, according to $n_{5-12}$, for which we have properly characterized the extinction and which show no evidence of an envelope; this amounts to 44 of the 57 objects identified as disk dominated (8 FS and 36 Class II).
\subsection{Vertical Structure of Dust} \label{vert}
When a protostar forms it is surrounded by an infalling cloud of gas and dust which eventually lands as a flared accretion disk around the star. 
As the disk evolves, small dust grains 
coagulate, become larger, and eventually settle out of 
the upper layers of the disk down to the midplane. This 
decreases the amount of dust in the upper layers of the 
disk and reduces the heating and associated flaring which causes the disk to become flatter \citep{dalessio06, dullemond07}.

The continuum emission from circumstellar accretion disks can be characterized through 
the use of spectral indices which sample silicate-feature-free emission, such as $n_{6-13}$ and $n_{13-31}$ \citep{dalessio06, furlan06, watson09,furlan09}. 
The majority of the emission characterized by $n_{6-13}$ and $n_{13-31}$ is shown by \citet{dalessio06} to originate in two distinct parts 
of the disk. Emission from the inner parts of a radially continuous disk is characterized by $n_{6-13}$, which generally corresponds to 0.1-3 AU for 
the least settled disks and 0.1-1 AU for the most settled disks. The middle to outer regions of radially continuous disks can be characterized by $n_{13-31}$, around 3-25 AU in the least 
settled disks 1-10 AU for the most settled disks \citep{dalessio06}. 

We calculate $n_{6-13}$ and $n_{13-31}$ 
for the our sample of disks in NGC 1333, as described above, using Equation~\ref{eq1} with fluxes at 6, 13, and 31$\mu$m 
derived by integrating the extinction corrected IRS spectra from 5.4-6.0$\mu$m, 
12.8-14.0$\mu$m, and 30.3-32.0$\mu$m respectively. 
The uncertainties in these spectral 
indices are propagated from the uncertainty in the original spectra. 
Some objects in the sample have significant uncertainty in their spectrum redward of 30$\mu$m 
resulting in a large uncertainty in $n_{13-31}$. The increase in noise seen in some spectra at wavelengths longer than 30$\mu$m can be attributed to the data being taken late in the {\it Spitzer} mission (Campaigns 44, 48 and 49) after the detector bias was decreased in the LL array.
 The degree of settling in the disk can be inferred through comparison 
of these continuum indices with disk models which take into account settling.

We compare our sample with predictions from models of irradiated 
accretion disks following the formalism of \citet{dalessio98,dalessio99, dalessio01, dalessio05, dalessio06}.
In the models used in this work, the surface density (and hence mass) of the disk is proportional to the mass accretion rate, $\dot{M}$, over the Shakura-Sunyaev viscosity parameter, $\alpha$.  We chose to fix $\alpha$ at a typically adopted value of 0.01 and vary the mass accretion rate.  The reverse could have also been performed.  Therefore, varying $\alpha$ is equivalent to varying the mass accretion rate.
These models probe the effects of turbulence on the heating of the disk
and thus the shape of the SED. Turbulent mixing has a non negligible
effect over the vertical structure of the disk (i.e., \citet{zsom11},
\citet{ciesla10}). However, studying the role of the turbulence as a mixing
mechanism and its effects on the emission of the disk is beyond the
scope of this paper.
Following these models, a grid of disks around stars with 
$M_*=0.2$ and $0.5$~M$_\odot$ was constructed \citep{espaillat09}. The disks 
have $\alpha=0.01$, mass accretion rate $\dot{M}$ ranging from 
$10^{-10}$ to $10^{-7}$~M$_{\odot}$ yr$^{-1}$, and a dust depletion 
parameter $\epsilon$ of 0.001, 0.01, 0.1 and 1 (no settling). 
The continuum indices $n_{6-13}$ and $n_{13-31}$ were computed for 
disk inclinations of $20^{\circ}$, $40^{\circ}$, $60^{\circ}$ and $80^{\circ}$. 

We compare these models 
with our observations in Figure~\ref{n613_n1331}.
Also in Figure~\ref{n613_n1331} 
are the observed data for Taurus \citep[middle panel;][]{furlan06} and our disk sample in NGC 1333 
(bottom panel).
\citet{furlan05, furlan06} and \citet{dalessio06} demonstrate as the settling parameter 
$\epsilon$ decreases, $n_{6-13}$ and $n_{13-31}$ also decrease 
for the models described above,
therefore lower values of $n_{6-13}$ and $n_{13-31}$ represent 
more settled disks overall. As discussed previously, it is important to remember these 
indices characterize the settling of dust in the upper layers of different spatial regions of the disk. 
Overall the objects in NGC 1333 appear to be well explained by models with the $\epsilon$ = 0.01 to 0.001. 
Dust depletion factors of 0.01 or 0.001 means 99 to 99.9\% of the small grains, presumably from the ISM, have grown and sunk to the disk midplane, where they cannot be easily seen.  This leads to an optically effectively flatter disk, with a bluer spectrum, as discussed by \citet{furlan05, furlan06, furlan09, furlan11, dalessio06}.
We find no statistically significant difference in the distribution of these continuum indices, as discussed in \S~\ref{contind}.

\subsection{Radial Structure} \label{radial}
\subsubsection{Continuum indices and disk geometry} \label{indices_geometry}
Another continuum index, $n_{2-6}$, characterizes 
the excess emission from the inner most region of the disk \citep[$\lesssim$ 1 AU;][]{mcclure10}.
A disk which displays a large excess at longer wavelengths but has a very negative $n_{2-6}$ is 
similar to the SED of CoKu Tau/4 in Taurus, which has been modeled to 
have the geometry of an inwardly truncated disk with a completely cleared out inner hole with radius $\sim$10 AU  \citep{dalessio05,calvet05}. 
TDs, as defined in this analysis, have little to no excess from a disk at short wavelengths ($<$ 8$\mu$m) and show IR
excess emission from an optically thick disk at longer wavelengths.
PTDs, like UX Tau A, have an inner disk at $\lesssim$1 AU which would result in an short wavelength excess similar to that of a full disk though they posses one or more gaps and an outer optically thick disk. It is important to note that $n_{2-6}$ does not necessarily characterize the amount of excess above the photosphere.

The $n_{13-31}$ continuum index, as discussed previously, probes properties of dust in a radially continuous disk about 1-10 AU from the star \citet{dalessio06}. 
For objects with interrupted radial structure, 
such as TDs, $n_{13-31}$ characterizes the excess originating from the outer disk wall, 
as the SED will show a steep rise at longer wavelengths which is essentially the superposition of a blackbody at the temperature 
of the disk wall onto the spectrum of the star \citep{dalessio06,espaillat07, espaillat09}. 
For TDs whose outer disk wall is heated due to stellar irradiation and accretion shocks on the stellar surface, we expect this warmer optically thick dust to manifest itself in the spectrum as a high $n_{13-31}$ \citep{dalessio06,furlan06}.

In this way, we can identify objects which have low $n_{2-6}$ and high $n_{13-31}$ as TDs \citep{mcclure10, manoj11}.
PTDs in this analysis are disks with one or more gaps, resulting in at least an inner disk and an outer disk. 
Like TDs, PTDs with large gaps will have an increase in the continuum from 13-31$\mu$m from the wall of the outer optically thick disk. 
For this reason, a large $n_{13-31}$ can identify PTDs with large (10s of AU) gaps, though PTDs with one or more smaller gaps would appear to have an $n_{13-31}$ similar to a radially continuous disk \citep{espaillat07,espaillat09,furlan09,mcclure10}.

In Figure~\ref{n26_n1331}, we plot $n_{13-31}$ vs. $n_{2-6}$ for the
sample of disks as done in \citet{mcclure10} to separate out objects whose radial structure is 
broken up by holes or gaps. As in 
Ophiuchus, most of the disks in NGC 1333 fall within -2 $<$
$n_{2-6}$ $<$ -0.7 and -1.33 $<$ $n_{13-31}$ $<$ 0.3. 
In Ophiuchus and Taurus, 
PTDs and TDs are found in the upper octile in $n_{13-31}$ with TDs in the lower octile of $n_{2-6}$ and PTDs 
between the lower octile and the median \citep{mcclure10}. Applying this classification 
to NGC 1333, \#137, the outlier in the TD range of $n_{12-20}$, also stands out here in $n_{13-31}$ and has a very steep slope from 2-6$\mu$m placing it in the TD range as defined by Ophiuchus and Taurus of the $n_{13-31}$ vs. $n_{2-6}$ plot.
Six objects [\#55, 75, 101, 110, 131, and 134] fall within or very close to the supposed PTD range of $n_{2-6}$, as they are outliers in $n_{13-31}$. All of these objects appear to be quite faint from 1-35$\mu$m with $\nu$F$_{\nu}$ $\sim$ 10$^{-11}$ to 10$^{-12}$ ergs s$^{-1}$ cm$^{-2}$. Two objects, \#55 and 75, do not appear to have a rising SED at longer wavelengths but instead have a $n_{13-31}$ outlier as a result of noise in the spectrum $>$ 30$\mu$m.
The three objects in the PTD range with the largest values of $n_{13-31}$, \#101, 110, and 131, appear to have SEDs similar to TDs, as they show steep slopes from 2-6$\mu$m and flux excess from 13-31$\mu$m. However, \#131 
has an excess $<$8$\mu$m similar to that of a disk and is likely not a TD. These objects do have a steep slope from 2-6$\mu$m, with $n_{2-6}$ $\sim$ -1.7. A small amount of dust in a central clearing would cause the slope to be slightly flatter, steeper than a full disk but flatter than a photosphere. For these reasons we consider \#101 and 110 as possible TDs.

The object with the lowest value of $n_{13-31}$ in the PTD range, \#134, does not look like any known PTDs \citep{espaillat09, espaillat10} but it shows a short wavelength excess similar to a disk, a significant dip in flux between 10 and 20$\mu$m and starts to rise out to about 30$\mu$m. We do not classify \#134 as a TD or PTD though note the strong dip in the middle of the IRS spectrum possibly indicates interrupted radial structure and makes this object a good candidate for further study. 
\citet{winston10} identified \#136 as a TD based on its IRAC and MIPS photometry. Though it is not an outlier in $n_{13-31}$, \#136 has the steepest $n_{2-6}$ (-2.6) which supports its classification as a TD. We will discuss \#136 further in \S~\ref{W10}. As mentioned previously $n_{2-6}$ is a slope and does not necessarily characterize the amount excess above the photosphere. For this reason it may not be able to identify some types of PTDs and TDs.

There are two blue outliers [\#53 and 123] in $n_{13-31}$ with an $n_{13-31}$
below -4/3, the limit for an infinite geometrically thin, optically 
thick disk. Low values of $n_{13-31}$ have been found in models 
of outwardly truncated disks \citep{mcclure08} and for this reason we suggest \#53 and 
\#123 have outwardly truncated disks. Important to note is objects may also have an $n_{13-31}$ below -4/3 if they are no longer optically thick, a sign of a more evolved state. However, we do not believe this is the case for these objects as they are not identified as Class III from \S~\ref{SEDclass} or photosphere dominated from \S~\ref{extfree}.
The presence of a companion is known to truncate a disk \citep[e.g.][]{artymowicz94} as in 
the case of SR20 in Ophiuchus \citep{mcclure08}. 
The lack of established
multiplicity information for the sample does not allow us to address definitively 
the multiplicity for these truncated disks in NGC 1333. 

\subsubsection{Silicate emission and disk geometry} \label{W10}
The broad silicate emission features in the spectrum of disks at 10 and 20$\mu$m arise from optically thin dust at the disk surface, and can be used to probe the dust properties in the upper layers of the 
disk \citep{malfait98,natta00, dalessio06}. For objects whose structure deviates from being radially continuous, such as TDs, the emission seen in these features comes
from optically thin dust within the gap or hole, such as in GM Aur, or from optically thin dust in the surface 
layers of the outer disk wall, such as in CoKu Tau/4 \citep{dalessio05}.
 Therefore measuring the 
strength of the 10 and 20$\mu$m silicate emission can also be used to probe the radial structure of the disk. 

The equivalent 
width of the silicate emission features in the IRS spectra is a measure of the optically thin emission per unit area of optically thick emission
from dust in the disk. The equivalent width is characterized by the sum of the excess flux from the silicate feature at each wavelength, $F_{\lambda}$, with respect to the flux of the underlying continuum, $F_{\lambda,cont}$. Equivalent width is given by
\begin{equation}
EW_\lambda = \int_{\lambda_1 }^{\lambda_2} \frac{F_{\lambda} - F_{\lambda,cont}}{F_{\lambda,cont}} d \lambda.
\end{equation}
In our analysis we adopt $\lambda_1$ = 8$\mu$m and $\lambda_2$ = 13$\mu$m and $\lambda_1$ = 16$\mu$m and $\lambda_2$ = 28$\mu$m in order to measure the 
10 and 20$\mu$m silicate emission features respectively. We adopt positive values of EW$_\lambda$ for emission features.

While the equivalent width expresses the strength of the silicate 
feature in units of its underlying continuum, the integrated continuum subtracted
flux of the silicate feature allows us a probe how much optically 
thin dust is actually producing this emission. For this reason, we 
also calculate the flux of the 10 and 20$\mu$m silicate features using
\begin{equation}
 F_\lambda = \int_{\lambda_1 }^{\lambda_2} (F_{\lambda} - F_{\lambda,cont}) d \lambda 
\end{equation}
with the same limits as above. The EW and integrated flux are calculated in the same way as \citet{watson09, furlan09, mcclure10, manoj11}.

We determine the continuum for the calculation of both properties by fitting 
the spectrum with a 3rd to 5th order polynomial with the fit anchored at 
5.61-7.94$\mu$m, 13.02-13.50$\mu$m, 14.32-14.80$\mu$m, 30.16-32.19$\mu$m 
and 35.07-35.92$\mu$m which \citet{watson09} showed to be free of distinct silicate 
emission features. For objects where a single polynomial did not produce
 an adequate fit, we used two polynomials to fit the continuum from 
5-14$\mu$m and 14-36$\mu$m.
 For a few cases a 
smoothly varying fit which did not exceed the extinction corrected flux of the spectrum could not be found. For these objects 
the equivalent width and integrated flux from the portion of the spectrum for which a satisfactory fit could not found 
are left out of our analysis.
 We calculate 
EW$_{10}$, EW$_{20}$, F$_{10}$ and F$_{20}$ as well as their 
uncertainties for 44 of the 57 disk dominated objects with no evidence of an envelope as discussed in \S~\ref{vert}; these values are listed in Table~\ref{tbl4}. The main source of 
uncertainty in these calculations results from the 
continuum fit, the uncertainties noted in Table~\ref{tbl4} are 
calculated assuming an uncertainty in the continuum fit of 10\%, though in some cases the uncertainty is likely to be larger. A continuum fit 
successfully extracting the 33$\mu$m crystalline forsterite feature could not be found 
for all but a few of the spectra as a result of significant uncertainty 
at wavelengths $\ga$ 30$\mu$m as discussed in \S~\ref{vert}, therefore we do not discuss this feature in our analysis.

The EW can be used to identify PTDs which have gaps which result in a deficit of 
optically thick emission \citep{furlan09}. Equivalent width measures optically thin emission with respect to optically thick emission, therefore for the same amount of optically thin emission, a lack of optically thick emission would manifest itself in a large EW. If the gaps have some amount of optically thin 
dust this will further enhance their ability to be identified by having a 
large EW. In Figure~\ref{W10_n1331} we plot $n_{13-31}$ vs. EW$_{10}$ for our sample and the Taurus sample from \citet{furlan06}, along with a polygon indicating the boundaries in these indices 
which correspond to boundaries found by the models of radially continuous, irradiated disks described in \S~\ref{vert}. 
These boundaries are a function of the parameters varied in the models (e.g., $\dot{M}$, M$_\odot$ and disk inclination). We also note that the EW$_{10}$ depends on the settling parameter,$\epsilon$ which we also varied in the models.
Objects with 
EW$_{10}$ and $n_{13-31}$ which fall outside of this polygon have an interrupted radial 
structure, meaning the disk has one or more gaps or is inwardly or outwardly truncated. This is reflected the positions of the well modeled TDs and PTDs in Taurus such as CoKu Tau/4, GM Aur and LkCa 15. Figure~\ref{polygon} demonstrates the areas outside the polygon of radially continuous disk models which generally correspond to disks with altered radial structure in Taurus, Chamaeleon I and Ophiuchus which have been modeled \citep{espaillat07, mcclure08, kim09}.

First, we 
examine the disks which are extreme outliers in EW$_{10}$ through the criterion that their EW$_{10}$ is in the top 12.5\%. Using this restrictive criterion we identify \#50, 52, 57, 73 and 116 as PTDs.
As discussed previously, objects 
which are likely to be TDs possess a large $n_{13-31}$, greater than that of the radially
continuous disk models. While \#110 falls within the PTD range in Figure~\ref{n26_n1331}, it is certainly not an outlier in EW$_{10}$, though it is an outlier in $n_{13-31}$ reinforcing its previous classification as a TD. Curiously, \#110 has a position in Figure~\ref{W10_n1331} similar to UX Tau A, a PTD which has been modeled to have no optically thin dust in the gap resulting in a small EW$_{10}$ \citep{espaillat07}. We cannot confirm nor exclude this possibility without detailed modeling. 
All disks which were in the PTD range of Figure~\ref{n26_n1331}, except for \#110, fall within the polygon of the models in Figure~\ref{W10_n1331}. The other TD in the PTD range from Figure~\ref{n26_n1331}, \#101, falls just within the polygon of radially continuous, irradiated disk models in Figure~\ref{W10_n1331} though its uncertainty in $n_{13-31}$ crosses the top boundary. Of the other disks in the PTD range in Figure~\ref{n26_n1331}, \#75 and 134 have EW$_{10}$ greater than the median, though are not in the upper 12.5\%, therefore are not large enough to be considered significant outliers. The classification of \#137 as a TD from Figure~\ref{n512_n1220} and Figure~\ref{n26_n1331}, is bolstered here as its position in Figure~\ref{W10_n1331} places it outside the polygon of radially continuous disk models and in the TD range. Our analysis agrees with that of \citet{merin10} which classifies \#137 (known therein as source 5 or SSTc2d J032929.3+311835) as a 'cold disk.'

The TD suggested by \citet{winston10} which did not fall in the TD range of Figure~\ref{n26_n1331}, is also not unique here in it falls within the polygon of models. 
Though this object has a `normal' $n_{13-31}$, examination of it SED shows it is a TD. The SED of \#136 is photospheric $<$ 9$\mu$m and the 
continuum slightly rises toward the long wavelength side of the 
spectrum. We interpret this as a sign of a large central hole size, relatively cool outer disk wall, and a settled low mass outer disk. Both the large inner hole and the low-mass settled outer disk prevent \#136 from being identified as a TD based on $n_{13-31}$.

 The case of \#136 suggests $n_{13-31}$ can pick 
out some TDs though others can have an $n_{13-31}$ similar to that of 
a radially continuous disk. The spectral indices $n_{13-31}$ and $n_{2-6}$ are only good at identifying TDs with a certain range of hole sizes. If the hole is too small ($\la$1 AU) or too big ($\ga$50 AU) the indices will not identify them as TDs. It is also possible $n_{13-31}$ can improperly identify radially continuous disks as TDs therefore individual SED inspection is key for proper classification. Further discussion of the classification TDs and PTDs in NGC 1333 is found in \S~\ref{disc}.

\subsubsection{Analysis and comparison of silicate emission features}
With respect to this classification we further analyze the radial structure of disks in NGC 1333 using F$_{10}$, EW$_{10}$, and EW$_{20}$.
Figure~\ref{W10_F10_W20} shows F$_{10}$ (top) and EW$_{20}$ (bottom) plotted 
against EW$_{10}$.
As expected, F$_{10}$ and EW$_{10}$ are well correlated, with Pearson's linear correlation coefficient $r$ = 0.65, and a probability $p_{rand}$ = 0.0024\% that this correlation could have been generated from a random distribution (see \citet{watson09}). A small value of $p_{rand}$, namely $\le$ 1\%, indicates the quantities are significantly correlated.
The F$_{10}$ vs. EW$_{10}$ plot shows most of the PTDs have large values of F$_{10}$ as well as an enhanced EW$_{10}$. A handful of disks have similar F$_{10}$ values as the PTDs but have much smaller EW$_{10}$.
 Objects with enhanced EW$_{10}$ in Ophiuchus \citep{mcclure10} and Chamaeleon I \citep{manoj11} show similar values of F$_{10}$ to radially continuous disks, which suggests these objects have a deficit in optically thick emission as a result of a gap in their disk. Interestingly, the disk with the highest F$_{10}$ also has the highest EW$_{10}$ in NGC 1333, which is not seen in Ophiuchus and Chamaeleon I. The TDs in NGC 1333 have low F$_{10}$, also the case for TDs in Ophiuchus and Chamaeleon I. The flux at 10$\mu$m comes from the inner regions of the disk, therefore a significantly cleared inner disk, as is the case for TDs, would result in a lack of emission at 10$\mu$m.

The bottom panel of Figure~\ref{W10_F10_W20}
compares EW$_{10}$ and EW$_{20}$ and shows the two trace each other well, with $r$ = 0.32 and $p_rand$ = 6.1\%. All of the PTDs, which have enhanced EW$_{10}$, fall above the median for EW$_{20}$. However most of the PTDs have a lower EW$_{20}$ than EW$_{10}$ which is also seen in Chamaeleon I. A large gap would affect the underlying optically thick emission contributing to EW$_{20}$ as well as EW$_{10}$. This trend in lower EW$_{20}$ with EW$_{10}$ outliers possibly suggests most of these EW$_{10}$ outliers have small gaps in the inner disk which only affect the EW$_{10}$. As discussed in \S~\ref{indices_geometry}, the small nature of these gaps would cause them to be more difficult to identify as PTDs with the $n_{2-6}$ vs. $n_{13-31}$ analysis (Figure~\ref{n26_n1331}). As was the case in Ophiuchus and Chamaeleon I, the trends of the continuum indices, fluxes and flux ratios with EW$_{20}$ are weaker versions of the trends with EW$_{10}$. Most of the scatter in EW$_{20}$ is due to the uncertainty in the continuum fit for the 20$\mu$m feature.
 
\subsection{Dust Processing}
One probe of the degree of crystallization and grain growth in the 
inner 1-2 AU of the disk is the flux density ratio F$_{11.3}$/F$_{9.8}$ 
\citep{przygodda03, vanboekel03, vanboekel05, kessler06, honda06, bouwman08, olofsson09}. F$_{11.3}$/F$_{9.8}$ quantifies the shape of the 
10$\mu$m silicate feature and as such tracks the degree of processing of
the dust which creates the silicate emission features. The higher 
the ratio, the more the dust has been processed (more grain growth 
and/or crystallization), with a `pristine' profile defined as one similar to the optically-thin interstellar-grain profile with
F$_{11.3}$/F$_{9.8}$ = 0.34. We calculate 
F$_{11.3}$ and F$_{9.8}$ according to Equation 4 with $\lambda_1$ = 10.9 
and 9.4$\mu$m and $\lambda_2$ = 11.7 and 10.2$\mu$m respectively. 
F$_{11.3}$/F$_{9.8}$ for the sample is listed in Table~\ref{tbl4}. 
An F$_{11.3}$/F$_{9.8}$ representative of the processing of dust in the disk of object \#110 was impossible to obtain due to contamination of F$_{9.8}$ by residual molecular hydrogen emission not removed by sky subtraction.

In Figure~\ref{W10_F11F9} we plot F$_{11.3}$/F$_{9.8}$ against 
EW$_{10}$. The objects with the highest values of 
F$_{11.3}$/F$_{9.8}$ correspond to the lowest EW strength, which is characteristic of radially continuous disks. This inverse relationship has Pearson $r$ = -0.34 and $p_{rand}$ = 2.4\%. 
We also 
notice the trend that as EW$_{10}$ 
decreases the frequency of disks with high F$_{11.3}$/F$_{9.8}$ increases. This trend is seen in Taurus \citep{furlan06}, Ophiuchus \citep{mcclure10} and Chamaeleon I \citep{manoj11}, as well as in other regions \citep{vanboekel05,kessler06,bouwman08, apai05,pascucci09}. 
Also, the PTDs have relatively low 
F$_{11.3}$/F$_{9.8}$, with most of them falling below the median and close to the pristine line. 
This would imply in general the dust seen in the PTDs has experienced 
less grain growth and/or crystallization than the radially continuous 
disks. Therefore the optically thin dust at 1-2 AU in PTDs 
is more pristine than radially continuous disks. The trend in low F$_{11.3}$/F$_{9.8}$ (more pristine shape)
in outliers in EW$_{10}$ is also seen in models of PTDs \citep{espaillat09} and observation of PTDs in Taurus \citep{sargent09}, Ophiuchus \citep{mcclure10} and Chamaeleon I \citep{kim09}. Modeling suggests the emission responsible for the shape and strength 
of the 10$\mu$m silicate feature in these disks comes from the optically 
thin dust within the gaps or in the surface layers of 
the outer disk wall \citep{espaillat09}. 
As also seen in Chamaeleon I, the disks with $n_{13-31}$ below -4/3 in NGC 1333 show a slight preference toward higher values of 
F$_{11.3}$/F$_{9.8}$ which implies these truncated or optically thin disks have a higher degree of dust processing than TDs or PTDs. 

\subsection{Statistical comparison to Taurus, Ophiuchus and Chamaeleon I \label{contind}} 
We use a one dimensional two-sided Kolmogorov-Smirnov (K-S) test to determine if the distribution of 
$n_{2-6}$, $n_{6-13}$, and $n_{13-31}$ in NGC 1333 is drawn from a significantly different population
from that of Taurus \citep{furlan06}, Chamaeleon I \citep{manoj11}, or the L1688 cloud in Ophiuchus \citep{mcclure10}. 
To do this we took our sample of 36 Class II disk dominated objects 
and did not include objects which were identified previously as TDs, 
truncated disks or PTDs, leaving us with 27 objects 
representative of the `normal' radially continuous disk population in
 NGC 1333. We culled the data from Taurus, Chamaeleon I and L1688 in the same way. The distribution of these indices is show in Figure~\ref{hist_indices} and p-values inset from a 1D two-sided K-S test comparing NGC 1333 to each 
region.

We find no statistically significant difference between NGC 1333 and 
Chamaeleon I, Taurus, or L1688 in the distribution of $n_{6-13}$ and 
$n_{13-31}$. 
These statistical tests imply there is a similar dispersion in these continuum indices in all four regions. The distribution implies a heavy concentration in the $\epsilon$ = 0.001-0.01 range and a tail that reaches 0.1, but very few objects that demand a well mixed disk.
We do, also, find a difference between NGC 1333 and 
Taurus and L1688 in the distribution of $n_{2-6}$ (p-value = 0.07 and 0.01, respectively). This would imply 
there are significant differences in the slope of the excess emission in the 
innermost regions of disks in these regions. 
NGC 
1333 and Chamaeleon I have distributions of $n_{2-6}$ skewed toward 
bluer (smaller) values. In Chamaeleon I, this is attributed to a spectral type difference between these two regions \citep{manoj11}. 
The same argument can be made for NGC 1333 which has many more objects 
in the M3-M8 than K5-M2 spectral type range and a median 
spectral type of M3. Taurus and Ophiuchus both have a median 
spectral type of M0. The median $n_{2-6}$ value for M3-M8 objects in NGC 
1333 is -1.64 which when compared to the median $n_{2-6}$ for K5-M2 objects 
of -1.50 shows the prevalence of M3-M8 stars in or sample skews the $n_{2-6}$ distribution of NGC 1333 to the smaller side.

Trends with model parameters within the polygon from Figure~\ref{W10_n1331} are shown with arrows in Figure~\ref{polygon} \citep{furlan09, dalessio06, espaillat09}. 
 To determine if there are difference in the scatter of points within this polygon a 2-dimensional two-sided K-S test was performed for 
only the disks in Taurus, Chamaeleon I, Ophiuchus, and NGC 1333 which 
fall within this polygon and have a spectral type later than M0.
 We find no statistically significant 
difference in the 2D distribution of points within this polygon 
between NGC 1333 and Taurus, Chamaeleon I, or L1688, with p-values 
of 0.36, 0.20, and 0.21 respectively.

\subsection{Median Spectra \label{medians_disc}}
We construct a representative median for our sample in NGC 1333 from the 35 objects classified as both Class II and disk dominated observed in program 40525. The objects in our median span a spectral type range from A3 to M8, with 7 between K4 and M2 and 20 between M3 and M8 and a median spectral type of M4. We also include 7 objects with unknown spectral type. The extinction measured in $A_V$ (derived from the $R_V$ = 5 version of \citet{mathis90} extinction 
curve) ranges from 0 to 22.3, with a median of 3.1 and includes seven objects with $A_V$$>$ 10. We follow the same procedure as \citet{furlan09} to construct our median. We normalized the dereddened IRS spectrum of all 35 Class II disks through multiplication by a scalar factor equal to the median dereddened H-band flux of the 35 objects (0.039 Jy) divided by the $H$-band flux of each object, computed the median of the sample at each wavelength and converted to $\nu$F$_{\nu}$. This method effectively normalizes the excess emission for differences in stellar luminosity for stars with spectral type K-M \citep{dalessio99}, which comprise the majority of our median sample (27/28 objects with known spectral type). 

In Figure~\ref{medians} we compare our median to the medians of K5 to M2 stars in Taurus, L1688, and Chamaeleon I previously presented in \citet{furlan09} each scaled to the median $H$-band flux of Taurus. The shape of the NGC 1333 median is quite similar to that of the other medians, across the IRS range. After scaling to the median H-band flux of Taurus, the NGC 1333 median is still fainter than the others. This can be attributed to the spectral type differences between these samples. The Taurus, L1688, and Chamaeleon I medians are for objects with K5 to M2 spectral types while our sample has six objects in the K5 to M2 range and 20 have spectral types later than M2. Late type stars heat their disks less as they are less luminous and therefore irradiate their disks less as well as have lower mass accretion rate resulting in weaker accretion heating. The cooler nature of late type stars manifests itself as fainter IR excess emission.

We compare the shape of the 10$\mu$m silicate features of the medians of Taurus, L1688, Chamaeleon I, and NGC 1333 in Figure~\ref{silicatefeatures}. We do this in the exact same way as described in \citet{furlan09}. The only difference between our Figure~\ref{silicatefeatures} and \citet{furlan09}'s Figure 7 is the addition of the 10$\mu$m silicate feature of NGC 1333 to the plot.
Also plotted in Figure~\ref{silicatefeatures} is a `pristine', ISM-like profile which is represented by the average of the continuum-subtracted and normalized 10$\mu$m silicate feature of LkCa 15 and GM Aur in Taurus \citep{sargent09}. This pristine profile is scaled to the mean 9.8$\mu$m flux of the four regions. All four regions show evidence for grain growth and/or crystallization as they show a silicate profile wider than the `pristine' profile for ISM-like small amorphous silicate grains (sub-$\mu$m).
 NGC 1333 shows a long wavelength wing similar to that of the other regions which suggests similar grain sizes in the optically thin dust. 

\section{Discussion of PTDs and TDs} \label{disc}
We have examined two ways of identifying TDs and PTDs from properties of their spectra in \S~\ref{indices_geometry} and \S~\ref{W10}. The first analysis, that of the $n_{13-31}$ vs. $n_{2-6}$ diagram (Figure~\ref{n26_n1331}), identified outliers in $n_{13-31}$, all but one of which fell within the PTD range as defined by the positions of TDs and PTDs in Taurus and Ophiuchus. Secondly, we analyzed the equivalent width of the 10$\mu$m silicate emission feature and identified strong outliers. An enhanced EW$_{10}$ is suggestive of a deficit in optically thick emission relative to optically thin emission, as would be the case if there are gaps in the disk. We also used the limits of model indices in the $n_{13-31}$ vs. EW$_{10}$ diagram (Figure~\ref{polygon}) to examine the PTDs and TDs from the first analysis. All of the outliers from the first analysis except \#110 and \#137 fell within the boundaries of model disks, though the uncertainty in $n_{13-31}$ for \#101 does cross out of the range of model disks. The TD suggested by \citet{winston10}, \#136, did not stand out in any of the previous analyses, except it had the most negative $n_{2-6}$ which is closest to that of a photosphere.

We now examine the TDs and PTDs in a third way, with respect to a representative median of the population of disks in NGC 1333 from \S~\ref{medians_disc}. 
In this analysis, we identify PTDs as having near-IR excess at or below the representative median of NGC 1333 scaled to the object's $H$-band flux, strong silicate emission with flux exceeding that of the scaled median, and flux comparable to the scaled median longwards of 24$\mu$m. While object \#73 exceeds the flux of the median at long wavelengths, its spectrum resembles known PTDs such as ROX 44 \citep{espaillat09}, as well as the ``kink" disks of \citet{sicilia2011}, for these reasons we classify it as a PTD. We identify TDs as having a flux deficit with respect to the scaled median in the near-IR and flux exceeding the median at wavelengths greater than 24$\mu$m. It is important to note that the dependence of these classifications on emission around 24$\mu$m restricts the range of hole sizes present in transitional objects identified in this manner. The TD identified by \citet{winston10}, \#136, notably does not conform to this definition as it falls below the scaled median for the entire IRS range, however it is clear that this spectrum is representative of a disk transitioning from a the spectrum of a full optically thick disk to that of an optically thin debris disk, and therefore we classify it as a TD. Object \#116 cannot be classified with respect to a scaled representative median of the disks in NGC 1333 as its flux is much greater than the scaled median.

Figure~\ref{median_comp_TDs} shows the dereddened SED of the 9 PTDs and TDs along with the representative median of NGC 1333 scaled to each object's $H$-band flux. Table~\ref{tbl5} shows what classification each analysis suggested and our final classification of these disks. We favored analysis of the SED with respect to the median as it has been reliably used to identify TDs and PTDs in many previous studies \citep{dalessio05, brown07, calvet05,espaillat07,espaillat08, kim09, merin10}. Also, as stated \S~\ref{indices_geometry} and \S~\ref{W10} analyses using $n_{2-6}$ and $n_{13-31}$ are only sensitive to a certain range of hole or gap sizes and will not effectively pick out small holes or gaps. In most cases, classification based on the EW$_{10}$ vs. $n_{13-31}$ agreed with the SED analysis. In all cases the $n_{13-31}$ vs. $n_{2-6}$ analysis, as defined by \citet{mcclure10} disagreed with the classification from the other two methods.

At least 9 of the 44 disks analyzed in our sample of NGC 1333, 20\%, show evidence for gaps or holes in their disks in various analyses. 
In our sample we identify 4 as TDs, which translates to 11\% of disks analyzed in our sample NGC 1333 (4/44). We do not have sufficient multiplicity information for our sample to address the contributions of companions on our transitional disk fraction.
NGC 1333's TD fraction is larger than the transitional disk fractions in the 1-2 Myr Taurus region (3.5\%), $\sim$2 Myr Chamaeleon I region (5.8\%) and L1688 $\sim$0.8 Myr (3.2\%)\citep{furlan09}. Though the sample of disks we analyzed is relatively small (N = 44), observations indicating gaps in disks are suggestive of giant planet formation regardless of sample size \citep{furlan09,kim09,mcclure10,manoj11}. At the young age of $\la$1 Myr, disks in NGC 1333 are forming planets and dissipating their disks. 
Furthermore, \citet{aspin03} cite age estimates for 4 of the PTDs identified in our analysis, \#50, 52, 57, and 116, with ages of 1 Myr, 0.5 Myr, 3 Myr and 1 Myr, respectively. It is important to remember that age estimates of young stars using isochrones are known to be uncertain as discussed in \S~\ref{intro}. However, the dispersion in these ages might suggest the PTD phase can occur more than once in the life of a disk.

The finding of PTDs at very young ages supports theories for the rapid formation of giant planets such as through core accretion with oscillatory migration \citep{rice03} or 
gravitational collapse of gas and dust in the disk \citep{gold73, boss01}. The quick timescale for planet formation demonstrated in NGC 1333 is not currently accounted for in some models of planet formation such as those in 
\citet{alexander09} or in the older core accretion models of giant planet formation \citep{pollack96}. However, our results do not discount theories of ``born again" disks which suggest giant planets could form more than once in the life of a disk \citep{sargent06, watson09}.

\section{Summary and Conclusions} \label{conclusion}
We presented the IRS spectra of 79 YSOs in NGC 1333. We supplemented the IRS spectra with published photometry and constructed SEDs. We used the SEDs, before extinction correction, to classify our sample into SED classes as well as evolutionary states, using ``extinction-free" indices. Based on this classification, our sample in NGC 1333 consists of 21 objects with envelope dominated emission, 57 objects with disk dominated emission and 1 object with photosphere dominated emission. Further, we analyzed various spectral properties of 44 objects which show disk dominated emission and had no evidence of an envelope in their observed SEDs. We compared the shape of the continuum emission as well as silicate emission features to models of radially continuous irradiated accretion disks in an effort to understand the vertical and radial structure of these disks in NGC 1333. We used two different methods in order to identify disks with altered radial structure. With these analyses, as well as comparison with a representative median spectrum, we classified TD and PTDs in our sample. We also compared the properties of disks in NGC 1333 to disks in the Taurus, Ophiuchus, and Chamaeleon I star forming regions in order to investigate potential differences with age and population. Our main results are summarized below:

\begin{itemize}
\item 
We find no difference between the distribution of dust settling continuum indices, $n_{6-13}$ and $n_{13-31}$ in NGC 1333, Taurus, L1688 in Ophiuchus, and Chamaeleon I. However, through comparison of class fractions, we find NGC 1333 is likely to be even younger than the youngest of those three regions, Ophiuchus. This similarity between star forming regions of significantly different ages implies advanced dust settling ($\epsilon$ $\sim$ 0.001) can happen very early in the life of a disk. 
Alternatively, few disks are well mixed even at the young age of $\lesssim$ 1 Myr.

\item
Of the disks analyzed in NGC 1333, 20\% show evidence of radial gaps and clearings and 11\% are identified as transitional disks. These fractions are significantly larger than those of Taurus, Ophichus and Chamaeleon I. We find these observations of NGC 1333 to be evidence of giant planet formation resulting in disk dissipation at a very young age, $\lesssim$ 1 Myr.

\item
Most PTDs, which are large outliers in EW$_{10}$, were not large outliers in EW$_{20}$. This suggests the gaps in these disks which produce an enhanced EW$_{10}$ are not large enough to affect EW$_{20}$. The smaller gaps in these PTDs likely results in an $n_{13-31}$ similar to radially continuous disks and makes them difficult to identify in analyses which use a high $n_{13-31}$ as a criterion.

\item
Multiple analyses are needed to identify PTDs from spectra without detailed modeling. The $n_{2-6}$ vs. $n_{13-31}$ was not effective in selecting PTDs in NGC 1333 and in some cases contradicted the classification via EW$_{10}$ vs. $n_{13-31}$ and comparison to a scaled representative median.

\item
Comparison of a representative median of disks in NGC 1333 to Taurus, Chamaeleon and L1688 demonstrates the median spectrum of NGC 1333 had a similar shape to all three regions, though after scaling was still fainter. The prevalence of late type stars in our NGC 1333 sample can account for this flux discrepancy.

\item
PTDs in our sample NGC 1333 have more pristine optically thin dust in the inner parts of their disks based on analysis of the shape of their 10$\mu$m emission using F$_{11.3}$/F$_{9.8}$. The dust in these disks with radial gaps seems to have experienced less grain growth and crystallization than radially continuous disks.

\end{itemize}

\acknowledgements

This work is based on observations made with the {\it Spitzer Space Telescope}, which
is operated by the Jet Propulsion Laboratory (JPL), California Institute of Technology (Caltech), under the National Aeronautics and Space Administration (NASA) contract 1407.
In this work we make use 
of data products from the Two Micron All Sky Survey, which is a joint project of the 
University of Massachusetts and the Infrared Processing and Analysis Center/California 
Institute of Technology, funded by NASA
and the National Science Foundation. 
This research also utilizes the Vizier and SIMBAD databases,
operated at CDS, Strasbourg, France, NASA's Astrophysics Data System Abstract Service. Support
for this work was provided by NASA through contract number 1257184 issued by JPL/Caltech, JPL contract 960803 to Cornell University, and Cornell subcontracts 31419-5714 to the University of Rochester.

{\it Facilities:} \facility{Spitzer (IRS)}

{}

\clearpage

\clearpage

\clearpage

\clearpage

\clearpage

\clearpage

\clearpage

\clearpage
\begin{figure}
\figurenum{8}
\epsscale{1}
\plotone{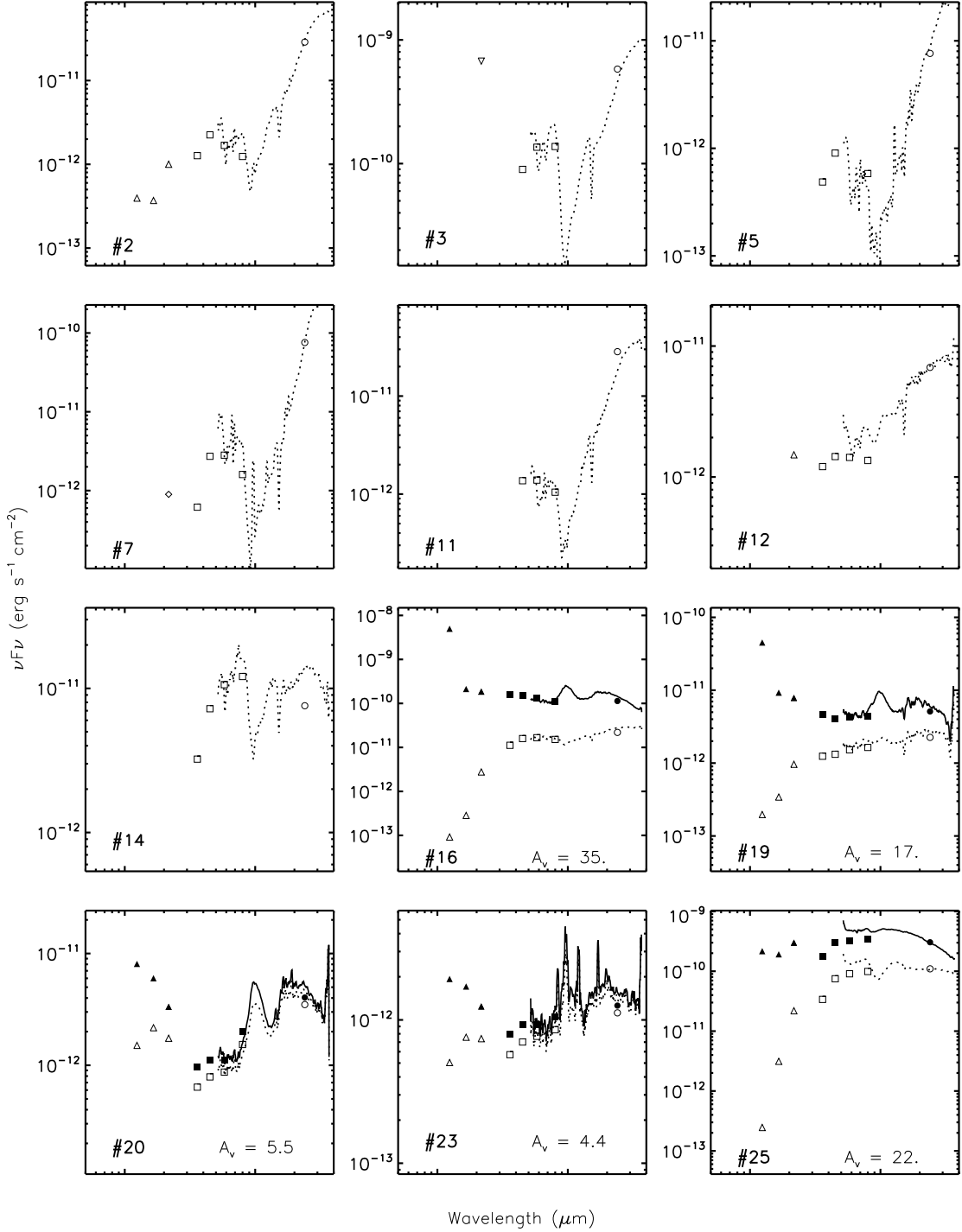}
\caption{SEDs of objects dominated by envelope emission, 
as defined in the $n_{5-12}$ analysis in \S~{\ref{extfree}}. 
The solid line represents the {\it Spitzer} IRS spectrum dereddened using the extinction laws described in \S~\ref{EC} with the $A_V$ calculated therein. The dotted line indicates the reddened, observed spectrum. The dashed line indicates a photosphere for the temperature of a given spectral type from \citet{kenyon95} scaled to the $H$ band flux of the object. The dereddened photometry is solid and reddened photometry is open. The photometry points, if available, are from \citet{aspin94} ($R$, $I$ and $K$; diamond), \citet{lada96} ($K$; upside-down triangle), 2MASS ($J$, $H$, $K_s$;  triangle), IRAC (3.6, 4.5, 5.8 and 8$\mu$m; square) and MIPS (24$\mu$m; circle) from \citet{gut08}. \label{envelope}}
\end{figure}

\clearpage
\begin{figure}
\figurenum{8}
\epsscale{1}
\plotone{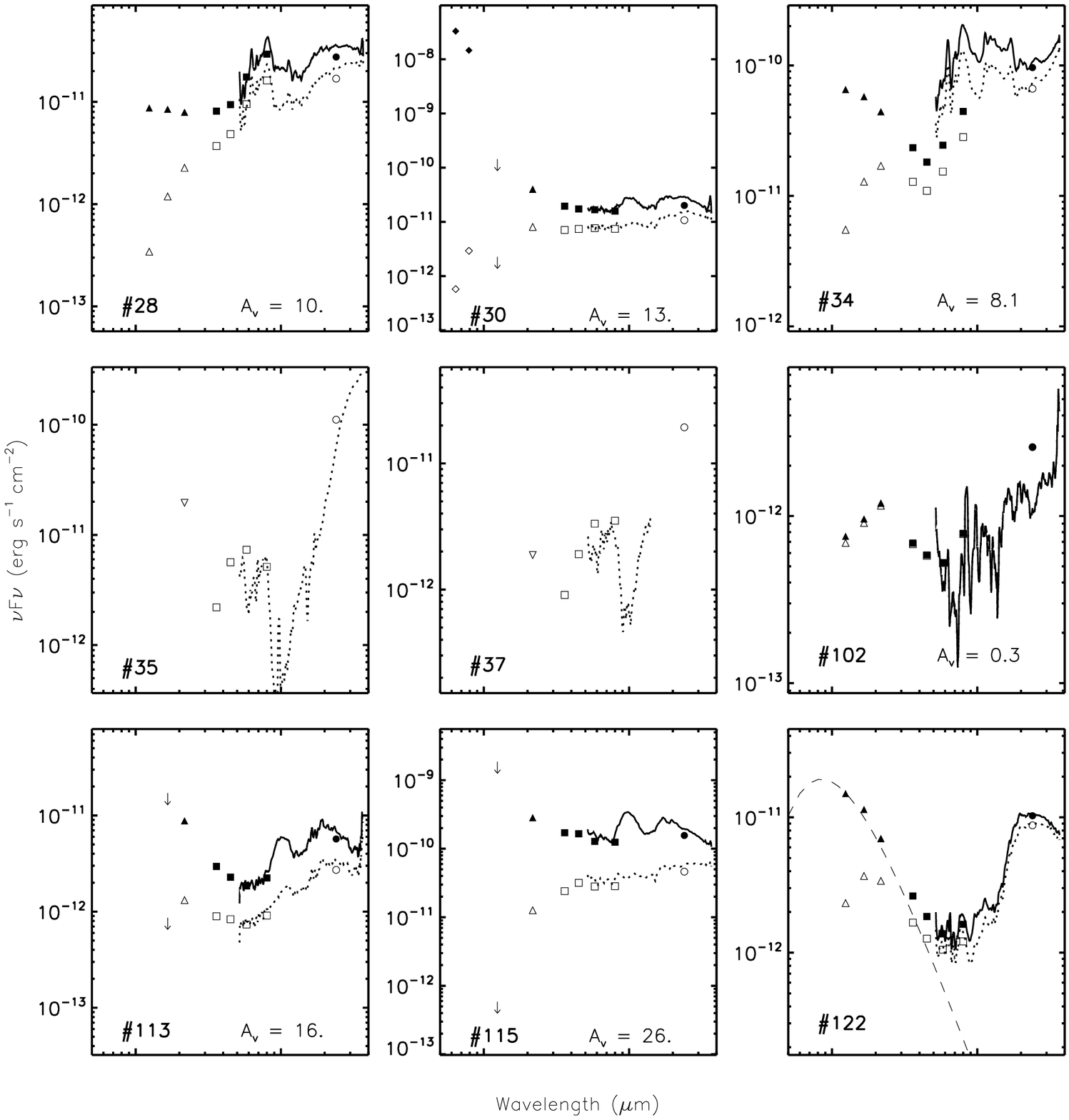}
\caption{Continued...}
\end{figure}

\clearpage
\begin{figure}
\figurenum{9}
\epsscale{1}
\plotone{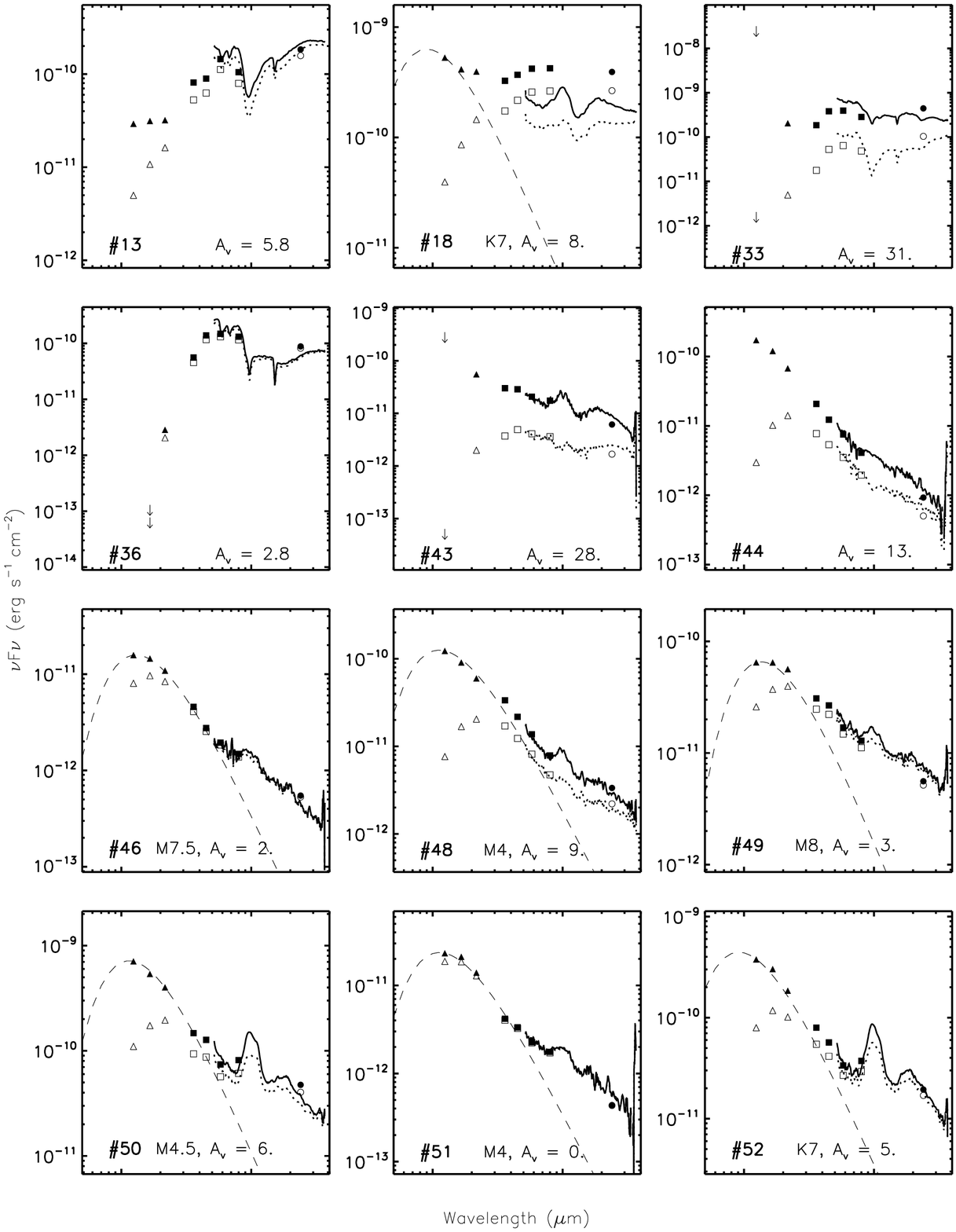}
\caption{SEDs of objects dominated by disk emission, as defined in the $n_{5-12}$ analysis in \S~{\ref{extfree}}. Symbols and lines are the same as in Figure~\ref{envelope}.\label{disk}}
\end{figure}

\clearpage
\begin{figure}
\figurenum{9}
\epsscale{1}
\plotone{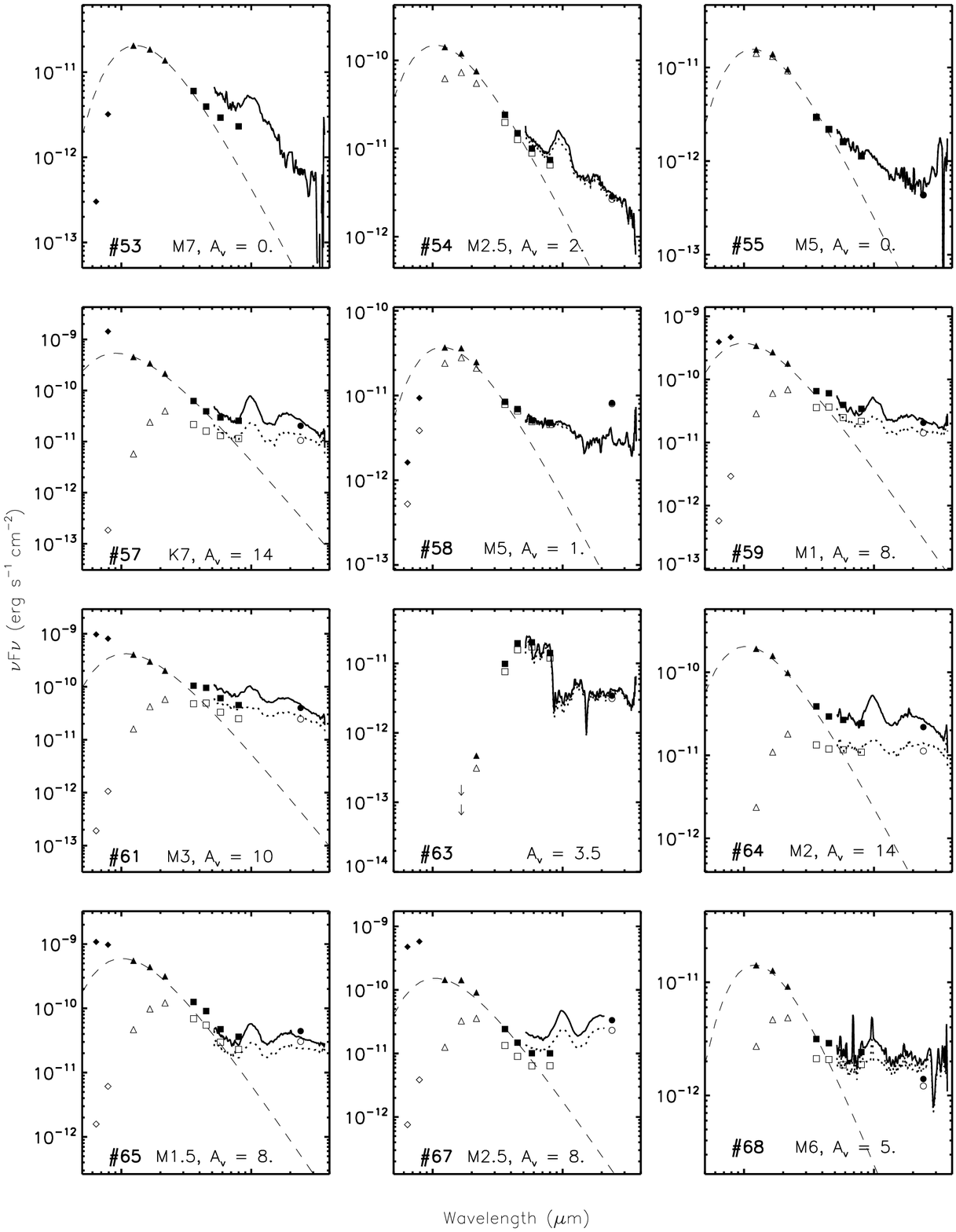}
\caption{Continued...}
\end{figure}

\clearpage
\begin{figure}
\figurenum{9}
\epsscale{1}
\plotone{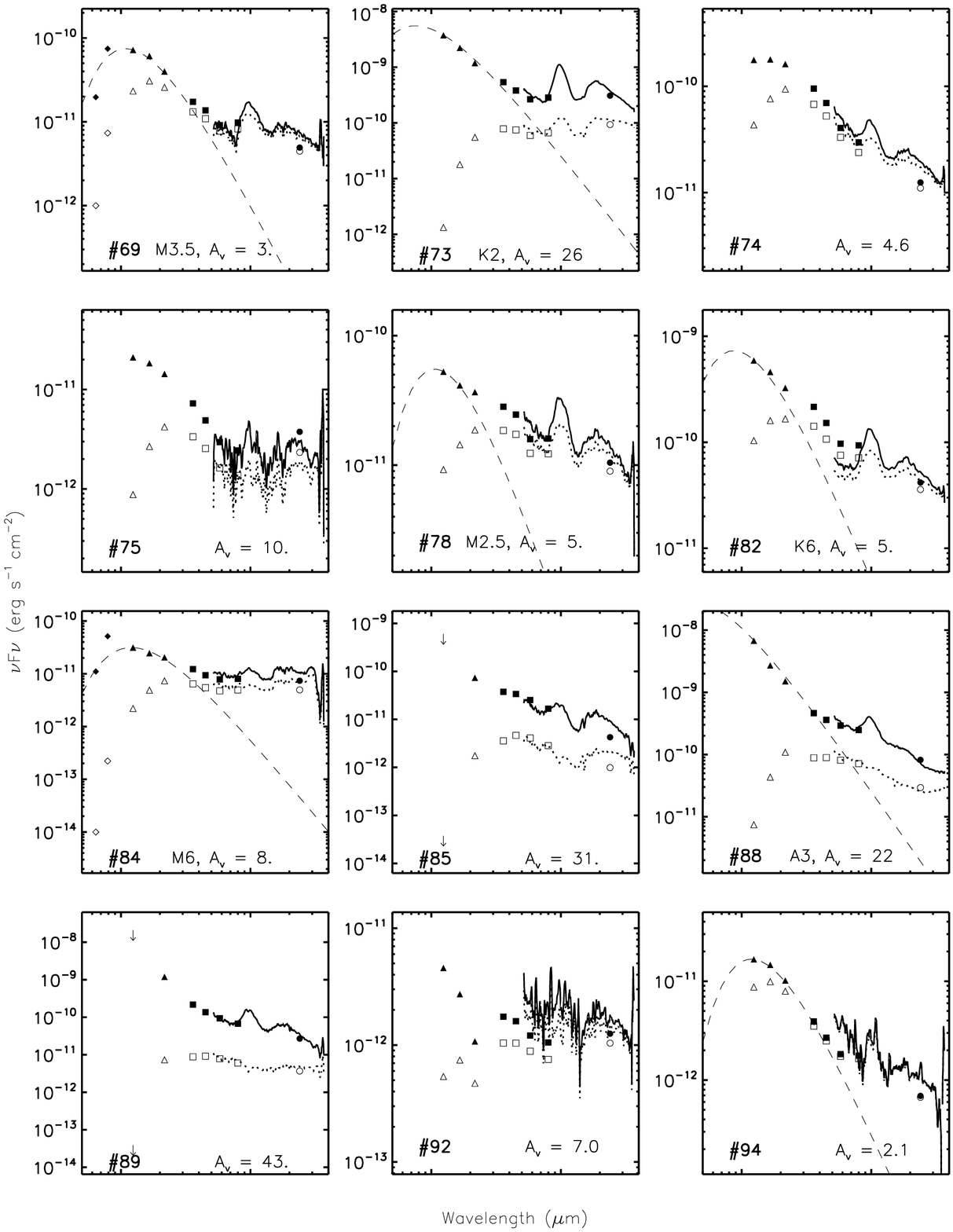}
\caption{Continued...}
\end{figure}

\clearpage
\begin{figure}
\figurenum{9}
\epsscale{1}
\plotone{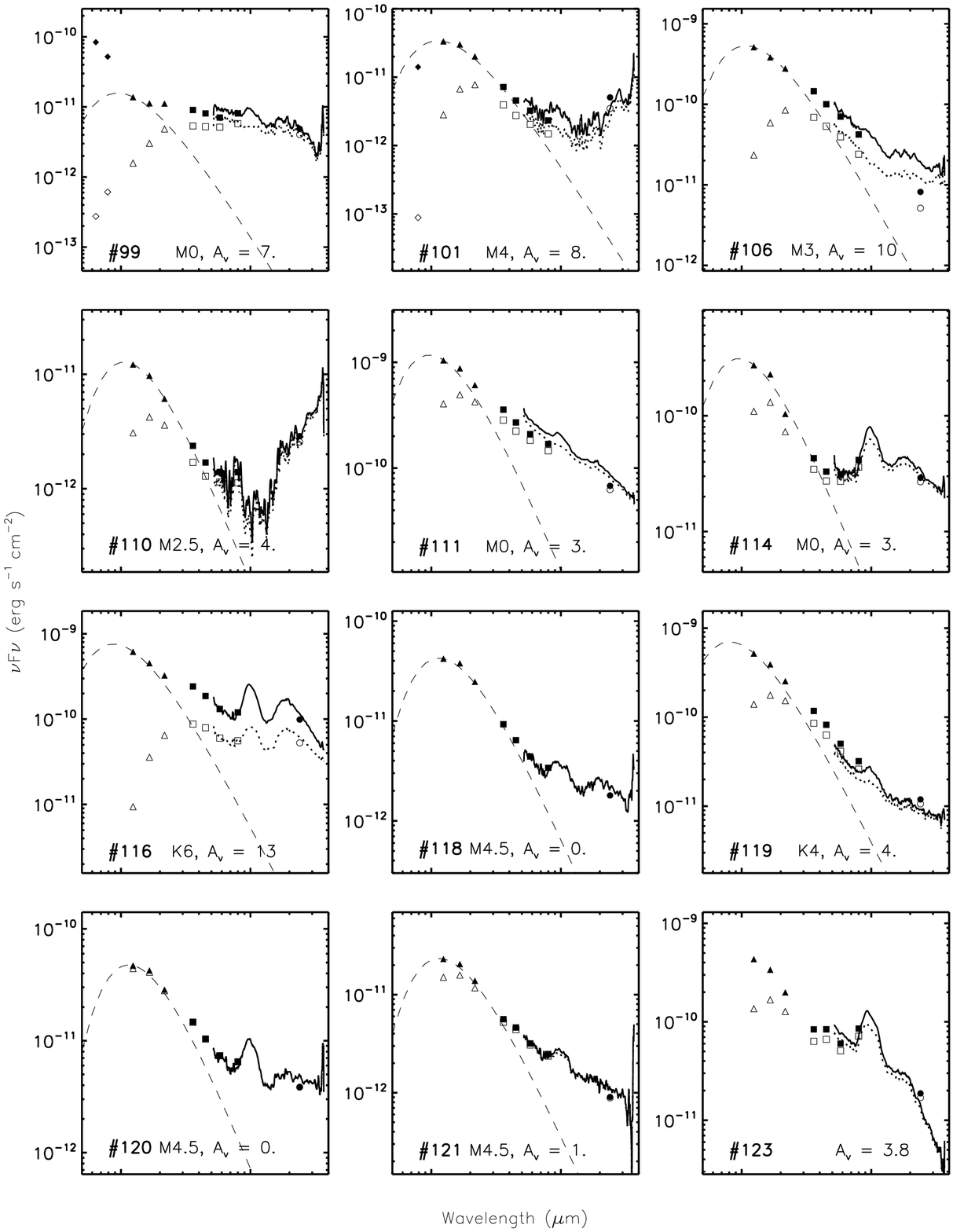}
\caption{Continued...}
\end{figure}

\clearpage
\begin{figure}
\figurenum{9}
\epsscale{1}
\plotone{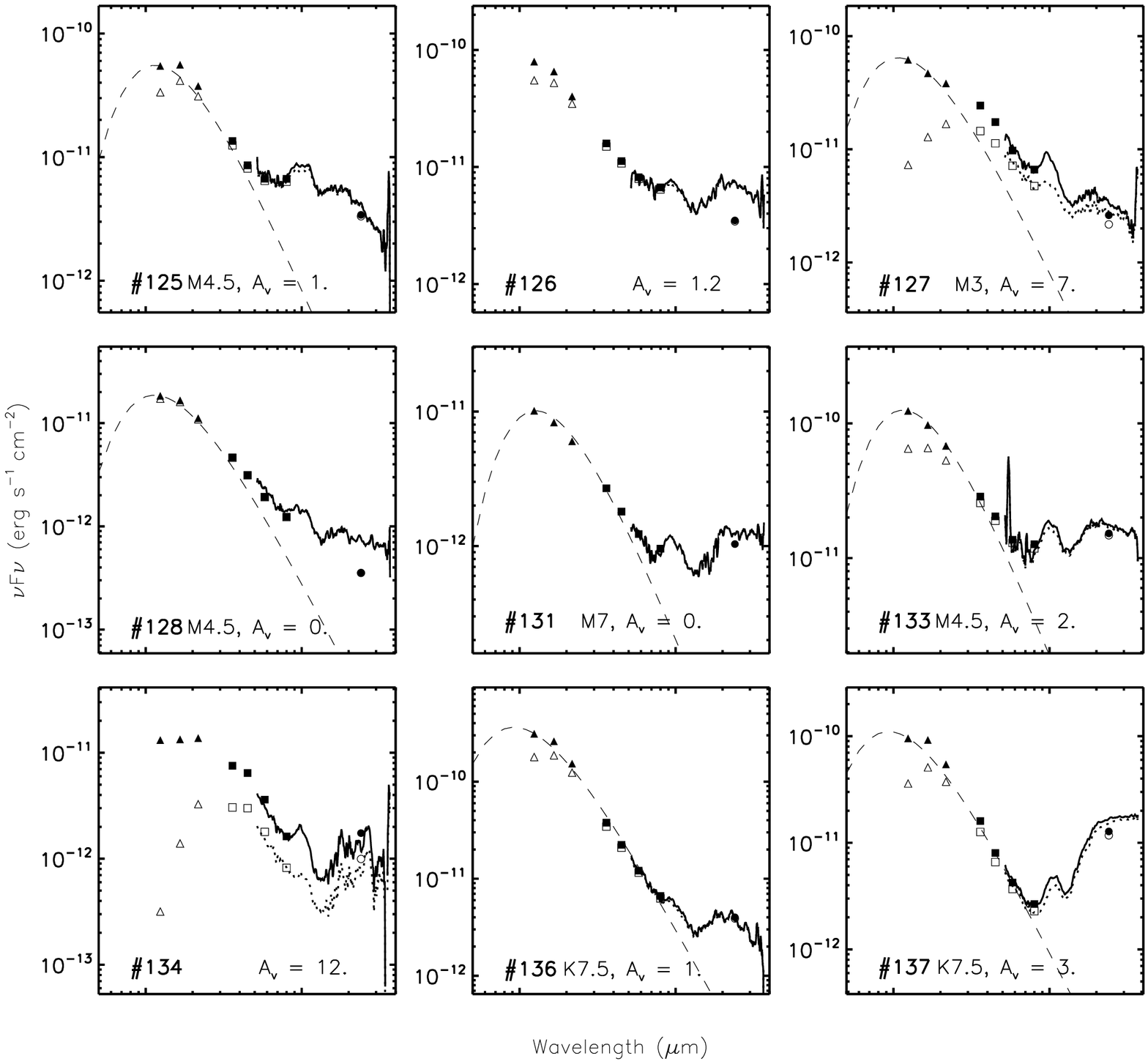}
\caption{Continued...}
\end{figure}

\clearpage
\begin{figure}
\figurenum{10}
\epsscale{1}
\plotone{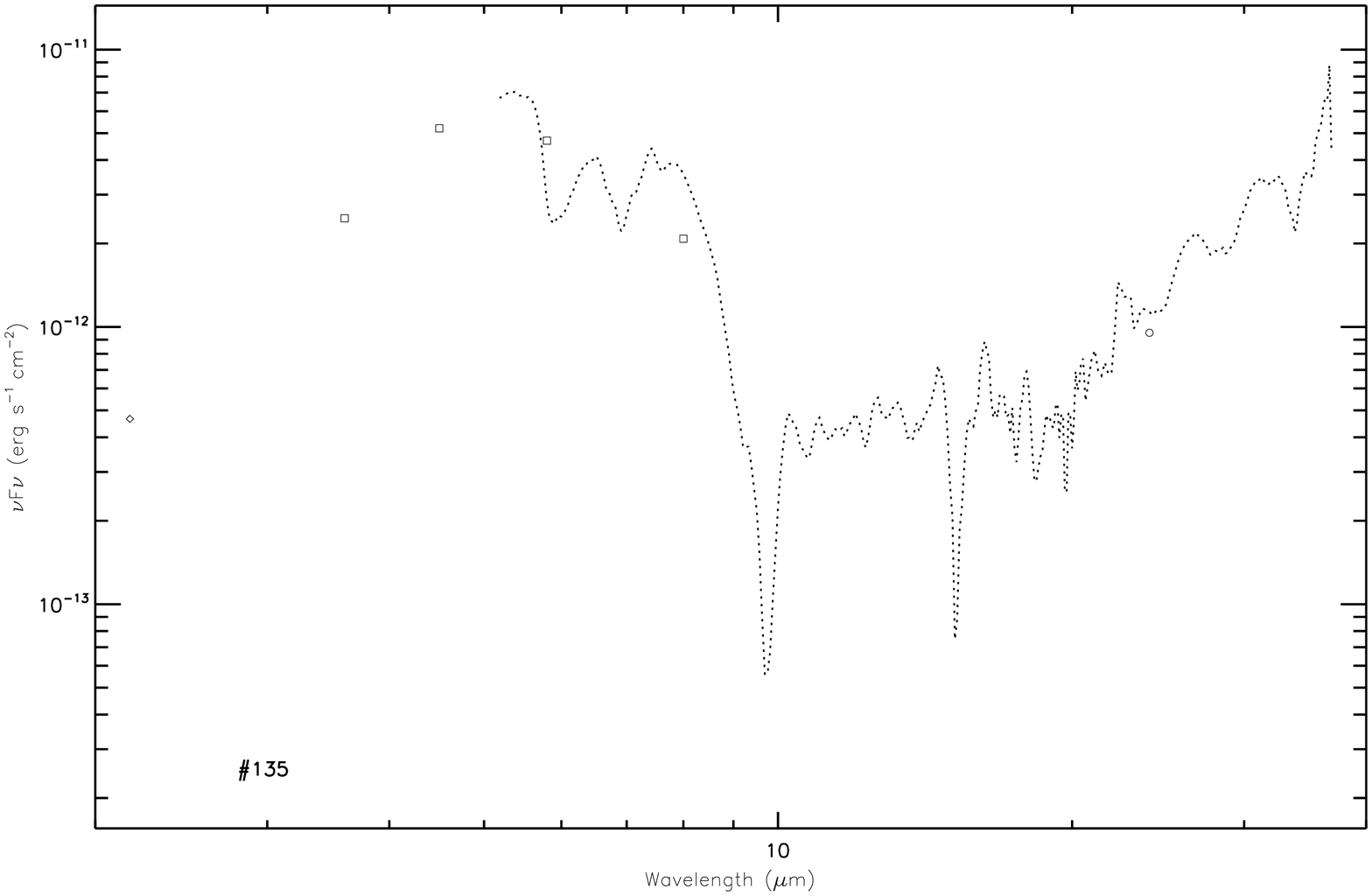}
\caption{Observed SED of photospheric object aka ASR 54. Symbols and lines are the same as in Figure~\ref{envelope}. \label{phot}}
\end{figure}

\clearpage
\begin{figure}
\figurenum{11}
\epsscale{1}
\plotone{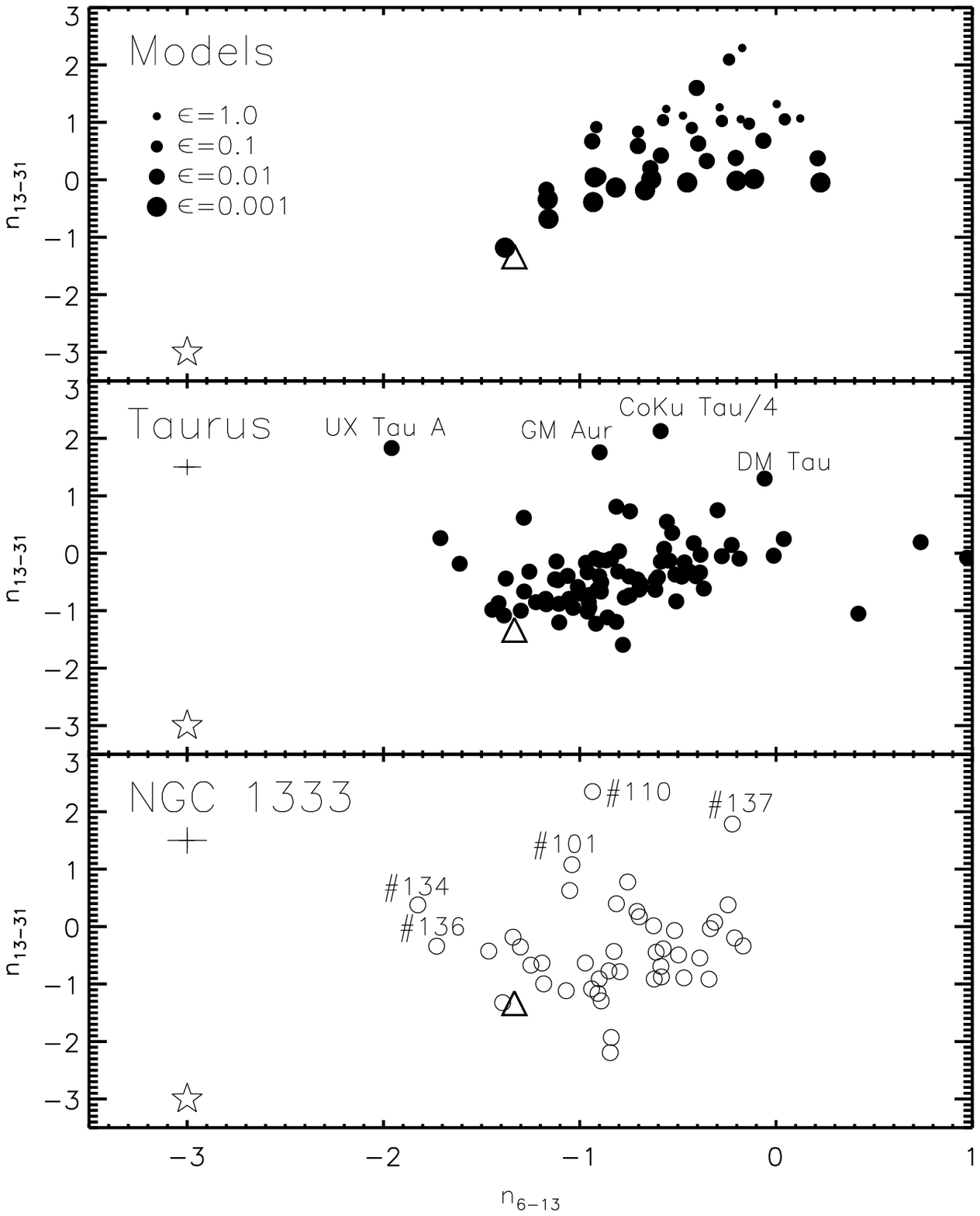}
\caption{$n_{6-13}$ vs. $n_{13-31}$ indices for radially continuous model accretion disks (see \S~\ref{vert} for details), Taurus (indices from P. Manoj, personal communication using spectra from \citet{furlan06}) and NGC 1333. The triangle indicates where $n_{6-13}$ and $n_{13-31}$ are that of a infinite geometrically flat optically thick disk. The value for a photosphere for $n_{6-13}$ and $n_{13-31}$ is -3 and plotted with a star. Similar to Figure 22 from \citet{mcclure10}. Typical uncertainties for Taurus and NGC 1333 are shown below the region's name. \label{n613_n1331}}
\end{figure}

\clearpage
\begin{figure}
\figurenum{12}
\epsscale{1}
\plotone{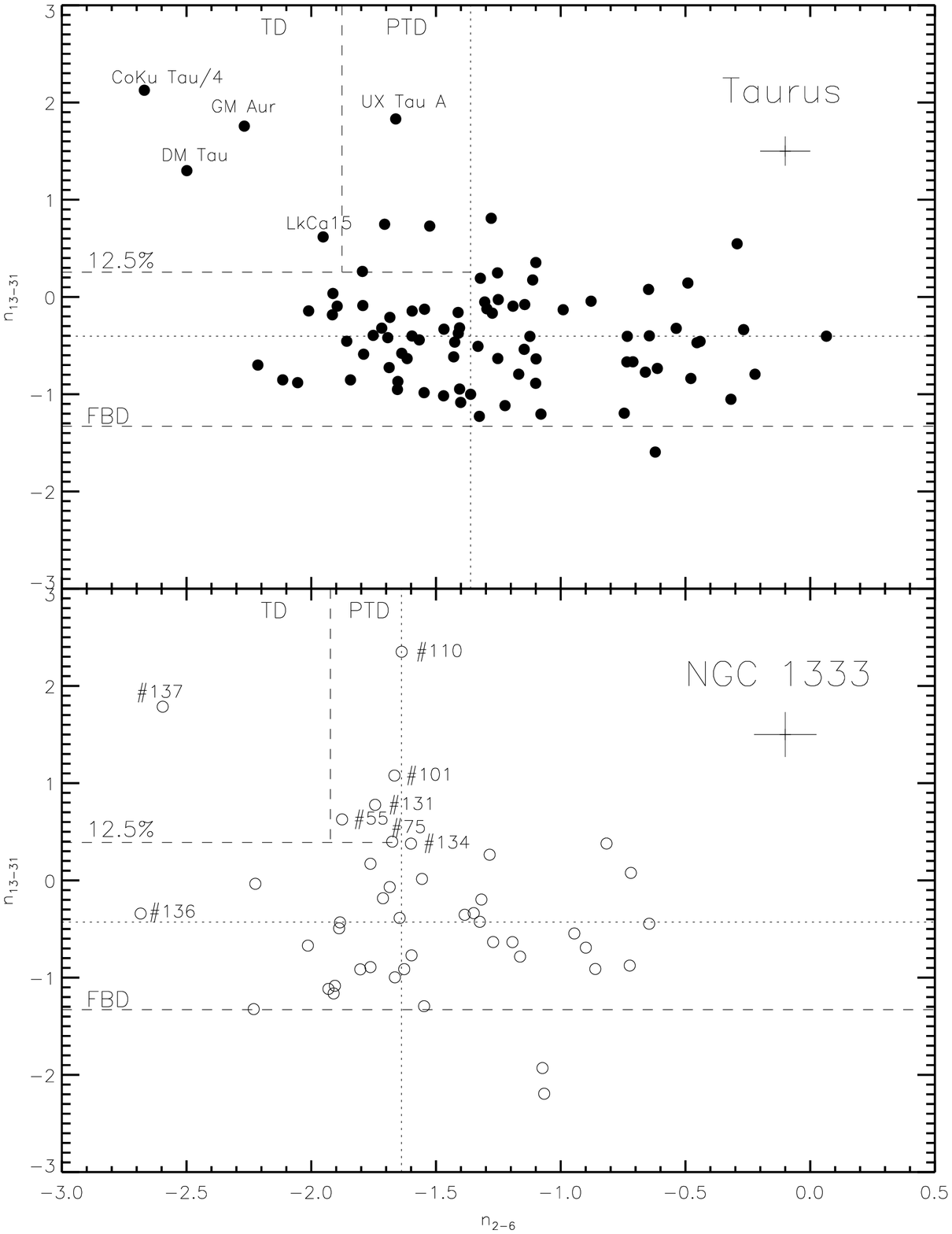}
\caption{\label{n26_n1331} Distribution of $n_{2-6}$ and $n_{13-31}$ for objects in NGC 1333 and Taurus (indices from P. Manoj, personal communication using spectra from \citet{furlan06}). Typical uncertainties for NGC 1333 are shown on right hand side of the plot. The dotted lines indicated the median value for each region. The dashed lines indicate the lowest octile in $n_{2-6}$ and the highest octile in $n_{13-31}$.}
\end{figure}

\clearpage
\begin{figure}
\figurenum{13}
\epsscale{1}
\plotone{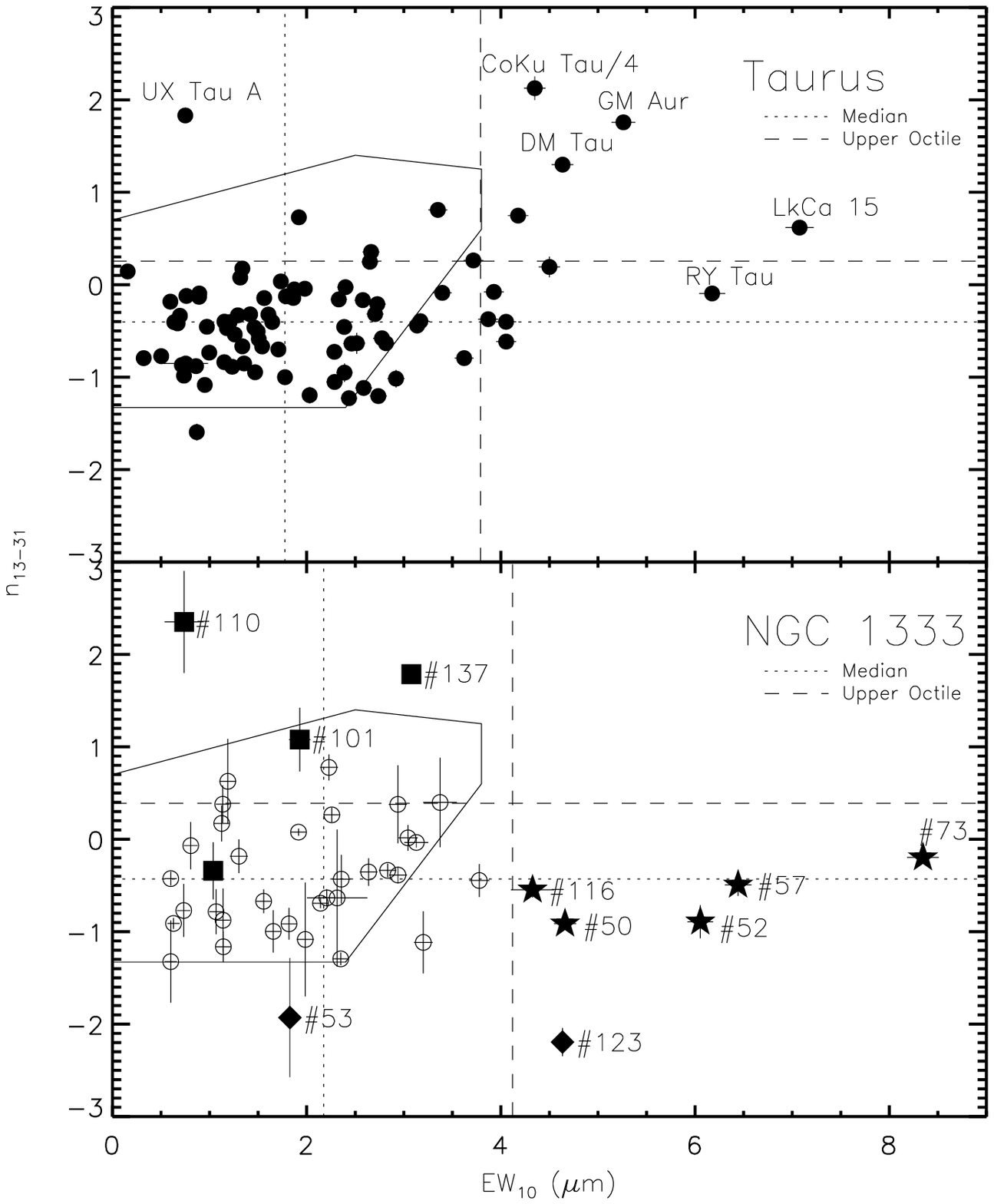}
\caption{ \label{W10_n1331} EW$_{10}$ vs. $n_{13-31}$ for our sample of disks in Taurus (top, indices from P. Manoj, personal communication using spectra from \citet{furlan06}) and NGC 1333 (bottom). TDs are plotted here with squares, PTDs with stars and truncated disks with diamonds. The dotted line indicates the median value and the dashed line indicates the upper octile. The polygon denotes the boundaries of these indices in models of radially continuous irradiated accretion disks (see \S~\ref{vert} for details). }
\end{figure}

\clearpage
\begin{figure}
\figurenum{14}
\epsscale{0.75}
\plotone{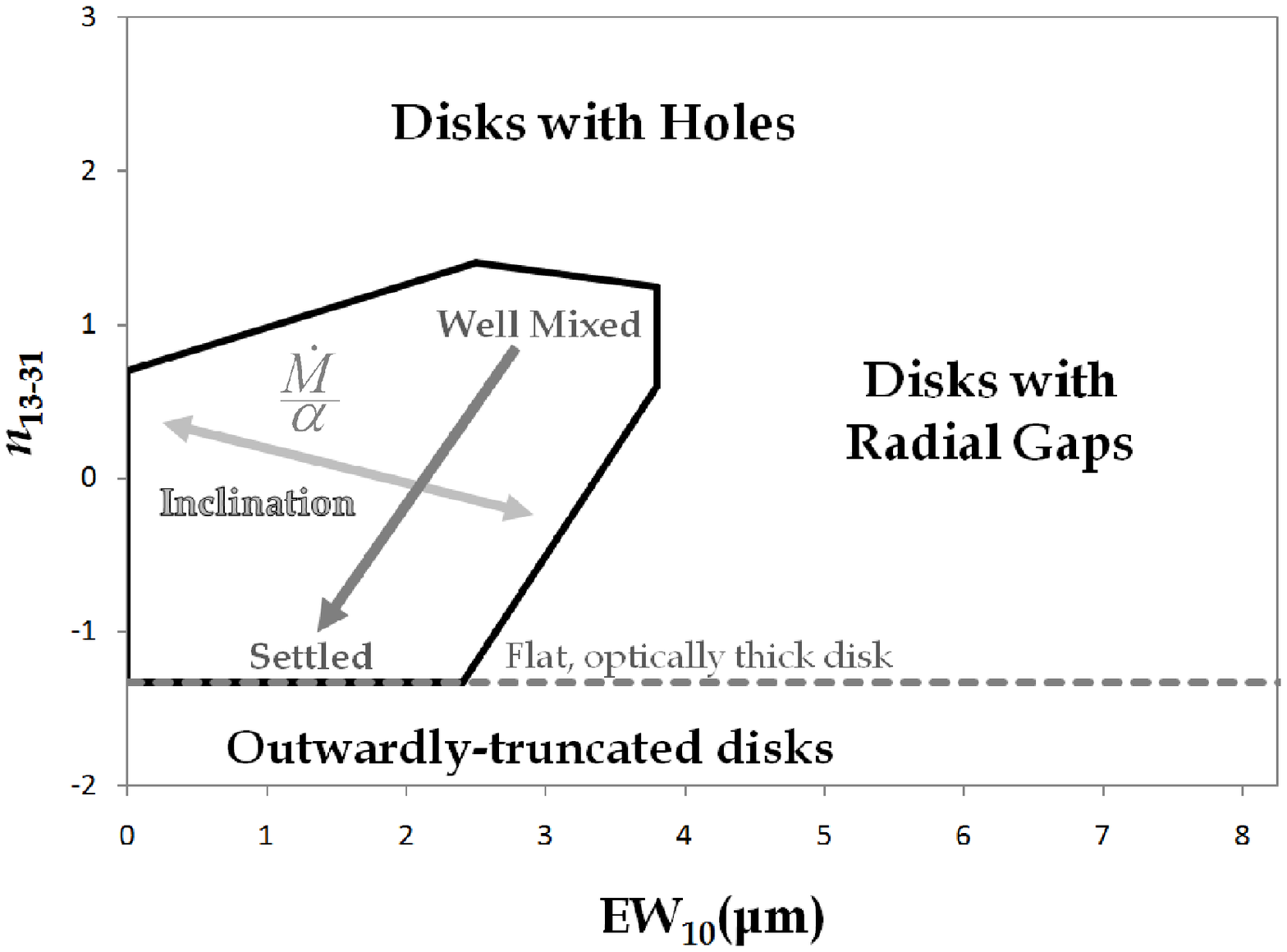}
\caption{ \label{polygon} Areas of EW$_{10}$ vs. $n_{13-31}$ diagram indicating properties of the radial and vertical structure of disks which is based on model indices. The polygon indicates the boundaries in values of EW$_{10}$ and $n_{13-31}$ for radially continuous disk models, described in \S~\ref{vert}. The dark gray arrow shows a trend from the upper right hand corner to the bottom of the polygon which corresponds an increase in settling (decreasing $\epsilon$), approaching the $n_{13-31}$ value of 
a geometrically flat optically thick disk (-4/3, indicated with a dashed line). This trend demonstrates areas of the polygon which correspond to disks which are well mixed or settled. The light gray arrow shows how varying the disk inclination angle and $\frac{\dot{M}}{\alpha}$ will affect the indices of the model SED.}
\end{figure}

\clearpage
\begin{figure}
\figurenum{15}
\epsscale{1}
\plotone{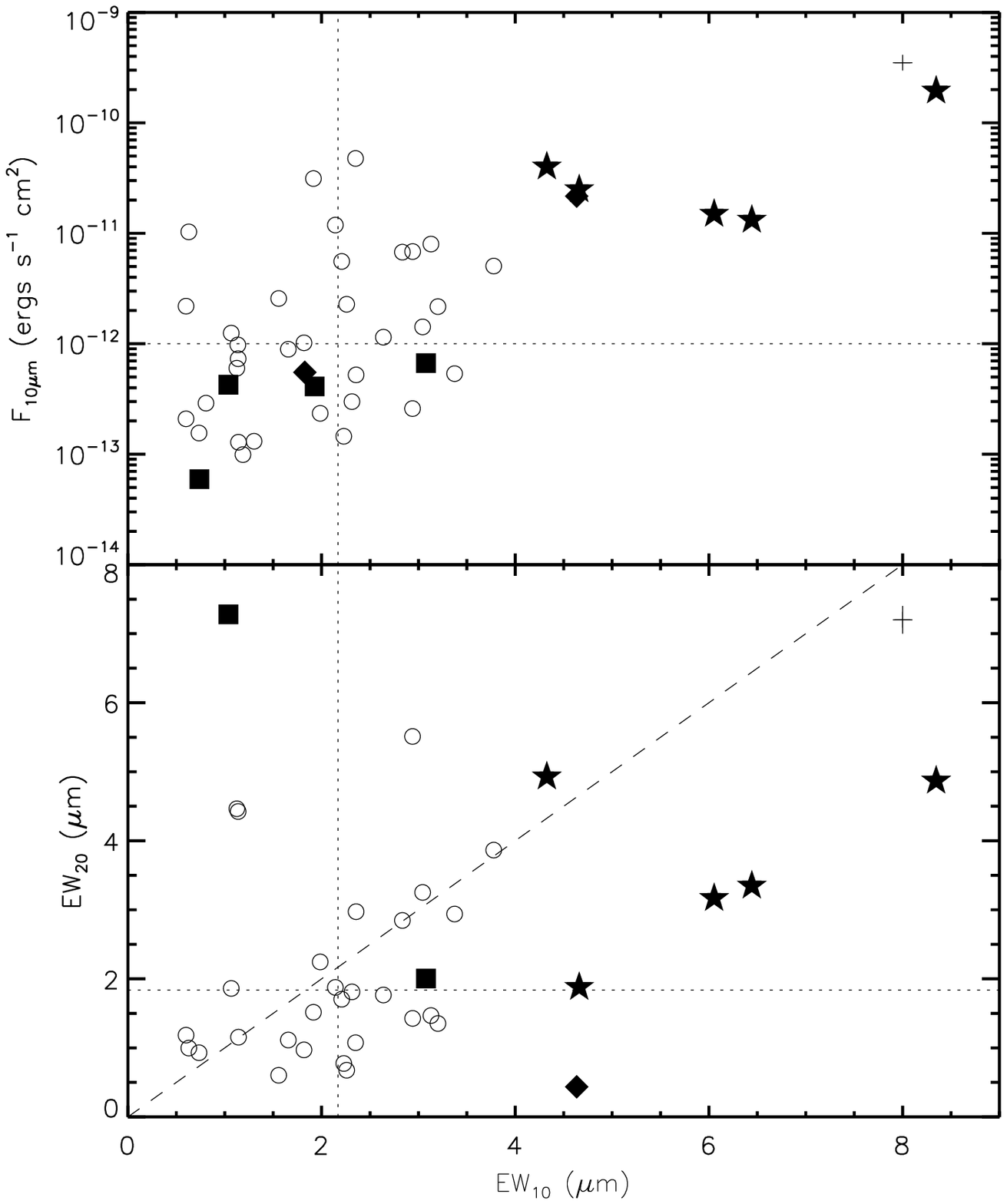}
\caption{ \label{W10_F10_W20} Distribution of F$_{10}$ and EW$_{20}$ vs. EW$_{10}$ for objects in NGC 1333, symbols are the same as Figure~\ref{W10_n1331}. Typical uncertainties are shown in the upper right hand corner of each plot. The dotted lines indicate the median value. The dashed line in the bottom plot indicates where EW$_{10}$ and EW$_{20}$ are equal.}
\end{figure}

\clearpage
\begin{figure}
\figurenum{16}
\epsscale{1}
\plotone{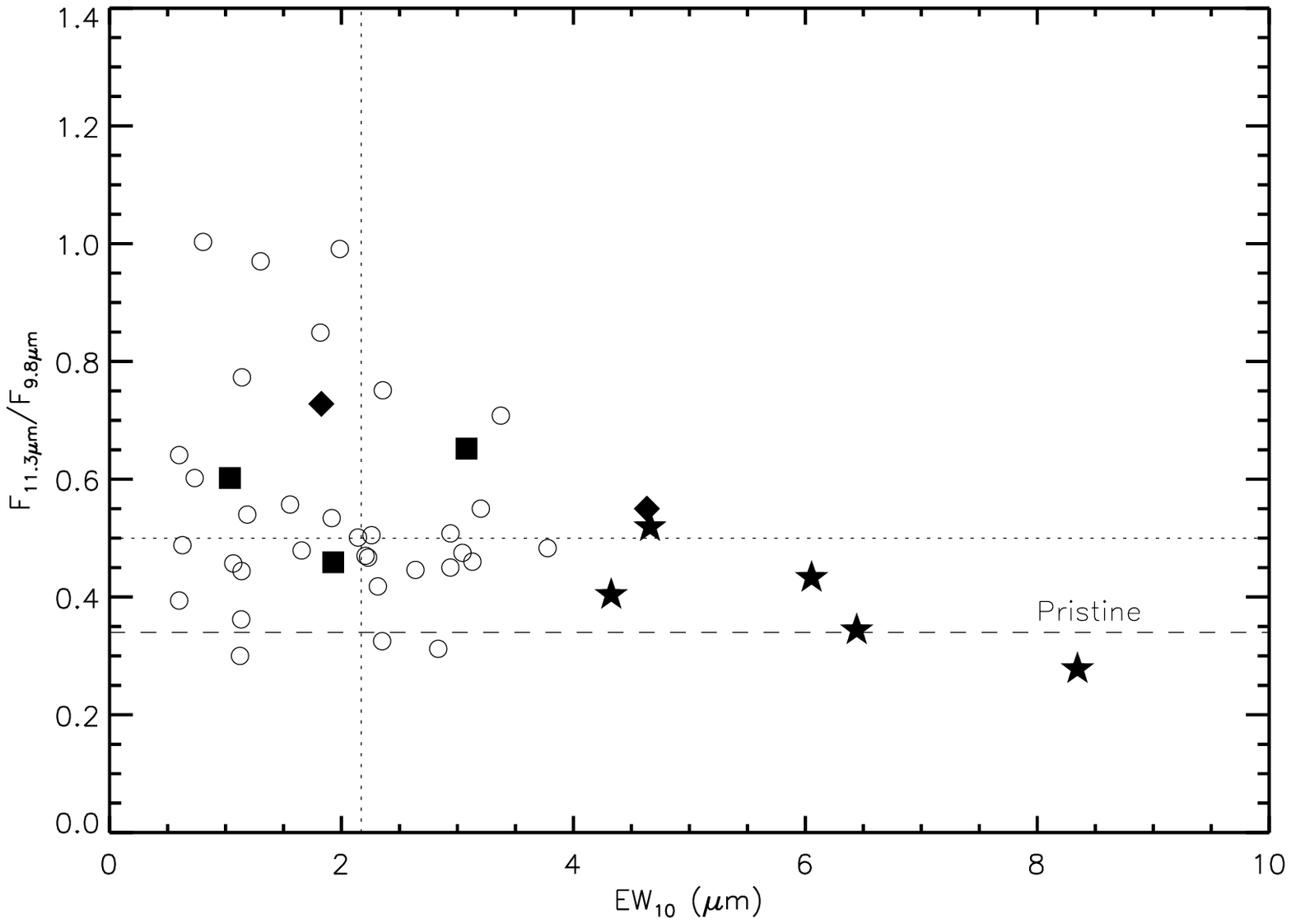}
\caption{ \label{W10_F11F9} Distribution of the flux density ratio F$_{11.3}$/F$_{9.8}$ vs. EW$_{10}$ for objects in NGC 1333, symbols are the same as Figure~\ref{W10_n1331}. The dotted line indicates the median value and the dashed line indicates 0.34, the value of F$_{11.3}$/F$_{9.8}$ of pristine dust.}
\end{figure}

\clearpage
\begin{figure}
\figurenum{17}
\epsscale{1}
\plotone{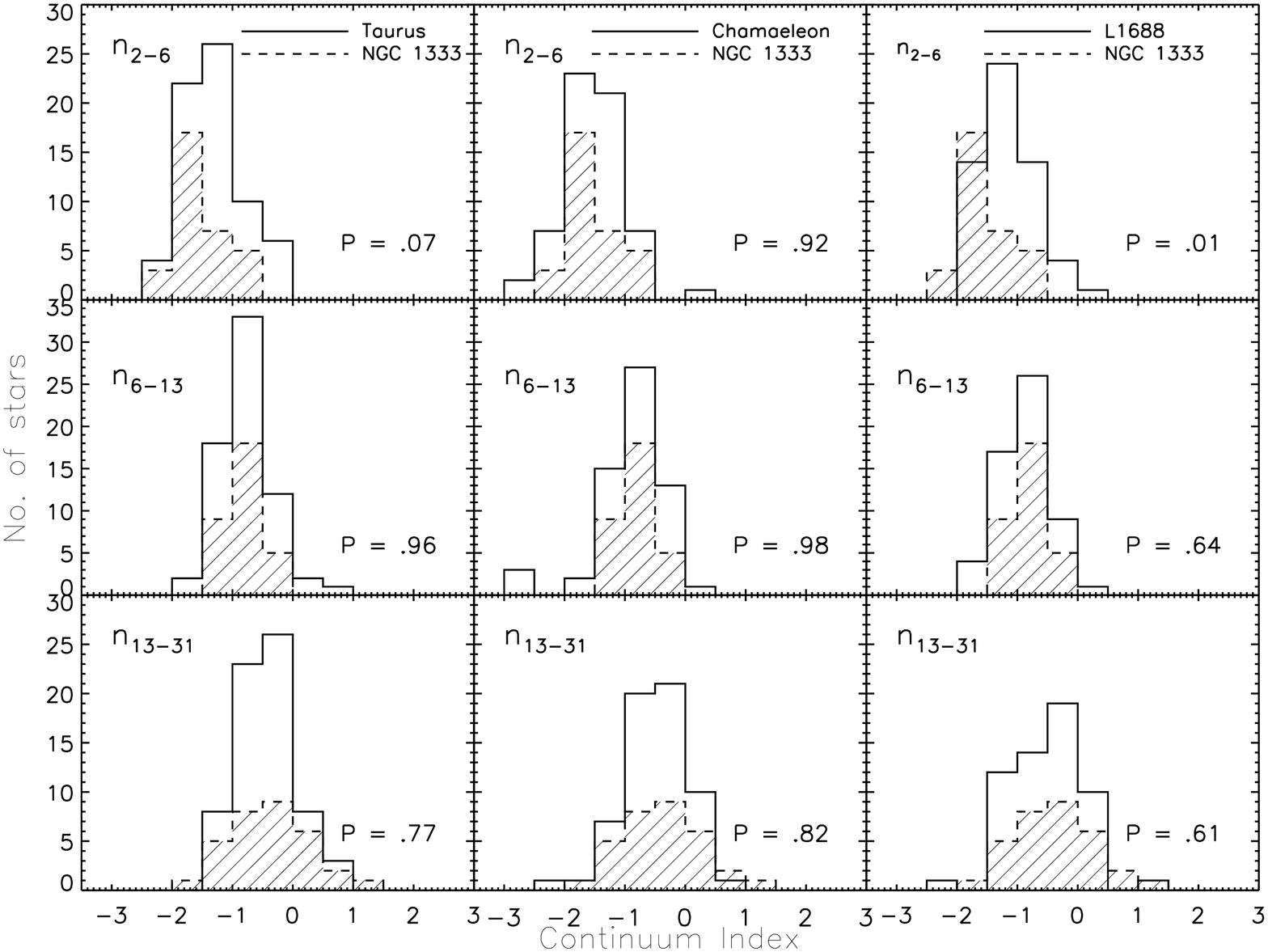}
\caption{Histograms comparing the distribution of $n_{2-6}$, $n_{6-13}$ and $n_{13-31}$ values of objects in NGC 1333 (N = 33; shaded) to those in Taurus (N = 58; left), Chamaeleon I (N = 61; middle) and L1688 (N = 56; center). P-values from one-dimensional two-sided Kolmogorov-Smirnov test are inset in each plot. \label{hist_indices} }
\end{figure}

\clearpage
\begin{figure}
\figurenum{18}
\epsscale{1}
\plotone{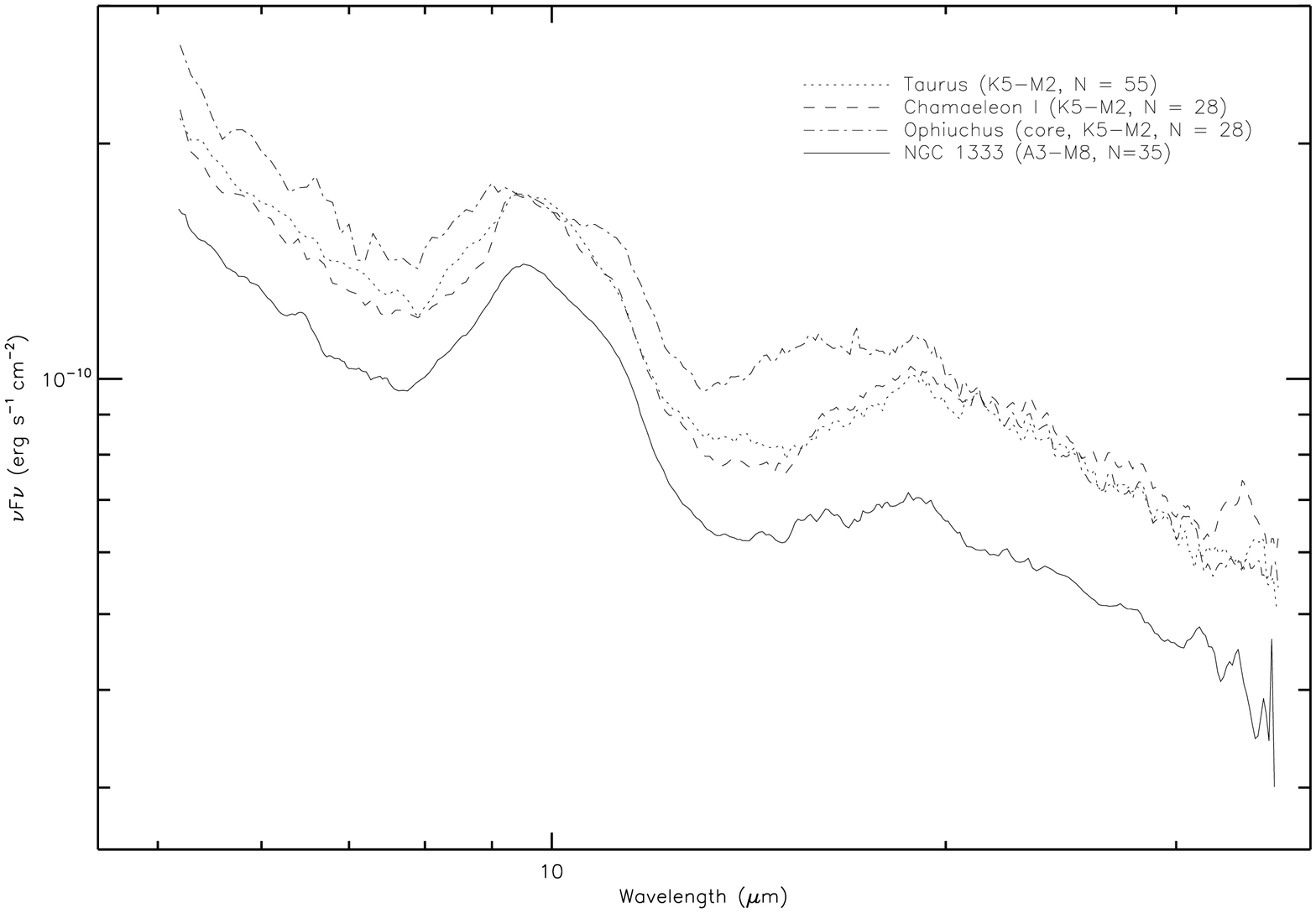}
\caption{Median spectrum of Class II disks in NGC 1333, along with the median spectra of K5 - M2 objects in Taurus, Chamaeleon I and Ophiuchus (core) from \citet{furlan09}. \label{medians}}
\end{figure}

\clearpage
\begin{figure}
\figurenum{19}
\epsscale{1}
\plotone{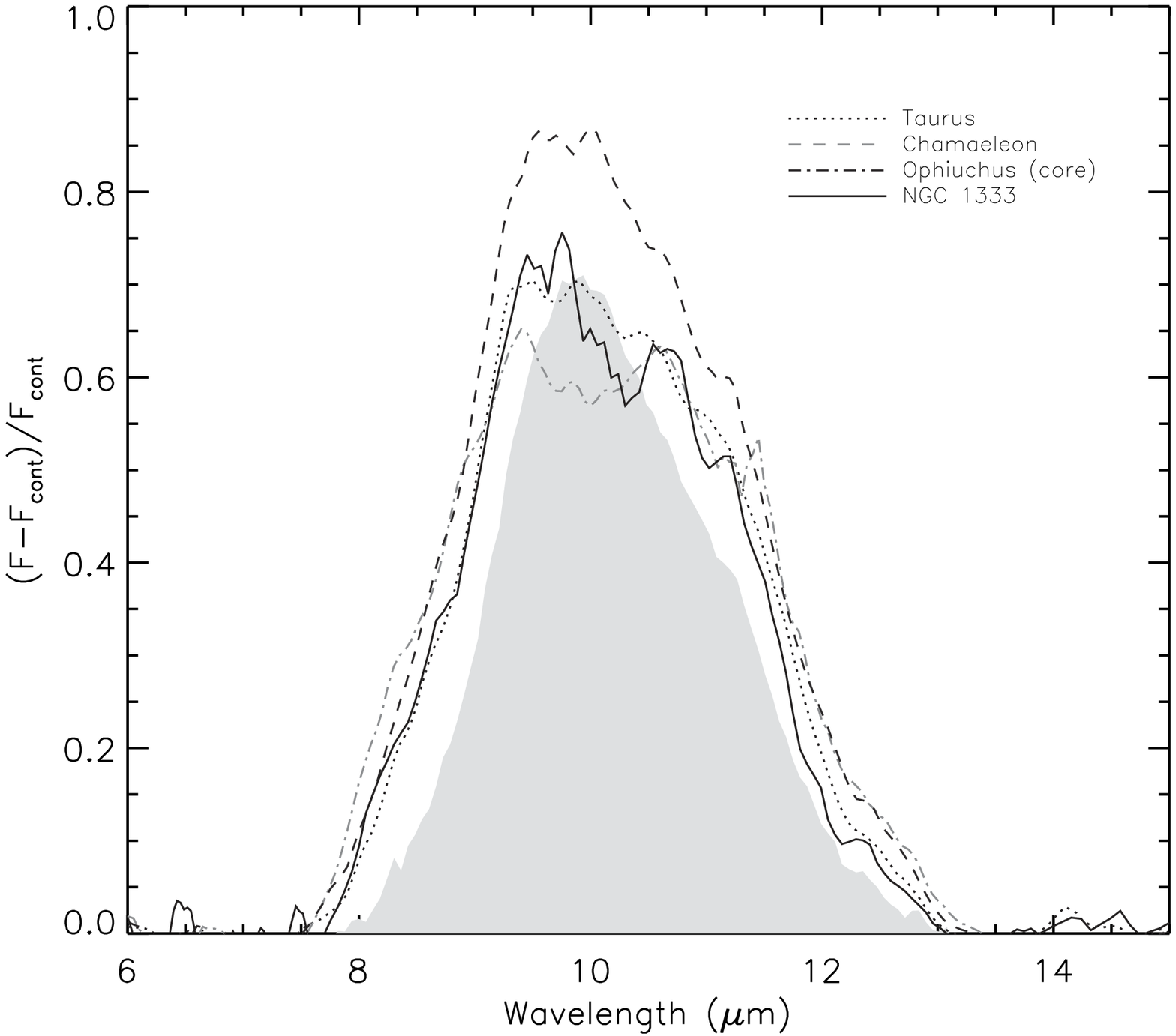}
\caption{Continuum-subtracted and -normalized 10$\mu$m silicate feature profile for the medians shown in Figure~\ref{medians}. The profile of pristine dust is scaled to the mean 9.8$\mu$m flux of the four regions and is shown by the shaded region. \label{silicatefeatures}}
\end{figure}

\clearpage
\begin{figure}
\figurenum{20}
\epsscale{1}
\plotone{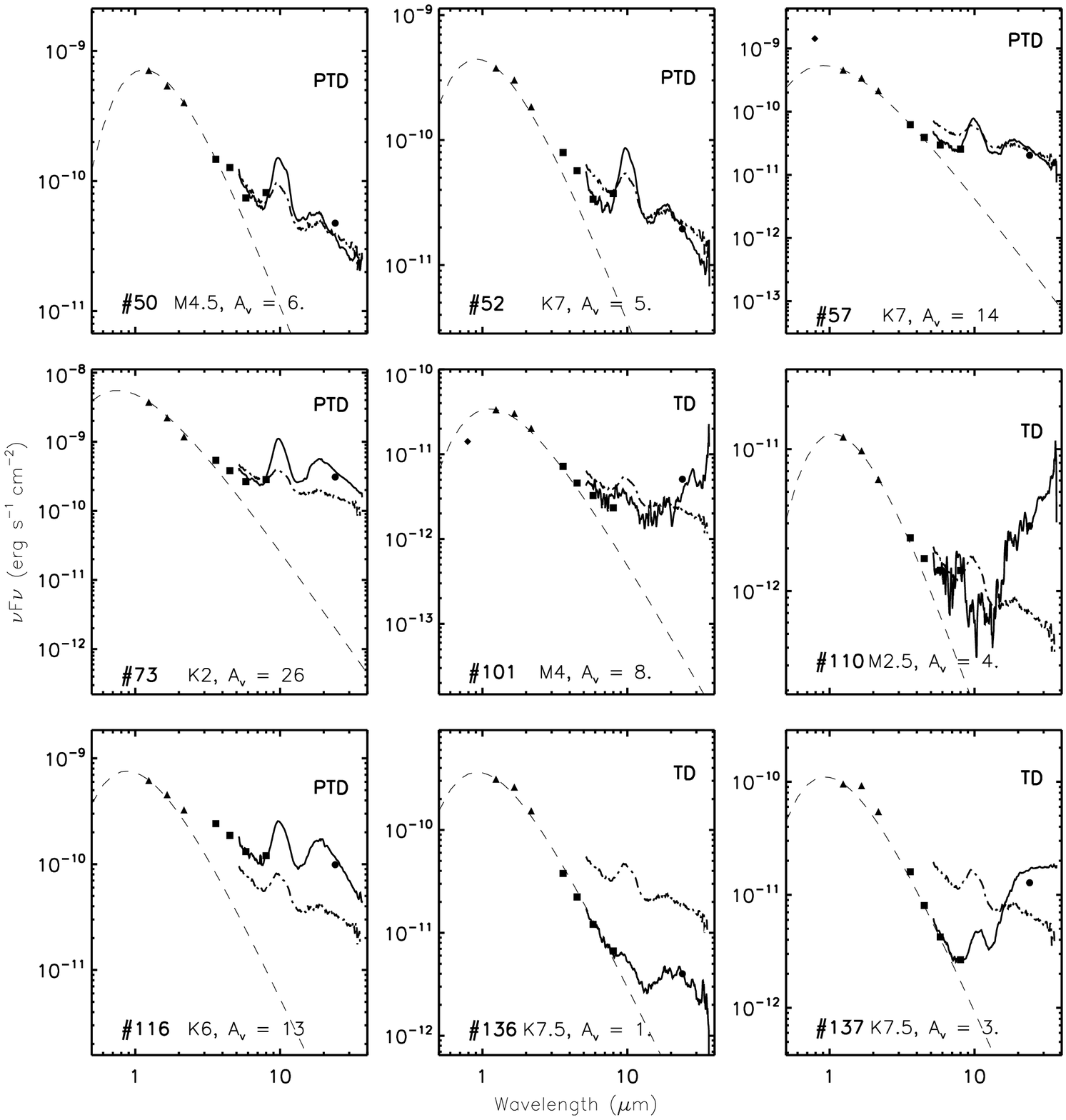}
\caption{ \label{median_comp_TDs} Dereddened SEDs of TDs and PTDs in NGC 1333 as well as the median of disks in NGC 1333, as described in \S~\ref{medians_disc}, plotted with the dot-dash line. The photosphere for a given spectral type scaled to $J$ band is plotted with a dashed line.}
\end{figure}

\begin{deluxetable}{l c c c c}

\tabletypesize{\scriptsize}
\rotate
\tablecaption{Basic Data\label{tbl1}}
\tablewidth{0pt}
\tablehead{
\colhead{Index} & \colhead{RA J2000} & \colhead{Dec J2000} & \colhead{AOR ID} & \colhead{{Alternate Name(s)}} \\
}
\startdata
2    &    03 28 45.30    &    +31 05 42.1    &    22024192    &    IRAS 03256+3055    \\
3    &    03 28 57.37    &    +31 14 16.2    &    22037248    &    IRAS 2b, [RAC97] VLA 10, SK 7, LAL 157, SVS 19   \\
5    &    03 29 04.06    &    +31 14 46.7    &    22019584    &    SK 14  \\
7    &    03 29 11.25    &    +31 18 31.7    &    22024704    &    ASR 32    \\ 
8    &    03 29 12.04    &    +31 13 02.0    &    22024448    &    IRAS 4b (Southern ouflow cavity)    \\
11    &    03 29 23.48    &    +31 33 29.4    &    22021632    &    IRAS 03262+3123    \\
12    &    03 28 32.55    &    +31 11 04.8    &    22023936    &    LAL 38, [HFR2007] 74    \\
13    &    03 28 34.49    &    +31 00 51.2    &    22034944    &    IRAS 03254+3050    \\
14    &    03 28 34.52    &    +31 07 05.5    &    22035200    &    [HFR2007] 69  \\
16    &    03 28 39.69    &    +31 17 32.0    &    22035456    &    LAL 68    \\
18    &    03 28 43.27    &    +31 17 33.1    &    22035712    &    SVS 9; IRAS 5; ASR 126; LAL 79*, HJ 27, HBC 340, CoKu NGC 1333 3, IRAS 03256+3107, [GMW2007] 9    \\
19    &    03 28 48.76    &    +31 16 08.9    &    22022912    &    ASR 67    \\
20    &    03 28 51.27    &    +31 17 39.5    &    22023680    &    ASR 41; LAL 111    \\
23    &    03 28 57.11    &    +31 19 11.9    &    22020352    &    ASR 64; MBO 148    \\
25    &    03 28 58.42    &    +31 22 17.7    &    22036480    &    LAL 166; MBO 38, [GMW2007] 38    \\
28    &    03 29 01.54    &    +31 20 20.7    &    22036992    &    SVS 12; ASR 114; LAL 181; MBO 19, [HFR2007] 45; SK 24; IRAS 6   \\
30    &    03 29 03.76    &    +31 16 04.0    &    22029568    &    SVS 13; ASR 1; LAL 196; HBC 346; SK 13; [HFR2007] 43    \\
33    &    03 29 08.95    &    +31 22 56.3    &    22036224    &    LAL 221; MBO 33    \\
34    &    03 29 09.07    &    +31 21 29.1    &    22036736    &    LAL 223; MBO 51    \\
35    &    03 29 09.09    &    +31 23 05.7    &    22026496    &    LAL 222; MBO 36, [GMW2007] 63    \\
36    &    03 29 10.71    &    +31 18 21.2    &    22035968    &      \\
37    &    03 29 12.95    &    +31 18 14.6    &    22026752    &    ASR 30; LAL 261, [GMW2007] 71    \\
38    &    03 29 18.25    &    +31 23 19.9    &    22019840    &    Background Galaxy \\
43    &    03 28 38.76    &    +31 18 06.7    &    22024960    &    LAL 63    \\
44    &    03 28 43.23    &    +31 10 42.7    &    22018816    &    LAL 80    \\
46    &    03 28 44.08    &    +31 20 52.9    &    22018560    &    LAL 83, [GMW2007] 11    \\
48    &    03 28 47.63    &    +31 24 06.1    &    22025216    &    LAL 93, [GMW2007] 15; MBO 34  \\
49    &    03 28 47.82    &    +31 16 55.3    &    22031104    &    SVS 17; ASR 111; LAL 97; HJ 32, [GMW2007] 16    \\
50    &    03 28 51.02    &    +31 18 18.5    &    22031616    &    SVS 10; ASR 122; LAL 106; MBO 6; Pr 6; LkH$\alpha$ 352A; [GMW2007] 19    \\
51    &    03 28 51.07    &    +31 16 32.6    &    22019072    &    ASR 44; LAL 107; HJ 33; [GMW2007] 20    \\
52    &    03 28 51.19    &    +31 19 54.9    &    22032384    &    ASR 125; LAL 110; MBO 14; [GMW2007] 21; HJ 9; Pr 7    \\
53    &    03 28 52.13    &    +31 15 47.2    &    22020608    &    ASR 45; LAL 125; HJ 42; [GMW2007] 22    \\
54    &    03 28 52.15    &    +31 22 45.4    &    22028544    &    LAL 120; MBO 18; HJ 109; [GMW2007] 23    \\
55    &    03 28 52.90    &    +31 16 26.6    &    22017536    &    ASR 46; LAL 128; HJ 34    \\
57    &    03 28 53.93    &    +31 18 09.3    &    22031360    &    ASR 40; LAL 129; [GMW2007] 25     \\
58    &    03 28 54.07    &    +31 16 54.5    &    22025984    &    SVS 18; ASR 42; LAL 131; [GMW2007] 26; HJ 37    \\
59    &    03 28 54.61    &    +31 16 51.3    &    22030848    &    ASR 43; LAL 136; HJ 38; [GMW2007] 27; [OTS2008] 1    \\
61    &    03 28 55.07    &    +31 16 28.8    &    22030336    &    ASR 107; LAL 141; [GMW2007] 28; HJ 35; [OTS2008] 3    \\
63    &    03 28 56.09    &    +31 19 08.6    &    22029312    &    MBO 146  \\
64    &    03 28 56.31    &    +31 22 28.0    &    22029056    &    LAL 147; MBO 37   \\
65    &    03 28 56.64    &    +31 18 35.7    &    22032128    &    ASR 120; LAL 150, MBO 11; HBC 344; SVS 11; [GMW2007] 30; HJ 21; Pr 8   \\
67    &    03 28 56.95    &    +31 16 22.3    &    19056128    &    SVS 15; ASR 118; LAL 154 \\
68    &    03 28 57.17    &    +31 15 34.6    &    22023168    &    ASR 17; LAL 156; [GMW2007] S1-5    \\
69    &    03 28 57.70    &    +31 19 48.1    &    22028800    &    ASR 113; MBO 35; HJ 20; [GMW2007] 35   \\
73    &    03 28 59.32    &    +31 15 48.5    &    22030080    &    SVS 16; ASR 106; LAL 171; Pr 11; [GMW2007] 39   \\
74    &    03 28 59.56    &    +31 21 46.8    &    22032896    &    LAL 173; MBO 12; [GMW2007] 41; HJ 8; LkH$\alpha$ 353    \\
75    &    03 29 00.15    &    +31 21 09.3    &    22022656    &    LAL 175; MBO 78; [GMW2007] 42    \\
78    &    03 29 03.13    &    +31 22 38.2    &    22033408    &    LAL 189; [GMW2007] 50; HJ 12; MBO 31    \\
82    &    03 29 03.86    &    +31 21 48.8    &    22033152    &    LAL 195; HJ 11; Pr 14; MBO 5; SVS 8; [GMW2007] 52    \\
84    &    03 29 04.67    &    +31 16 59.2    &    22026240    &    ASR 105; LAL 203; [GMW2007] S2-9    \\
85    &    03 29 04.73    &    +31 11 35.0    &    22021888    &    ASR 99; LAL 205; [GMW2007] 56     \\
88    &    03 29 05.76    &    +31 16 39.7    &    22030592    &    SVS 14; ASR 7; LAL 207; [GMW2007] 58; [RAC97] VLA 22    \\
89    &    03 29 06.32    &    +31 13 46.5    &    22027008    &    ASR 53; LAL 208    \\
92    &    03 29 08.95    &    +31 26 24.1    &    22018048    &    LAL 220; MBO 108    \\
94    &    03 29 09.47    &    +31 27 21.0    &    22020864    &    MBO 60  \\
99    &    03 29 10.82    &    +31 16 42.7    &    22027264    &    ASR 23; LAL 235; [GMW2007] S3A-1    \\
101    &    03 29 11.64    &    +31 20 37.7    &    22023424    &    ASR 85; LAL 245; MBO 58    \\
102    &    03 29 11.77    &    +31 26 09.7    &    22019328    &    LAL 246; MBO 118    \\
106    &    03 29 13.13    &    +31 22 52.9    &    22033920    &    LAL 262; MBO 8; [GMW2007] 72; HJ 15    \\
110    &    03 29 16.81    &    +31 23 25.4    &    22021376    &    LAL 279; MBO 79    \\
111    &    03 29 17.66    &    +31 22 45.2    &    22033664    &    SVS 2; LAL 283; HJ 103; Pr 17; LkH$\alpha$ 270; [GMW2007] 79; MBO 2; HBC 12; VSS 4     \\
113    &    03 29 18.65    &    +31 20 17.9    &    22021120    &    ASR 81; MBO 109    \\
114    &    03 29 18.72    &    +31 23 25.5    &    22034176    &    LAL 293; HJ 102; Pr 18; MBO 16; [GMW2007] 81    \\
115    &    03 29 20.05    &    +31 24 07.6    &    22034432    &    LAL 296; MBO 46; [HFR2007] 67; [GMW2007] 84    \\
116    &    03 29 20.42    &    +31 18 34.3    &    22031872    &    SVS 5; IRAS 03262+3108; ASR 112; LAL 300; [GMW2007] 86; MBO 17    \\
118    &    03 29 21.55    &    +31 21 10.5    &    22025728    &    LAL 304; MBO 32; HJ 17; [GMW2007] 88    \\
119    &    03 29 21.87    &    +31 15 36.4    &    22029824    &    ASR 123; LAL 307; [GMW2007] 90; HJ 46; VSS 27; LkH$\alpha$ 271; SVS 20; HBC 13; StH$\alpha$ 20    \\
120    &    03 29 23.15    &    +31 20 30.5    &    22027776    &    ASR 78; LAL 310; HJ 2; LkH$\alpha$ 355; MBO 30; [GMW2007] 92    \\
121    &    03 29 23.23    &    +31 26 53.1    &    22022144    &    LAL 308; MBO 54    \\
122    &    03 29 24.08    &    +31 19 57.8    &    19052544    &    ASR 79; LAL 313; SSTc2d J032924.1+3119 \\
123    &    03 29 25.92    &    +31 26 40.1    &    22034688    &    SVS 4; LAL 318; VSS IX-2; MBO 9; [GMW2007] 98   \\
125    &    03 29 29.79    &    +31 21 02.8    &    22028032    &    LAL 333; HJ 53; MBO 27; [GMW2007] 103    \\
126    &    03 29 30.39    &    +31 19 03.4    &    22027520    &    LAL 336; HJ 49; MBO 28; LkH$\alpha$ 356    \\
127    &    03 29 32.56    &    +31 24 37.0    &    22025472    &    LAL 342; HJ 105; MBO 40; LkH$\alpha$ 357    \\
128    &    03 29 32.86    &    +31 27 12.7    &    22017792    &    LAL 344; MBO 53; [GMW2007] 106    \\
131    &    03 29 37.72    &    +31 22 02.6    &    22020096    &    MBO 65  \\
133    &    03 29 54.03    &    +31 20 53.1    &    22032640    &      \\
134    &    03 29 02.69    &    +31 19 05.8    &    22018304    &    ASR 62; LAL 187; MBO 84    \\
135    &    03 29 09.41    &    +31 14 14.1    &    22022400    &    ASR 54    \\
136    &    03 29 26.79    &    +31 26 47.7    &    22028288    &    SVS 6; LAL 321; Pr 19; MBO 15; VSS IX-1; [GMW2007] 100    \\
137    &    03 29 29.26    &    +31 18 34.8    &    19053312    &    LAL 331; SSTc2d J032929.3+3118 \\
\enddata

\tablecomments{VLA - [RAC97] - \citet{rodriguez99}, SK - [SK2001] - \citet{sandell01}, HJ - [HJ83] - \citet{herbig83} , HBC - \citet{herbig88}, ASR - \citet{aspin94}, LAL - [LAL96] \citet{lada96}, MBO - \citet{wilking04}, [GFT2002] - \citet{getman02}, Pr - [P97] - \citet{preibisch97}, [GMW2007] - \citet{greissl07}, [OTS2008] - \citet{oasa08}, [HFR2007] - \citet{hatchell07} , IRAS - [JCC87] -  \citet{jennings87}, SVS - [SVS76] NGC 1333 - \citet{strom76}, VSS - \citet{vrba76}. SSTc2d - \citet{merin10}}
\end{deluxetable}

\begin{deluxetable}{l c c c c c c c c c}
\tabletypesize{\scriptsize}
\tablecaption{Spectral Type and Class Information for Sample\label{tbl2}}
\tablewidth{0pt}
\tablehead{
\colhead{Index} & \colhead{SpT$^a$} & \colhead{$A_V$} & \colhead{Method$^b$} & \colhead{Obs $n_{12-20}$} & \colhead{Obs $n_{5-12}$} & \colhead{State$^c$} & \colhead{Obs ${n_{2-25}}^d$} & \colhead{Class} \\
}
\startdata
2   &   ...   &   0   &   neg*   &   2.39   &   0.12   &   envelope   &   1.43   &   I   \\
3   &   ...   &   ...   &   ...   &   1.26   &   -0.62   &   envelope$^c$   &   ...   &   ...   \\
5   &   ...   &   ...   &   ...   &   1.79   &   0.23   &   envelope   &   ...   &   ...   \\
7   &   ...   &   ...   &   ...   &   5.16   &   -2.05   &   envelope$^c$   &   ...   &   ...   \\
11   &   ...   &   ...   &   ...   &   2.91   &   0.05   &   envelope   &   ...   &   ...   \\
12   &   ...   &   ...   &   ...   &   1.09   &   0.31   &   envelope   &   0.62   &   I   \\
13   &   ...   &   5.8   &   JH   &   -0.09   &   -0.32   &   disk   &   0.97   &   I   \\
14   &   ...   &   0   &   ...   &   -0.70   &   -0.08   &   envelope   &   ...   &   ...   \\
16   &   ...   &   35.7   &   HK*   &   -0.02   &   0.06   &   envelope   &   0.92   &   I   \\
18   &   K7.0   &   8.5   &   JH   &   0.14   &   -0.44   &   disk   &   -0.07   &   FS   \\
19   &   ...   &   17.8   &   HK*   &   0.20   &   0.05   &   envelope   &   0.38   &   I   \\
20   &   ...   &   5.5   &   JH   &   1.44   &   0.68   &   envelope   &   0.33   &   I   \\
23   &   ...   &   4.4   &   JH   &   0.78   &   0.05   &   envelope   &   0.22   &   FS   \\
25   &   ...   &   22.2   &   JH*   &   -1.30   &   -0.15   &   envelope   &   0.62   &   I   \\
28   &   ...   &   10.6   &   JH   &   0.72   &   0.64   &   envelope   &   0.89   &   I   \\
30   &   ...   &   13.6   &   HK*   &   0.25   &   0.15   &   envelope   &   0.23   &   FS   \\
33   &   ...   &   31.6   &   HK   &   -0.48   &   -0.96   &   disk   &   1.02   &   I   \\
34   &   ...   &   8.1   &   JH   &   -1.50   &   1.31   &   envelope   &   0.59   &   I   \\
35   &   ...   &   ...   &   ...   &   3.72   &   -0.75   &   envelope$^c$   &   ...   &   ...   \\
36   &   ...   &   2.8   &   JH*   &   -1.16   &   -1.67   &   disk   &   1.34   &   I   \\
37   &   ...   &   ...   &   ...   &   -0.27   &   0.48   &   envelope   &   ...   &   ...   \\
43   &   ...   &   28.2   &   HK*   &   -0.02   &   -1.05   &   disk   &   0.05   &   FS   \\
44   &   ...   &   13.3   &   JH   &   -1.66   &   -1.5   &   disk   &   -1.3   &   II   \\
46   &   M7.5   &   2.2   &   JH   &   -1.35   &   -0.77   &   disk   &   -1.28   &   II   \\
48   &   M4.0   &   9.1   &   JH   &   -1.09   &   -1.42   &   disk   &   -1.01   &   II   \\
49   &   M8.0   &   3   &   JH   &   -0.85   &   -0.87   &   disk   &   -0.76   &   II   \\
50   &   M4.5   &   6.1   &   JH   &   -0.73   &   -0.71   &   disk   &   -0.81   &   II   \\
51   &   M4.0   &   0.7   &   HK   &   -1.76   &   -1.04   &   disk   &   -1.32   &   II   \\
52   &   K7.0   &   5.1   &   JH   &   -0.52   &   -0.64   &   disk   &   -0.78   &   II   \\
53   &   M7.0   &   0   &   neg   &   -2.62   &   -0.9   &   disk   &   -1.27   &   II   \\
54   &   M2.5   &   2.7   &   HK   &   -1.58   &   -1.07   &   disk   &   -1.26   &   II   \\
55   &   M5.0   &   0.3   &   HK   &   -1.30   &   -1.12   &   disk   &   -1.16   &   II   \\
57   &   K7.0   &   14.3   &   JH   &   0.22   &   -0.71   &   disk   &   -0.49   &   II   \\
58   &   M5.0   &   1.4   &   HK   &   -1.72   &   -0.51   &   disk   &   -0.86   &   II   \\
59   &   M1.0   &   8.1   &   JH   &   -0.31   &   -0.83   &   disk   &   -0.65   &   II   \\
61   &   M3.0   &   10.6   &   JH   &   -0.42   &   -0.73   &   disk   &   -0.35   &   II   \\
63   &   ...   &   3.5   &   JH*   &   -1.43   &   -1.72   &   disk   &   0.99   &   I   \\
64   &   M2.0   &   14.4   &   HK   &   -0.24   &   -0.24   &   disk   &   -0.15   &   FS   \\
65   &   M1.5   &   8.1   &   JH   &   0.18   &   -0.6   &   disk   &   -0.7   &   II   \\
67   &   M2.5   &   9.7   &   HK   &   0.99   &   -0.09   &   disk$^c$   &   ...   &   ...   \\
68   &   M6.0   &   5.4   &   HK   &   -0.61   &   -0.14   &   disk$^c$   &   -0.42   &   II   \\
69   &   M3.5   &   3.7   &   HK   &   -0.45   &   -0.13   &   disk$^c$   &   -0.6   &   II   \\
73   &   K2.0   &   26   &   JH   &   0.89   &   -0.41   &   disk   &   0.25   &   FS   \\
74   &   ...   &   4.6   &   JH   &   -0.59   &   -0.99   &   disk   &   -0.74   &   II   \\
75   &   ...   &   10.4   &   JH   &   -0.30   &   -0.53   &   disk   &   -0.39   &   II   \\
78   &   M2.5   &   5.7   &   JH   &   -0.10   &   -0.61   &   disk   &   -0.24   &   FS   \\
82   &   K6.0   &   5.7   &   JH   &   -0.48   &   -0.11   &   disk$^c$   &   -0.58   &   II   \\
84   &   M6.0   &   8.7   &   JH   &   0.11   &   -0.33   &   disk   &   -0.05   &   FS   \\
85   &   ...   &   31.7   &   HK*   &   0.75   &   -1.57   &   disk   &   0.04   &   FS   \\
88   &   A3.0   &   22.3   &   JH   &   -1.64   &   -0.93   &   disk   &   -0.63   &   II   \\
89   &   ...   &   43.2   &   JH*   &   0.02   &   -0.98   &   disk   &   -0.22   &   FS   \\
92   &   ...   &   7   &   JH*   &   -0.23   &   -0.28   &   disk   &   0.41   &   I   \\
94   &   ...   &   2.1   &   HK   &   -1.04   &   -1.4   &   disk   &   -0.84   &   II   \\
99   &   M0.0   &   7.1   &   JH   &   -0.91   &   -0.61   &   disk   &   -0.13   &   FS   \\
101   &   M4.0   &   8.1   &   HK   &   -0.13   &   -1.11   &   disk   &   -0.44   &   II   \\
102   &   ...   &   0.3   &   JH   &   1.22   &   -0.09   &   envelope   &   -0.02   &   FS   \\
106   &   M3.0   &   10.1   &   JH   &   -0.90   &   -1.58   &   disk   &   -0.85   &   II   \\
110   &   M2.5   &   4.5   &   JH   &   1.59   &   -0.59   &   disk   &   -0.14   &   FS   \\
111   &   M0.0   &   3.1   &   JH   &   -1.22   &   -0.96   &   disk   &   -0.73   &   II   \\
113   &   ...   &   16.1   &   JH   &   0.74   &   0.89   &   envelope   &   0.32   &   I   \\
114   &   M0.0   &   3   &   JH*   &   -0.60   &   0.15   &   disk$^c$   &   -0.4   &   II   \\
115   &   ...   &   26.4   &   HK*   &   0.39   &   0.02   &   envelope   &   0.6   &   I   \\
116   &   K6.0   &   13.7   &   JH   &   0.64   &   -0.52   &   disk   &   -0.07   &   FS   \\
118   &   M4.5   &   0   &   neg   &   -0.21   &   -0.96   &   disk   &   -1.04   &   II   \\
119   &   K4.0   &   4.3   &   JH   &   -0.97   &   -1.31   &   disk   &   -1.24   &   II   \\
120   &   M4.5   &   0.2   &   HK   &   -0.07   &   -0.68   &   disk   &   -0.78   &   II   \\
121   &   M4.5   &   1.4   &   HK   &   -1.76   &   -1.01   &   disk   &   -1.02   &   II   \\
122   &   K5.0   &   6.1   &   JH   &   2.90   &   0.69   &   envelope   &   0.37   &   I   \\
123   &   ...   &   3.8   &   JH   &   -1.67   &   -0.86   &   disk   &   -1.04   &   II   \\
125   &   M4.5   &   1.6   &   HK   &   -0.60   &   -0.43   &   disk   &   -0.91   &   II   \\
126   &   ...   &   1.2   &   JH   &   0.59   &   -0.67   &   disk   &   -0.69   &   II   \\
127   &   M3.0   &   7   &   JH   &   -0.49   &   -1.37   &   disk   &   -0.78   &   II   \\
128   &   M4.5   &   0.2   &   HK   &   -0.62   &   -1.4   &   disk   &   -1.15   &   II   \\
131   &   M7.0   &   0   &   neg   &   0.93   &   -0.88   &   disk   &   -0.68   &   II   \\
133   &   M4.5   &   2.1   &   JH   &   0.33   &   -0.45   &   disk   &   -0.5   &   II   \\
134   &   ...   &   12.2   &   JH   &   0.53   &   -1.99   &   disk   &   -0.56   &   II   \\
135   &   ...   &   ...   &   ...   &   -2.52   &   -3.03   &   photosphere   &   ...   &   ...   \\
136   &   K7.5   &   1.8   &   HK   &   0.39   &   -1.93   &   disk   &   -1.5   &   II   \\
137   &   K7.5   &   2.96   &   HK   &   2.66   &   -0.51   &   disk   &   -0.38   &   II   \\
\enddata

\tablenotetext{a} {Spectral types for \#68, 84 and 99 from \citet{greissl07}, \#110 from \citet{wilking04}, \#135 from \citet{gut08}, \#122 from \citet{merin10}, the rest are from \citet{winston09}}

\tablenotetext{b} {Objects with null value in Column 3 had no $J$, H, or $K_s$ band measurements, objects with $A_V$ $=$ 0.00 had negative values of $A_V$, asterix denote objects $A_V$ values which have large uncertainty because their colors are outside of the reddened classical T Tauri locus.}

\tablenotetext{c} {The state for these objects was determined upon careful examination of their SED and does not correspond with the state associated with its $n_{5-12}$ value}

\tablenotetext{d} {Objects with upper limit for K are indicated with ...}
\end{deluxetable}

\begin{deluxetable}{l c c c c c c}
\tabletypesize{\scriptsize}
\tablecaption{Continuum indices\label{tbl3}}
\tablewidth{0pt}
\tablehead{
\colhead{Index} & \colhead{${n_{2-6}}^a$} & \colhead{$\sigma_{n_{2-6}}$}& \colhead{$n_{6-13}$}  
& \colhead{$\sigma_{n_{6-13}}$} & \colhead{$n_{13-31}$} & \colhead{$\sigma_{n_{13-31}}$} \\
}
\startdata
18    &    -0.72    &    0.04    &    -0.31    &    0.05    &    0.08    &    0.04     \\
44    &    -2.23    &    0.12    &    -1.39    &    0.14    &    -1.32    &    0.44     \\
46    &    -1.91    &    0.12    &    -0.91    &    0.18    &    -1.16    &    0.17     \\
48    &    -1.66    &    0.08    &    -1.18    &    0.14    &    -1.00    &    0.23     \\
49    &    -1.16    &    0.13    &    -0.80    &    0.15    &    -0.78    &    0.24     \\
50    &    -1.63    &    0.06    &    -0.62    &    0.07    &    -0.91    &    0.04     \\
51    &    -1.91    &    0.09    &    -0.94    &    0.14    &    -1.08    &    0.62     \\
52    &    -1.76    &    0.11    &    -0.47    &    0.14    &    -0.89    &    0.18     \\
53    &    -1.07    &    0.13    &    -0.84    &    0.15    &    -1.93    &    0.64     \\
54    &    -1.93    &    0.08    &    -1.07    &    0.12    &    -1.12    &    0.34     \\
55    &    -1.88    &    0.09    &    -1.05    &    0.18    &    0.63    &    0.46     \\
57    &    -1.89    &    0.11    &    -0.50    &    0.13    &    -0.49    &    0.12     \\
58    &    -1.69    &    0.14    &    -0.52    &    0.15    &    -0.07    &    0.26     \\
59    &    -1.65    &    0.06    &    -0.58    &    0.07    &    -0.39    &    0.09     \\
61    &    -0.90    &    0.05    &    -0.59    &    0.06    &    -0.69    &    0.05     \\
64    &    -1.35    &    0.19    &    -0.17    &    0.21    &    -0.34    &    0.07     \\
65    &    -2.22    &    0.10    &    -0.33    &    0.12    &    -0.04    &    0.06     \\
73    &    -1.32    &    0.08    &    -0.21    &    0.08    &    -0.20    &    0.06     \\
74    &    -1.27    &    0.06    &    -0.97    &    0.07    &    -0.63    &    0.09     \\
75    &    -1.68    &    0.19    &    -0.81    &    0.50    &    0.40    &    0.48     \\
78    &    -0.65    &    0.09    &    -0.61    &    0.12    &    -0.45    &    0.18     \\
84    &    -0.82    &    0.07    &    -0.24    &    0.08    &    0.38    &    0.20     \\
88    &    -1.55    &    0.07    &    -0.89    &    0.09    &    -1.29    &    0.07     \\
94    &    -1.19    &    0.16    &    -1.19    &    0.67    &    -0.64    &    0.74     \\
99    &    -0.72    &    0.13    &    -0.58    &    0.17    &    -0.88    &    0.34     \\
101    &    -1.67    &    0.17    &    -1.04    &    0.36    &    1.08    &    0.35     \\
106    &    -1.32    &    0.06    &    -1.46    &    0.08    &    -0.43    &    0.07     \\
110    &    -1.64    &    0.23    &    -0.94    &    0.59    &    2.35    &    0.55     \\
111    &    -0.86    &    0.06    &    -0.90    &    0.05    &    -0.91    &    0.06     \\
116    &    -0.95    &    0.07    &    -0.39    &    0.08    &    -0.55    &    0.08     \\
118    &    -1.88    &    0.10    &    -0.83    &    0.15    &    -0.43    &    0.26     \\
119    &    -2.01    &    0.13    &    -1.25    &    0.16    &    -0.67    &    0.13     \\
120    &    -1.56    &    0.16    &    -0.62    &    0.19    &    0.02    &    0.14     \\
121    &    -1.60    &    0.11    &    -0.85    &    0.13    &    -0.77    &    0.29     \\
123    &    -1.07    &    0.06    &    -0.84    &    0.09    &    -2.19    &    0.15     \\
125    &    -1.80    &    0.07    &    -0.34    &    0.09    &    -0.92    &    0.17     \\
126    &    -1.76    &    0.12    &    -0.70    &    0.18    &    0.17    &    0.19     \\
127    &    -1.39    &    0.13    &    -1.30    &    0.15    &    -0.35    &    0.15     \\
128    &    -1.71    &    0.08    &    -1.34    &    0.16    &    -0.18    &    0.18     \\
131    &    -1.74    &    0.09    &    -0.76    &    0.14    &    0.78    &    0.14     \\
133    &    -1.29    &    1.01    &    -0.71    &    1.10    &    0.26    &    0.08     \\
134    &    -1.60    &    0.11    &    -1.82    &    0.12    &    0.38    &    0.42     \\
136    &    -2.68    &    0.08    &    -1.73    &    0.15    &    -0.34    &    0.31     \\
137    &    -2.60    &    0.07    &    -0.22    &    0.12    &    1.79    &    0.09     \\
\enddata
\tablenotetext{a} {Objects with upper limit for K are indicated with ...}
\tablecomments{Continuum indices for disk dominated objects in NGC 1333} 
\end{deluxetable}
\begin{deluxetable}{l c c c c c c c c c c}
\tabletypesize{\scriptsize}
\tablecaption{Dust Processing Indicators \label{tbl4}}
\tablewidth{0pt}
\tablehead{
\colhead{{Index}} & \colhead{${EW_{10}}^a$} & \colhead{${\sigma_{EW_{10}}}^a$}& \colhead{${EW_{20}}^a$}  
& \colhead{${\sigma_{EW_{20}}}^a$} & \colhead{${F_{10}}^b$} & \colhead{${\sigma_{F_{10}}}^b$} 
& \colhead{${F_{20}}^b$} & \colhead{${\sigma_{F_{20}}}^b$} & \colhead{$F_{11.3}/F_{9.8}$} & \colhead{$\sigma_{F_{11.3}/F_{9.8}}$}\\
}
\startdata
18    &    1.92    &    0.08    &    1.52    &    0.14    &    31.30    &    0.89    &    14.04    &    0.89    &    0.53    &    0.04     \\
44    &    0.60    &    0.06    &    ...    &    ...    &    0.21    &    0.02    &    0.27    &    0.02    &    0.64    &    0.14     \\ 
46    &    1.14    &    0.08    &    1.16    &    0.27    &    0.13    &    0.01    &    0.03    &    0.01    &    0.77    &    0.09     \\
48    &    1.66    &    0.08    &    1.11    &    0.17    &    0.89    &    0.03    &    0.17    &    0.03    &    0.48    &    0.04     \\
49    &    1.07    &    0.07    &    1.86    &    0.18    &    1.25    &    0.07    &    0.59    &    0.07    &    0.46    &    0.06     \\
50    &    4.66    &    0.11    &    1.88    &    0.15    &    24.97    &    0.31    &    3.77    &    0.31    &    0.52    &    0.03     \\
51    &    1.99    &    0.09    &    2.25    &    0.35    &    0.23    &    0.01    &    0.08    &    0.01    &    0.99    &    0.07     \\
52    &    6.05    &    0.13    &    3.17    &    0.17    &    14.97    &    0.14    &    3.07    &    0.14    &    0.43    &    0.02     \\
53    &    1.83    &    0.08    &    ...   &    ...   &    0.55    &    0.02    &    ...   &    ...   &    0.73    &    0.06     \\
54    &    3.20    &    0.10    &    1.35    &    0.22    &    2.17    &    0.04    &    0.25    &    0.04    &    0.55    &    0.03     \\
55    &    1.19    &    0.08    &    ...   &    ...   &    0.10    &    0.01    &    ...   &    ...   &    0.54    &    0.11     \\
57    &    6.44    &    0.14    &    3.35    &    0.16    &    13.23    &    0.11    &    3.92    &    0.11    &    0.34    &    0.02     \\
58    &    0.81    &    0.07    &    ...   &    ...   &    0.29    &    0.02    &    ...   &    ...   &    1.00    &    0.14     \\
59    &    2.94    &    0.09    &    1.43    &    0.15    &    6.83    &    0.13    &    1.53    &    0.13    &    0.51    &    0.03     \\
61    &    2.14    &    0.08    &    1.88    &    0.15    &    11.87    &    0.31    &    4.46    &    0.31    &    0.50    &    0.04     \\
64    &    2.83    &    0.09    &    2.85    &    0.16    &    6.75    &    0.13    &    3.15    &    0.13    &    0.31    &    0.02     \\
65    &    3.13    &    0.12    &    1.47    &    0.15    &    7.99    &    0.28    &    2.53    &    0.28    &    0.46    &    0.03     \\
73    &    8.35    &    0.16    &    4.87    &    0.18    &    195.40    &    1.26    &    77.74    &    1.26    &    0.28    &    0.01     \\
74    &    2.21    &    0.08    &    1.71    &    0.15    &    5.56    &    0.14    &    1.62    &    0.14    &    0.47    &    0.03     \\
75    &    3.37    &    0.17    &    2.94    &    0.29    &    0.54    &    0.02    &    0.30    &    0.02    &    0.71    &    0.06     \\
78    &    3.78    &    0.10    &    3.87    &    0.19    &    5.05    &    0.08    &    2.15    &    0.08    &    0.48    &    0.03     \\
84    &    1.14    &    0.07    &    ...   &    ...   &    0.98    &    0.05    &    ...   &    ...   &    0.36    &    0.04     \\
88    &    2.35    &    0.08    &    1.08    &    0.14    &    47.59    &    1.07    &    5.60    &    1.07    &    0.33    &    0.03     \\
94    &    2.31    &    0.31    &    1.81    &    0.34    &    0.30    &    0.04    &    0.10    &    0.04    &    0.42    &    0.06     \\
99    &    1.14    &    0.07    &    4.42    &    0.24    &    0.73    &    0.04    &    0.91    &    0.04    &    0.44    &    0.05     \\
101    &    1.93    &    0.11    &    ...   &    ...   &    0.41    &    0.02    &    ...   &    ...   &    0.46    &    0.05     \\
106    &    0.60    &    0.06    &    1.18    &    0.15    &    2.19    &    0.20    &    1.34    &    0.20    &    0.39    &    0.10     \\
110    &    0.74    &    0.20    &    ...   &    ...   &    0.06    &    0.02    &    ...   &    ...   &    ...   &    ...    \\
111    &    0.63    &    0.06    &    1.00    &    0.14    &    10.30    &    0.91    &    4.71    &    0.91    &    0.49    &    0.09     \\
116    &    4.33    &    0.22    &    4.93    &    0.48    &    40.16    &    1.87    &    24.16    &    1.87    &    0.40    &    0.05     \\
118    &    2.36    &    0.09    &    2.97    &    0.25    &    0.52    &    0.02    &    0.24    &    0.02    &    0.75    &    0.06     \\
119    &    1.56    &    0.08    &    0.60    &    0.16    &    2.57    &    0.10    &    0.32    &    0.10    &    0.56    &    0.05     \\
120    &    3.04    &    0.10    &    3.25    &    0.20    &    1.42    &    0.03    &    0.59    &    0.03    &    0.48    &    0.03     \\
121    &    0.73    &    0.07    &    0.93    &    0.27    &    0.16    &    0.01    &    0.04    &    0.01    &    0.60    &    0.08     \\
123    &    4.63    &    0.11    &    0.44    &    0.15    &    21.68    &    0.26    &    1.05    &    0.26    &    0.55    &    0.03     \\
125    &    1.82    &    0.08    &    0.97    &    0.18    &    1.02    &    0.03    &    0.20    &    0.03    &    0.85    &    0.07     \\
126    &    1.12    &    0.08    &    4.46    &    0.25    &    0.60    &    0.03    &    1.02    &    0.03    &    0.30    &    0.05     \\
127    &    2.64    &    0.09    &    1.77    &    0.19    &    1.15    &    0.03    &    0.28    &    0.03    &    0.45    &    0.03     \\
128    &    1.30    &    0.09    &    ...   &    ...   &    0.13    &    0.01    &    ...   &    ...   &    0.97    &    0.15     \\
131    &    2.23    &    0.09    &    0.77    &    0.25    &    0.15    &    0.00    &    0.04    &    0.00    &    0.47    &    0.04     \\
133    &    2.26    &    0.09    &    0.68    &    0.15    &    2.28    &    0.06    &    0.58    &    0.06    &    0.51    &    0.04     \\
134    &    2.94    &    0.09    &    5.51    &    0.34    &    0.26    &    0.01    &    0.24    &    0.01    &    0.45    &    0.03     \\
136    &    1.04    &    0.08    &    7.28    &    0.35    &    0.43    &    0.03    &    0.82    &    0.03    &    0.60    &    0.09     \\
137    &    3.08    &    0.10    &    2.00    &    0.15    &    0.67    &    0.01    &    1.16    &    0.01    &    0.65    &    0.04     \\
\enddata
\tablenotetext{a} {EW$_{\lambda}$ values are in units of $\mu$m}
\tablenotetext{b} {F$_{\lambda}$ values are in units of 10$^{-12}$ $\times$ erg s$^{-1}$ cm$^{-2}$}
\tablecomments{Objects with unsatisfactory continuum fit are indicated with ...}
\end{deluxetable}
\begin{deluxetable}{lcccc}
\tabletypesize{\scriptsize}
\tablecaption{TDs and PTDs in NGC 1333 \label{tbl5}}
\tablewidth{0pt}
\tablehead{
\colhead{Index} & \colhead{$n_{2-6}$ vs. $n_{13-31}$} & \colhead{EW$_{10}$ vs. $n_{13-31}$} & \colhead{SED Analysis with Median} & \colhead{Classification} } 
\startdata
50  &  II &  PTD  & PTD  & PTD \\
52  &  II &  PTD  & PTD  & PTD \\
57  &  II &  PTD  & PTD  & PTD \\
73  &  II &  PTD  & PTD  & PTD \\
101  &  PTD  &  II & TD  & TD \\
110  &  PTD  &  TD  & TD  & TD \\
116  &  II &  PTD  & well above median  & PTD \\
136  &  small $n_{2-6}$  &  II & well below median  & TD \\
137  &  TD  &  TD & TD  & TD \\
\enddata
\tablecomments{Analysis which indicated the disk was unidentifiable from a radially continuous disk based on the given method is denoted with ``II"}
\end{deluxetable}

\end{document}